\documentclass[pdftex,epjc3,twocolumn]{svjour3} 
\RequirePackage{fix-cm} 
\smartqed  
\usepackage{subfigure}
\usepackage{color,amsmath,amssymb,amsfonts,amsbsy}
\usepackage{bbm}
\usepackage{bm}
\usepackage[final]{graphics}
\usepackage{epsfig}
\usepackage{sidecap}
\usepackage{hyperref}
\usepackage{enumerate}
\usepackage{lipsum}
\usepackage[sort&compress,numbers]{natbib}
\usepackage{xcolor}
\usepackage{cuted}
\usepackage{lineno}

\def\empile#1\over#2{\mathrel{\mathop{\kern 0pt#1}\limits_{#2}}}

\def\beq{\begin{equation}}
\def\eeq{\end{equation}}
\def\bea{\begin{eqnarray}}
\def\eea{\end{eqnarray}}
\definecolor{cyan}{rgb}{0.0, 0.255, 0.255}
\definecolor{darkcandyapplered}{rgb}{0.64, 0.0, 0.0}

\def\p{{\boldsymbol p}}

\def\pr{{^\prime}}

\def\d3p{\frac{d^3\p}{(2\pi)^3}E_\p}

\newcommand{\Lb}{\left(}
\newcommand{\Rb}{\right)}

\newcommand{\xpom}{{x_\mathbb{P}}}

\newcommand{\rtt}{ \mathbf{r}_{\mathbf{T}}^{2} }
\newcommand{\btt}{ \mathbf{b}_{\mathbf{T}}^{2} }
\newcommand{\rt}{{\mathbf{r_{T}}}}
\newcommand{\bt}{{\mathbf{b_{T}}}}

\newcommand{\Deltat}{{\boldsymbol{\Delta}}}

\newcommand{\ud}{\, \mathrm{d}}

\newcommand{\dint}{{\rm d}}

\definecolor{darkyellow}{rgb}{0.5, 0.5, 0.0}

\catcode`\@=11

\journalname{Eur. Phys. J.}

\begin{document}

\title{Scaling properties of exclusive vector meson production cross section from gluon saturation}

\author{Gregory Matousek\thanksref{e1,addr1,addr2} \and \\ Vladimir Khachatryan\thanksref{e2,addr1,addr2}
\and \\ Jinlong Zhang\thanksref{e3,addr3}}
\thankstext{e1}{e-mail:gregory.matousek@duke.edu}
\thankstext{e2}{vladimir.khachatryan@duke.edu}
\thankstext{e3}{jlzhang@email.sdu.edu.cn}

\institute{
\label{addr1} Physics Department, Duke University, Durham, NC 27708, USA \and
\label{addr2} Triangle Universities Nuclear Laboratory, Durham, NC 27708, USA \and
\label{addr3} Key Laboratory of Particle Physics and Particle Irradiation (MOE), Shandong University, Qingdao 266237, China 
}

\date{\today}

\maketitle

\begin{abstract}
It is already known from phenomenological studies that in exclusive deep-inelastic scattering off nuclei there appears to be a 
scaling behavior of vector meson production cross section in both nuclear mass number, $A$, and photon virtuality, $Q^{2}$, 
which is strongly modified due to gluon saturation effects. In this work we continue those studies in a realistic setup based 
upon using the Monte Carlo event generator Sar{\it t}re. We make quantitative predictions for the kinematics of the Electron-Ion
Collider, focusing on this $A$ and $Q^{2}$ scaling picture, along with establishing a small region of squared momentum transfer,
$t$, where there are signs of this scaling that may potentially be observed at the EIC. Our results are represented as pseudo-data
of vector meson production diffractive cross section and/or their ratios, which are obtained by parsing data collected by the
event generator through smearing functions, emulating the proposed detector resolutions for the future EIC.
\end{abstract}


\section{Introduction}
\label{sec:Intro}
Studies of the proton partonic structure using the most precise data provided by the H1 and ZEUS experiments at HERA 
facility \cite{Aaron:2009aa,Abramowicz:2015mha}, based upon deep inelastic scattering (DIS) measurements, have 
highly enriched our knowledge on small Bjorken-$x$ physics with findings of the rapid growth of gluon density at small 
longitudinal momentum fraction $x$. In particular, the HERA measurements in 
$e+p$ scattering at small $x \leq 10^{-2}$ have shown that the gluon number density seems to have an ``uncontrollably'' rising
nature because of which the gluonic part of the proton cross section dominates its total cross section. This strong growth 
of gluon density occurs in the space of high gluon occupancies of the order of $1/\alpha_{s}$, where $\alpha_{s}$ is the 
QCD coupling constant. It results in violation of the cross section unitarity bound, which can be tamed by introducing 
{\em gluon saturation} effects into the whole high energy scattering picture. 

The maximal gluon occupancy takes place for any $x$ value at small $x$, for which there is a corresponding saturation 
scale $Q_{s}(x)$, with $Q_{s}^{2} \gg \Lambda_{QCD}^{2}$ (where $\Lambda_{QCD}$ is the QCD intrinsic scale).
The saturation effects are nonlinear QCD phenomena that can be described by the Color Glass Condensate (CGC) 
effective theory \cite{GLR:1983,McVen:1994,AyJaMcVen:1996,Iancu:2001,Iancu:2002,Gelis:2007,Gelis:2010}. In particular, 
the data describing the proton structure function are well reproduced within CGC ansatz
\cite{Albacete:2010sy,Mantysaari:2018nng,Lappi:2013zma}\footnote{Refs.~\cite{Arsene:2004ux,Adams:2006uz,Braidot:2010ig,Adare:2011sc,Albacete:2010pg,Lappi:2012nh} 
discuss the first hints of the onset of gluon saturation stemming from $p(d)+A$ collisions at RHIC energies, meanwhile, there 
are also alternative explanations \cite{Strikman:2010bg,Kang:2011bp,Kang:2012kc} to consider.}. 

For direct studies of nonlinear QCD saturation phenomena it is necessary to have the $e+p$ systems being collided 
at center-of-mass energies far exceeding those reached at HERA because the proton's saturation scale, $Q_{s,p}^{2}(x)$, 
is not large enough at values of $x$ probed at HERA energies, and consequently the evidence for gluon saturation has not 
been very clear so far. Nonetheless, in the case of $e+A$ scattering one can look at higher gluon density effects, at energies 
by an order of magnitude lower than those used for $e+p$ scattering at HERA. In this case the nuclear saturation momentum 
and gluon density will scale as $Q_{s,A}^{2}(x) \sim Q_{s,p}^{2}(x)\!\times\!A^{1/3}$, by which the nonlinear effects in 
heavy nuclei shall be amplified efficiently. Probing a nucleus with large mass number $A$ is equivalent to probing the proton
at several times larger energy. Detailed studies like those accomplished in \cite{Kowalski:2007rw,Kowalski:2003hm} 
support the use of $A^{1/3}$ dependence in various phenomenological calculations and applications. The scale $Q_{s,A}^{2}(x)$ 
controls the nuclear dynamics at high energies.

Currently there are no available nuclear DIS data at small-$x$ region but the proposed Electron-Ion Collider (EIC) in the 
USA \cite{Boer:2011fh,Accardi:2012qut,Aschenauer:2017,AbdulKhalek:2021gbh} and Large Hadron Electron Collider (LHeC) at CERN 
\cite{AbelleiraFernandez:2012cc} will aim at directly measuring the saturation regime of large gluon densities in the upcoming 
high-precision EIC and LHeC era. In particular, the EIC White Paper~\cite{Accardi:2012qut} has already shown the potential 
of EIC to collide high energy electron and ion beams, providing unprecedented access to gluon dominated kinematic regions 
of nucleons and nuclei. Furthermore, strongly polarized electron and proton beams will unravel the spatial and spin structure of 
the proton. Ref.~\cite{Aschenauer:2017} scrutinizes the kinematc coverage of EIC and the energy dependence of key observables 
that are essential to assure solid and robust EIC program. The EIC Yellow Report \cite{AbdulKhalek:2021gbh} describes the 
program's physics case and the resulting detector requirements/concepts. 

But before these colliders come into their existence and operations, another possible way to study DIS on nuclei is provided by 
ultraperipheral $A+A$ and $p+A$ collisions (UPC), where relatively short-range strong interactions are suppressed by processes 
taking place in the nuclear periphery with large impact parameter. The data on {\it diffractive vector meson} production in such collisions 
\cite{Abbas:2013oua,Abelev:2012ba,Adam:2015gsa,Khachatryan:2016qhq,TheALICE:2014dwa,Chudasama:2016eck,Acharya:2018jua,Sosnov:2021ucz,Acharya:2021gpx,Acharya:2021ugn} 
demonstrate the sensitivity of this production process on nuclear effects at small $x$, since the pertinent 
measurements are quite sensitive to gluon distributions at saturation. The reason is that in perturbative QCD (pQCD), the hard 
diffractive cross section at leading order is proportional to gluon density squared \cite{Ryskin:1992ui}, which makes it the 
most sensitive probe to small-$x$ gluons, whereby the vector meson production becomes an extremely useful process to study the 
small-$x$ hadronic structure in general. Ref.~\cite{Bendova:2020hkp} shows that the contribution of diffractive events is 
enhanced in nuclear collisions, giving predictions for $e+p$ and $e+A$ collisions at the EIC/LHeC kinematics.

Phenomenological studies of \cite{Mantysaari:2017slo} (see also the references therein) have already demonstrated that 
the onset of gluon saturation is potentially observable in exclusive vector meson production off large nuclei in high-energy 
$e+A$ scattering. It is in particular shown that within CGC theory, the nuclear saturation effects significantly modify 
the $A$ and $Q^{2}$ scaling properties of the exclusive vector meson production cross section, if one passes from the pQCD regime 
($Q^{2} > Q_{s,A}^{2}$) to the saturation regime ($Q^{2} < Q_{s,A}^{2}$). In a diffractive scattering process (see the next 
section for more details) an electron probe scatters off a target proton or nucleus, where the exchanged virtual photon splits into a 
$q\bar{q}$ {\em dipole}. The dipole subsequently interacts with the target in the target's rest frame via a color-neutral vacuum excitation, 
{\em Pomeron}, which in pQCD is visualized as a colorless combination of two or more gluons. The parton longitudinal momentum fraction 
within color-neutral Pomeron (that is also transferred to the produced vector meson) is designated by $\xpom$, which in diffractive 
DIS is equivalent to the Bjorken $x$ for exclusive processes. 

Based on the aforementioned simple dipole interaction mechanism, advanced and elaborated dipole model frameworks have been 
developed in \cite{Kowalski:2003hm} and \cite{GolecBiernat:1998js,GolecBiernat:1999qd,Bartels:2002cj,Kowalski:2006hc}. Refs.~\cite{Kowalski:2003hm} 
and \cite{Kowalski:2006hc} have the impact parameter dependence introduced in their dipole models. The exclusive processes are also 
included in the dipole model of \cite{Kowalski:2006hc}, which goes by the name bSat or IPSat. A linearized dipole model (to the 
model of \cite{Kowalski:2006hc}) called bNonSat or IPNonSat, which separates and isolates the gluon saturation effects from other 
small-$x$ effects, is introduced in \cite{Kowalski:2003hm}. The IPSat and IPNonSat dipole models, for both protons and nuclei, are 
implemented into Monte Carlo event generator Sar{\em t}re \cite{Toll:2013gda,Toll:2012mb}, the purpose of which is to simulate 
diffractive exclusive vector meson production and deeply virtual Compton scattering (DVCS \cite{Bendova:2022}) events in $e+p$ and $e+A$ 
scatterings at EIC and LHeC center-of-mass energies\footnote{There is also another Monte Carlo generator, STARlight \cite{Klein:2016yzr}, 
which simulates a wide variety of vector meson final states, produced in $e+A$ scattering. The improved version of this generator, dubbed 
as eSTARlight, has been used for exclusive vector meson production studies at the EIC kinematics \cite{Lomnitz:2018juf}.}. 
The current (and upcoming) work on Sar{\em t}re includes simulations of events in $p+A$ and $A+A$ UPC \cite{Sambasivam:2019gdd}, 
simulations of inclusive processes, as well as simulations of diffractive exclusive events after {\em geometrical} and {\em saturation 
scale} fluctuations of gluon spatial distributions (see \cite{Mantysaari:2016ykx,Mantysaari:2016jaz,Mantysaari:2017dwh,Mantysaari:2020axf} 
for more details) are implemented into the generator's framework.

In this paper we partially continue the studies of Ref.~\cite{Mantysaari:2017slo} in a realistic EIC setup utilizing the Sar{\em t}re 
generator. The content of the paper is structured as follows. In Sec.~\ref{sec:IPSat} we describe the basics of diffractive scattering 
and IPSat dipole model, outlining also the IPNonSat model and phenomenological corrections to the diffractive scattering amplitude. 
In this regard,~\ref{sec:AppI} shows some details related to diffractive differential cross sections. 
In Sec.~\ref{sec:Scaling}, we first focus on a realistic view of the experimental measurement of coherent and incoherent cross 
sections, from exclusive vector meson production processes in diffractive $e+Au$ scattering, obtained with the integrated luminosity 
of $10\,{\rm fb}^{-1}/A$. Then, we discuss the $A$ and $Q^{2}$ scaling properties of exclusive vector meson production. 
We make quantitative predictions for the EIC kinematics by focusing on the scaling picture as a function of $Q^{2}$ 
at close-to-zero squared momentum transfer region of $|t| = [0.000 - 0.001]\,{\rm GeV^{2}}$, as well as at non-zero $t$ regions 
of $|t| = [0.003 - 0.004]\,{\rm GeV^{2}}$, $[0.006 - 0.007]\,{\rm GeV^{2}}$, $[0.009 - 0.010]\,{\rm GeV^{2}}$, and $[0.012 - 0.013]\,{\rm GeV^{2}}$. 
Throughout the paper unless specified otherwise, instead of these $t$ bins we will refer to their central values for simplicity: 
namely $|t| = 5\times10^{-4}\ {\rm GeV^{2}}$, $3.5\times10^{-3}\ {\rm GeV^{2}}$, $6.5\times10^{-3}\ {\rm GeV^{2}}$, 
$9.5\times10^{-3}\ {\rm GeV^{2}}$, and $1.25\times10^{-2}\ {\rm GeV^{2}}$. However, it should be always understood that we have 
performed Sar{\em t}re simulations in the corresponding $t$ bins, rather than at their central values. In the case of 
$|t| = 5\times 10^{-4}\,{\rm GeV^{2}}$, we consider two regimes of $Q^{2}$ as in \cite{Mantysaari:2017slo}, one for 
$Q^{2} > Q_{s,A}^{2}$ and another for $Q^{2} < Q_{s,A}^{2}$. 

We conclude on our results in Sec.~\ref{sec:outlook} afterwards, discussing the prospects of potential future developments as well.
We show our results in terms of pseudo-data on vector meson production cross sections and/or their ratios. We also calculate 
the uncertainties of the Sar{\em{t}}re-simulated pseudo-data based upon using detector resolutions and a smearing technique outlined 
in the EIC Detector Handbook \cite{Handbook} (see~\ref{sec:AppII}).

\section{IPSat dipole model in Monte Carlo event generator Sar{\bf\em{t}}re}
\label{sec:IPSat}

\subsection{Diffractive DIS picture}
\label{sec:DIS}
\begin{figure}[h!]
\begin{center}
\includegraphics[width=0.3\textwidth]{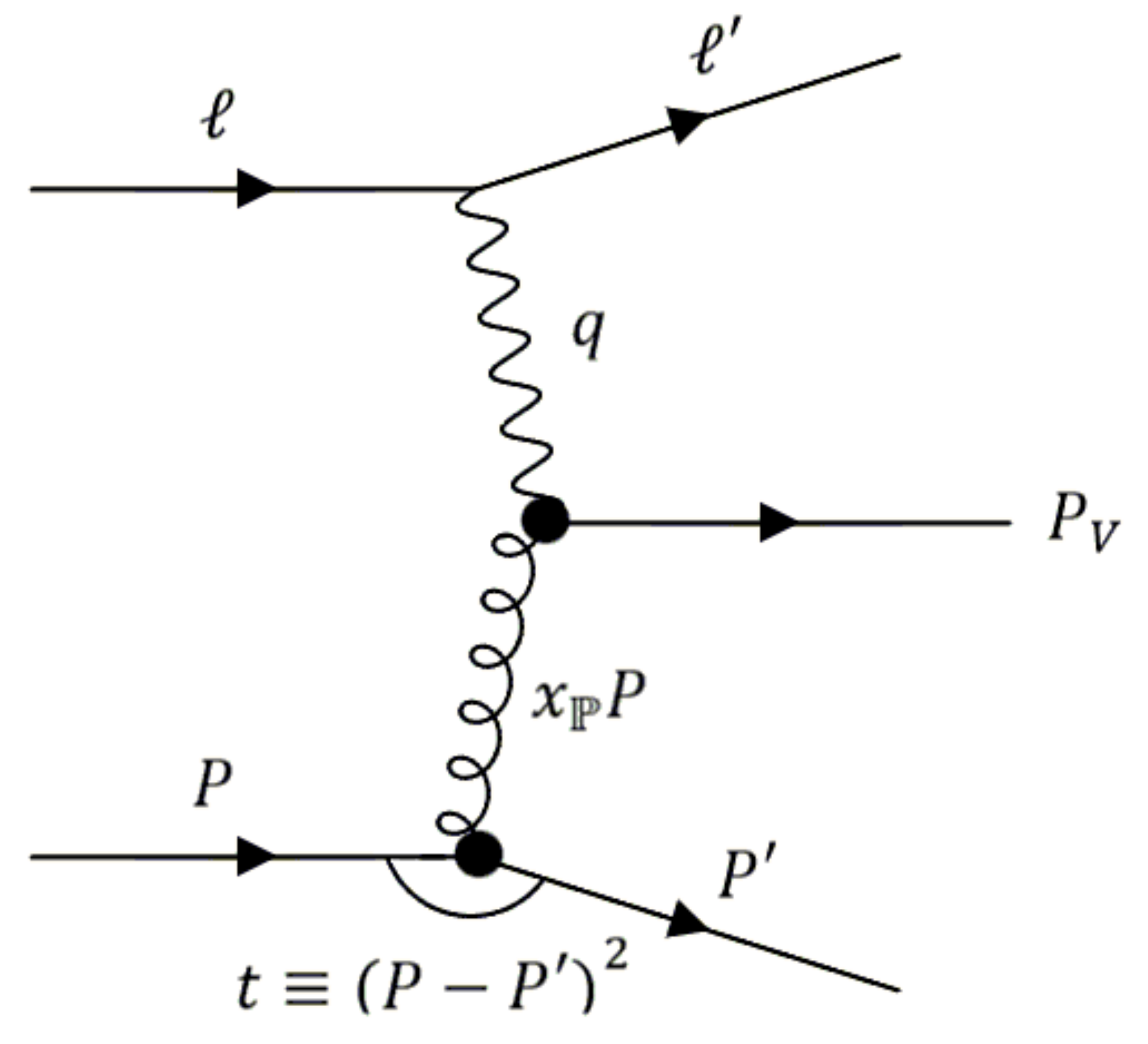} 
\end{center}
\caption{The ``curly'' line reflects a Pomeron exchange between the virtual photon and the target. The net color charge exchange with 
the target is zero, leading to a rapidity gap with the size equal to $\ln{\!(1/\xpom)}$ between the vector meson and the target, a gap in 
rapidity coverage, which is used to identify diffractive events experimentally. 
}
\label{fig:ddis}
\end{figure}

First let us briefly take a look at diffractive DIS kinematics of $e+p$ scattering. The scattering process of exclusive production of a 
vector meson $V$ with a momentum $P_{V}$ in DIS $e+p$ is given by
\beq
l(\ell) + p(P) \rightarrow l^{\prime}(\ell^{\prime}) + p^{\prime}(P^{\prime}) + V(P_{V}) ,
\eeq
where $\ell$ and $\ell^{\prime}$ are the electron's incoming and outgoing momenta, $P$ and $P^{\prime}$ are the proton's incoming 
and outgoing momenta (see Fig.~\ref{fig:ddis}). The scattering process is characterized by the following Lorentz invariant quantities:
\begin{displaymath}
Q^{2} \equiv -q^{2} = -(\ell-\ell^{\prime})^{2}, \,\,\,\,\,\,\,\,\,\,\,\,\,\,\,\,\,\,\,\, t \equiv (P - P^{\prime})^{2},
\end{displaymath}
\beq
\xpom \equiv \frac{(P - P^{\prime})\!\cdot\!q}{P\!\cdot\!q} = \frac{M_{V}^{2} + Q^{2} - t}{W^{2} + Q^{2} - m_{p}^{2}} ,
\eeq
where $q$ is the virtual photon momentum, $m_{p}$ is the proton mass, $M_{V}$ is the mass of the produced vector 
meson $V$, and $W^{2}=(P + q)^{2}$ is the total center-of-mass energy squared of the $\gamma^{\ast}$-$p$ scattering. The 
scattered proton with the momentum $P^{\prime}$ can either remain intact or break up, leading to {\it coherent} and {\it incoherent} 
diffractive events, respectively. 

The spatial distribution of gluons at small $x$ will be studied experimentally at EIC kinematics. In order to obtain this distribution one 
should measure the diffractive cross section, $d\sigma/dt$, at small-$x$ region over a large range of $t$. Then, a Fourier transform 
from momentum space to coordinate space will represent the gluon source distribution as a function of the impact parameter 
${\bf b_{T}}$. 
There can be an access to $t$ with sufficiently high precision from measurements of exclusive diffractive processes, such as exclusive 
vector meson production and DVCS. In this case $t$ can be calculated from the measured $V$ and $\gamma$, the scattered electron, and the 
known beam energies.

\subsection{Sar{\bf\em{t}}re and IPSat dipole model basics}
\label{sec:IPSat_Sartre}
The purpose of the dipole model Monte Carlo event generator Sar{\em{t}}re is to provide simulations of pseudo-data at kinematics in 
which future data are supposed to be taken at the proposed EIC~\cite{Boer:2011fh,Accardi:2012qut,Aschenauer:2017} and LHeC 
\cite{AbelleiraFernandez:2012cc} machines. With Sar{\em{t}}re one can simulate diffractive exclusive vector meson production and DVCS in the following processes:
\bea
& & e+p \rightarrow e^{\prime} + V(\gamma) + p^{\prime}~~~~
e+A \rightarrow e^{\prime} + V(\gamma) + A^{\prime} 
\nonumber\\
& & p+p \rightarrow p^{\prime} + V(\gamma) + p^{\prime}~~~~
p+A \rightarrow p^{\prime} + V(\gamma) + A^{\prime}
\nonumber\\
& & ~~~~~~~~~~~~~~~
A+A \rightarrow A^{\prime} + V(\gamma) + A^{\prime} ,
\label{eq:processes}
\eea
where all the processes are mediated by a virtual photon ($\gamma^{\ast}$) and/or a Pomeron. The vector meson $V$ can be a $J/\Psi$, 
$\phi$ or $\rho$ particle, $\gamma$ is the DVCS real photon. However, the generator is restricted for studying these processes at 
$\xpom< 10^{-2}$ and at large $\beta \equiv x_{I\!\!P}/x$ because the IPSat and IPNonSat dipole models, implemented in it, are only 
valid for small values of $x_{I\!\!P}$ and not for too small values of $\beta$. If $\beta$ becomes too small, the dipole becomes unphysically 
large \cite{Kowalski:2008sa}\footnote{In any case the program has an imposed cut-off on the dipole size, $r$, against its any unphysical
increase. For nuclei it is $r < 3R_{N}$, where $R_{N}$ is the nuclear radius given in the Woods-Saxon parametrization. For the proton it is 
$r <3$\,fm. This cut-off does not show any changes in final simulated cross sections, though it can be altered in a broad kinematic range.}.

As far as the IPSat dipole model is concerned, it has been very successful in describing the exclusive vector meson and photon production 
at HERA. In this section we represent some of the features of this dipole model \cite{Kowalski:2006hc} implemented in Sar{\em{t}}re 
\cite{Toll:2013gda,Toll:2012mb}, for simulating $V$ and $\gamma$. We will be following the line of discussions and argumentations of 
\cite{Toll:2013gda,Toll:2012mb,Klein:2016yzr,Lomnitz:2018juf,Sambasivam:2019gdd,Mantysaari:2016ykx,Mantysaari:2016jaz}.

\subsubsection{IPSat framework for $e+p$ and $e+A$ scatterings}
\label{sec:ep}
The DIS diffraction can be described in terms of states, which diagonalize the scattering matrix ($S$-matrix)~\cite{Good:1960ba}.  
In these states a virtual photon $\gamma^{\ast}$ at high energies fluctuates into a ${q\bar q}$ dipole, with a fixed dipole transverse 
size ${\bf r_{T}}$ and an impact parameter ${\bf b_{T}}$, along with a given specific configuration of the target. Then, the scattering 
cross section can be obtained by averaging over multiple target configurations. One can average on the level of scattering amplitude, 
where the cross section is proportional to the average target density, or otherwise stated, to the average gluon density. This 
averaging corresponds to coherent diffraction \cite{Kovchegov:1999kx}, where the target remains intact. One can also average 
on the level of scattering cross section that includes events in which the target breaks up. This averaging corresponds to total 
diffraction. Subtraction of the coherent from the total cross section gives the incoherent diffractive cross section, which describes 
only broken up target remnants. The incoherent diffraction is proportional to the target profile's variance 
\cite{Frankfurt:2008vi,Caldwell:2010zza,Miettinen:1978jb}.

\begin{figure}[h!]
\begin{center}
\includegraphics[width=0.485\textwidth]{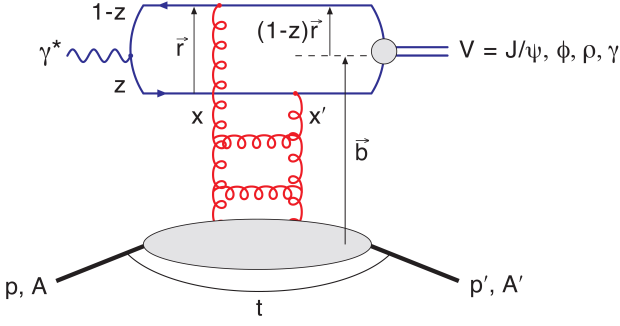} 
\end{center}
\caption{(Color online) A schematic view of the dipole-target scattering and its variables in $e+p$ and $e+A$ scatterings. 
In the figure $\vec{r} \equiv \rt$ and $\vec{b} \equiv \bt$. This figure is from \cite{Toll:2013gda}, and is a more 
detailed version of Fig.~\ref{fig:ddis}.}
\label{fig:dipole_target}
\end{figure}

\noindent
\paragraph{$e+p$ scattering.}
Let us now discuss this entire picture in more technical terms, starting with the amplitude for diffractively producing an exclusive 
vector meson $V$ (or a DVCS $\gamma$) in an interaction between the virtual photon $\gamma^{\ast}$ and proton $p$ 
\cite{Kowalski:2006hc,Klein:2016yzr,Lomnitz:2018juf,Sambasivam:2019gdd,Mantysaari:2016ykx,Mantysaari:2016jaz}:
\bea
& & \mathcal{A}_{T,L}^{\gamma^{*}p\,\rightarrow\,V p}(x_{I\!\!P},Q^{2},\Deltat) =
\nonumber\\
& & = i\int{\rm d}^{2}{\bf r_{T}} \int{\rm d}^{2}\bt \int\frac{{\rm d} z}{4\pi}
\left[\Lb \Psi\Psi_{V}^{*} \Rb_{T,L}\!(Q^{2},r_{T}, z)\right] \times
\nonumber \\
& & ~~~~~~~~~~~~~~\times e^{-i[\bt - (1 - z)\rt] \cdot \Deltat}\,\,\frac{{\rm d}\sigma_{q\bar q}^{p}}{{\rm d}^{2}\bt}(x_{I\!\!P}, r_{T}, b_{T}) ,
\label{eq:ttepamplitude}
\eea
where the subscripts $T/L$ refer to the transversely and longitudinally polarized $\gamma^{\ast}$; \,$Q^2$ is the photon virtuality;
$\Deltat = (P^{\prime} - P)_{T}$ is the momentum transfer in the scattering process assumed to be $|\Deltat| \approx \sqrt{-t}$; $z$ 
and $1-z$ are the longitudinal momentum fractions of the dipole taken by the quark and antiquark, respectively\footnote{The phase
factor $\exp\!{[i(1 - z) \rt \cdot \Deltat]}$ standing in Eq.~(\ref{eq:ttepamplitude}) stems from the difference between forward and 
non-forward wavefunctions. Although used in many phenomenological applications, the validity of the exponent $(1 - z) \rt \cdot \Deltat$ 
is challenged in \cite{Hatta:2017cte}, nevertheless, we keep it in our framework because it is anyway expected to have a negligible effect 
at small-$x$ region \cite{Mantysaari:2017slo}.}; 
$\Lb \Psi^{*}\Psi_{V} \Rb_{T,L}\!(Q^{2},r_{T}, z)$ is the wave-function overlap between the incoming $\gamma^{\ast}$ and outgoing 
$V$ or $\gamma$; \,$\sigma_{q\bar q}^{p}(x_{I\!\!P}, r_{T}, b_{T})$ is the ${q\bar q}$-$p$ cross section describing the dipole scattering 
off the target proton\footnote{This cross section is Fourier transformed into momentum space with $\Deltat$, which is the Fourier 
conjugate to the center-of-mass of the dipole, $\bt - (1-z)\rt$, relative to the proton's center 
(see Fig.~\ref{fig:dipole_target} for details).}, as given by \cite{Kowalski:2003hm}:
\bea
& & \frac{{\rm d}\sigma_{q\bar q}^{p}}{{\rm d}^{2}\bt}(x_{I\!\!P}, r_{T}, b_{T}) = 
\nonumber\\
& & ~~~~~~~~~~= 2\,\mathcal{N}^{p}(x_{I\!\!P}, r_{T}, b_{T})
= 2\left[1-\Re(S)\right] ,
\label{eq:epcs}
\eea
where $\mathcal{N}^{p}$ is the scattering amplitude (with $ r_{T} = |\rt|$ and $b_{T} = |\bt|$), and $\Re(S)$ is the real part of 
the $S$-matrix. The amplitude in turn is given by
\beq
\mathcal{N}^{p}(x_{I\!\!P}, r_{T}, b_{T}) = 1 - \exp\!\Lb -\rtt F(x_{I\!\!P}, \rtt)\,T_{p}(b_{T}) \Rb ,
\label{eq:epsa}
\eeq
where $T_{p}(b_{T})$ is the proton transverse density profile function assumed to be Gaussian:
\beq
T_{p}(b_{T}) = \frac{1}{2\pi B_{p}}\,e^{-\btt/(2B_{p})} .
\label{eq:Tp}
\eeq
The parameter $B_{p}$ is called proton width. The function $F(x_{I\!\!P}, \rtt)$ in Eq.~(\ref{eq:epsa}) is proportional to gluon 
distribution, which undergoes DGLAP evolution \cite{Bartels:2002cj}:
\beq
F(x_{I\!\!P}, \rtt) = \frac{\pi^{2}}{2N_{c}}\,\alpha_{\rm s}(\mu^{2})\,x_{I\!\!P}g(x_{I\!\!P}, \mu^{2}) .
\label{eq:F}
\eeq
For initial gluon density one uses a parametrization $x_{I\!\!P}g(x_{I\!\!P}, \mu^{2}) = A_{g}x_{I\!\!P}^{-\lambda_g}(1 - x_{I\!\!P})^{6}$, 
where $\mu^{2} = (4/\rtt) + \mu_{0}^{2}$ with $\mu_{0}^{2}$ being a cut-off scale in the gluon DGLAP evolution. 
The model parameters $A_{g}$ and $\lambda_{g}$ are determined by fitting the IPSat and IPNonSat models to HERA DIS data 
\cite{Mantysaari:2018nng}. $B_{p} = 4\,{\rm GeV}^{-2}$ and $\mu_0^{2} = 1.1$\,GeV$^2$ are fixed in the fitting process. For $A_{g}$ 
and $\lambda_{g}$ as well as for the used quark masses (treated as parameters in the model) we refer to Ref.~\cite{Mantysaari:2018nng}. 
The QCD running coupling $\alpha_{\rm s}$ takes into account some next-to-leading log effects but generally IPSat is a multiple 
two-gluon exchange model at leading log.

The total diffractive $\gamma^{\ast}$-$p$ cross section is given by averaging the absolute square of the scattering amplitude: 
\bea
& & \Lb \frac{\dint\sigma^{\gamma^{*}p\,\rightarrow\,V p}_{T, L} }{\dint t} \Rb_{\rm tot.\,diff.} = 
\nonumber\\
& & ~~~~~~~~~~= \frac{1}{16\pi} \left< \left| \mathcal{A}_{T,L}^{\gamma^{*}p\,\rightarrow\,V p}(x_{I\!\!P},Q^{2},t) \right|^{2} \right> .
\label{eq:gamma_p_tot}
\eea
The coherent diffractive $\gamma^{\ast}$-$p$ cross section is given by averaging the amplitude and taking the absolute square of it: 
\bea
& & \Lb \frac{\dint\sigma^{\gamma^{*}p\,\rightarrow\,V p}_{T, L} }{\dint t} \Rb_{\rm coh.\,diff.} = 
\nonumber\\
& & ~~~~~~~~~~= \frac{1}{16\pi} \left| \left< \mathcal{A}_{T,L}^{\gamma^{*}p\,\rightarrow\,V p}(x_{I\!\!P},Q^{2},t) \right> \right|^{2} .
\label{eq:gamma_p_cohdiff}
\eea
Then, the incoherent diffractive $\gamma^{\ast}$-$p$ cross section can be written as the following variance 
\cite{Frankfurt:2008vi,Caldwell:2010zza,Miettinen:1978jb} (as it was already mentioned before):
\bea
& & \Lb \frac{\dint\sigma^{\gamma^{*}p\,\rightarrow\,V p}_{T, L} }{\dint t} \Rb_{\rm incoh.\,diff.} =
\nonumber\\
& & = \Lb \frac{\dint\sigma^{\gamma^{*}p\,\rightarrow\,V p}_{T, L} }{\dint t} \Rb_{\rm tot.\,diff.} - 
\Lb \frac{\dint\sigma^{\gamma^{*}p\,\rightarrow\,V p}_{T, L} }{\dint t} \Rb_{\rm coh.\,diff.} \Rightarrow
\nonumber\\
& & ~~~ \Rightarrow
\frac{1}{16\pi} \left(\,\,\left< \left| \mathcal{A}_{T,L}^{\gamma^{*}p\,\rightarrow\,V p}(x_{I\!\!P},Q^{2},t) \right|^{2} \right> - \right.
\nonumber\\
& & ~~~~~~~~~~~~~~~
- \left. \left| \left< \mathcal{A}_{T,L}^{\gamma^{*}p\,\rightarrow\,V p}(x_{I\!\!P},Q^{2},t) \right> \right|^{2}\,\,\right) .
\label{eq:gamma_p_incohdiff}
\eea
Thereby, calculating the incoherent and coherent diffractive cross sections becomes a matter of finding the second and
first moments of the scattering amplitude, respectively.

\noindent
\paragraph{Extension to $e+A$ scattering.}
The explicit $\bt$ dependence of IPSat makes it possible to consider a nucleus as a collection of nucleons, and model it based upon a given 
nuclear transverse density distribution. At small-$x$ values the dipole has a large life-time such that it passes through the entire longitudinal
scope of the nucleus, where the nucleus is considered to be a two-dimensional object in the transverse plane. The exact position of each
nucleon within the nucleus, nevertheless, is not an observable quantity. That is why in order to calculate the total diffractive 
$\gamma^{\ast}$-$A$ cross section correctly, one has to average the squared amplitude over all possible states of nucleon configurations
(designated by $\Omega$):
\bea
& & \Lb \frac{\dint\sigma^{\gamma^{*}A\,\rightarrow\,V A}_{T, L} }{\dint t} \Rb_{\rm tot.\,diff.} =
\nonumber\\
& & ~~~~~~~~~~= \frac{1}{16\pi}
\left< \left|\mathcal{A}_{T,L}^{\gamma^{*}A\,\rightarrow\,V A}(x_{I\!\!P},Q^{2},t,\Omega)\right|^{2} \right>_{\Omega} .
\label{eq:gamma_A_tot}
\eea
Analogously to Eq.\,(\ref{eq:gamma_p_cohdiff}), the coherent diffractive $\gamma^{\ast}$-$A$ cross section will be given by
\bea
& & \Lb \frac{\dint\sigma^{\gamma^{*}A\,\rightarrow\,V A}_{T, L} }{\dint t} \Rb_{\rm coh.\,diff.} = 
\nonumber \\
& & ~~~~~~~~~~ = \frac{1}{16\pi}
\left| \left< \mathcal{A}_{T,L}^{\gamma^{*}A\,\rightarrow\,V A}(x_{I\!\!P},Q^{2},t,\Omega) \right>_{\Omega} \right|^{2} ,
\label{eq:gamma_A_cohdiff}
\eea
and analogously to Eq.\,(\ref{eq:gamma_p_incohdiff}), the incoherent diffractive $\gamma^{\ast}$-$A$ cross section will be represented 
as a dipole-nucleus variance given by

\bea
& & \Lb \frac{\dint\sigma^{\gamma^{*}A\,\rightarrow\,V A}_{T, L} }{\dint t} \Rb_{\rm incoh.\,diff.} =
\nonumber\\
& & = \Lb \frac{\dint\sigma^{\gamma^{*}A\,\rightarrow\,V A}_{T, L} }{\dint t} \Rb_{\rm tot.\,diff.} - 
\Lb \frac{\dint\sigma^{\gamma^{*}A\,\rightarrow\,V A}_{T, L} }{\dint t} \Rb_{\rm coh.\,diff.} \Rightarrow
\nonumber\\
& & ~~~ \Rightarrow
\frac{1}{16\pi} \left(\,\,\left< \left| \mathcal{A}_{T,L}^{\gamma^{*}A\,\rightarrow\,V A}(x_{I\!\!P},Q^{2},t,\Omega) \right|^{2} \right>_{\Omega} - \right.
\nonumber\\
& & ~~~~~~~~~~~~~~~
- \left. \left| \left< \mathcal{A}_{T,L}^{\gamma^{*}A\,\rightarrow\,V A}(x_{I\!\!P},Q^{2},t,\Omega) \right>_{\Omega} \right|^{2}\,\,\right) .
\label{eq:gamma_A_incohdiff}
\eea

For the ${q\bar q}$-$p$ cross section we have the scattering amplitude $\mathcal{N}^{p}$ from Eq.~(\ref{eq:epsa}). For the 
${q\bar q}$-$A$ cross section one should use the following approximation to construct the nuclear scattering amplitude from that of
the proton:
\bea
& & \mathcal{N}^{A}(x_{I\!\!P}, r_{T}, b_{T}) = 
\nonumber\\
& & 
~~~~~= 1 - \prod_{i=1}^{A} \left( 1 - \mathcal{N}^{p}(x_{I\!\!P}, r_{T}, |\bt - \bt_i|) \right) ,
\label{eq:eptoea}
\eea
where $\bt_i$ is the position of each nucleon within the nuclear transverse plane. The nucleon positions are treated according 
to projections of a three-dimensional Woods-Saxon function onto the transverse plane. Making use of
Eqs.~(\ref{eq:epcs}),~(\ref{eq:epsa}),~(\ref{eq:F})~and~(\ref{eq:eptoea}) gives us the ${q\bar q}$-$A$ cross section, which is
expressed by
\bea
& & \frac{{\rm d}\sigma_{q\bar q}^{A}}{{\rm d}^{2} \bt}(x_{I\!\!P}, r_{T}, b_{T}, \Omega_{j}) =
\nonumber\\
& & ~~~~~= 2 \left[ 1 - \exp\!\Lb -\frac{\pi^{2}}{2N_{c}} \rtt \alpha_{s}(\mu^{2})\,x_{I\!\!P}g(x_{I\!\!P},\mu^{2}) \times \right. \right.
\nonumber\\
& & ~~~~~~~~~~~~~~~~~~~~~~~ \left. \left. \times \sum_{i=1}^AT(|\bt - \bt_i|) \Rb \right] ,
\label{eq:bSateA}
\eea
where $\Omega_{j} = \{ |\bt_1, \bt_{2}, \dots, \bt_{A}| \}$ represents a specific Woods-Saxon nucleon configuration.
And here is a formula for the dipole cross-section average \cite{Kowalski:2003hm}, which should be used in calculations of 
the first moment of the amplitude:
\beq
\left< \frac{{\rm d}\sigma_{q\bar q}^{A}}{{\rm d}^{2} \bt} \right>_{\Omega} = 
2\left[ 1 - \left(1 - \frac{T_A(b_{T})}{2}\sigma^{p}_{q\bar q}\right)^{A} \right] ,
\label{eq:analytical}
\eeq
where $\sigma^p_{q\bar q}$ is integrated over $\bt$, and $T_{A}(b_{T})$ is the nuclear transverse density profile function, which is taken 
to be the Woods-Saxon potential in the transverse plane. We continue the discussion of this section in~\ref{sec:AppI}, where we generalize
and show how Sar{\em{t}}re calculates (simulates) the $\gamma^{\ast}$-$A$ diffractive differential cross sections.

Along with the IPSat framework, Sar{\em{t}}re also has the IPNonSat dipole model implemented in it. The purpose of this model is 
to separate the gluon saturation effects from other small-$x$ effects. Eq.~(\ref{eq:epsa}) has an exponential term through which 
the gluon saturation is introduced in IPSat. Its non-saturation version is constructed if the dipole-target cross section is linearized. 
This means that one should keep the first term in the expansion of the exponent in the IPSat dipole-target cross section, to 
obtain that for IPNonSat. It results in gluon density becoming unsaturated for small $x_{I\!\!P}$ as well as when the ratio 
$\beta = x_{I\!\!P}/x$ is also large. In IPSat the rise of the cross section at large $\rt$ is under control, whereas in IPNonSat there 
is no taming for such a rise. Fore more details on the IPNonSat framework, we refer to Refs.~\cite{Kowalski:2003hm,Toll:2013gda}.

\subsubsection{Phenomenological corrections to dipole-target cross sections}
\label{sec:corrections}
\paragraph{Real part of the diffractive amplitude.}
In derivation of the $\gamma^{\ast}$-$p$ ($\gamma^{\ast}$-$A$ in general) diffractive scattering amplitude shown in Eq~(\ref{eq:ttepamplitude}), 
the assumption is that the amplitude is imaginary. Meanwhile, $\mathcal{N}^{p}$ in Eq.~(\ref{eq:epcs}), where only $\Re(S)$ is included, is purely real. 
Nonetheless, the real part of the diffractive scattering amplitude can be taken into account in a dipole-target final calculated cross section, 
if the cross section is multiplied by a coefficient $(1+B^{2})$, where $B$ is the ratio of the real and imaginary parts of the scattering amplitude 
\cite{Kowalski:2006hc}. The resulting expression for $B$ is given by
\beq
B = \tan\!\Lb \frac{\pi \lambda}{2} \Rb,~\mbox{with}~
\lambda = \frac{\partial \ln\!\Lb \mathcal{A}_{T,L}^{\gamma^{*} p(A)\,\rightarrow\,V p(A)} \Rb}{\partial \ln\!\Lb 1/\xpom \Rb} .
\label{eq:realpart_beta}
\eeq

\paragraph{Skewness correction.}
In the IPsat model, the diagonal ({\it collinear factorization}) gluon distribution, $\xpom g(\xpom, \mu^2)$, should be corrected to 
correspond to the off-diagonal ({\it skewed}) gluon distribution, which depends on longitudinal momentum fractions $x_{1}$ 
and $x_{2}$ of two gluons in a two-gluon exchange at lowest order in a dipole-target  (proton or nucleus) scattering\footnote{In 
a dipole-target scattering there is no exchange of color charge, which is already mentioned in the introduction.}, where $x_{1}$ 
and $x_{2}$ satisfy the condition of $x_{1} - x_{2} = \xpom$. In the high energy limit, the $x_{1}$ gluon exchange brings the 
$q \bar q$ dipole mass close to $M_{V}$ because the dominant contribution of the diffractive scattering amplitude is obtained when 
the intermediate propagators are close to the mass shell. In this case the second gluon is left with a significantly smaller $x_{2}$, 
and the dominant kinematic regime will be at $x_{2} \ll x_{1} \approx \xpom$ \cite{Martin:1997wy,Shuvaev:1999ce,Martin:1999wb}. 

Thereby, in order to account for scenarios, where the gluons in the two-gluon exchange carry different longitudinal momentum fractions, 
the so-called skewness correction should be applied to a dipole-target final calculated cross section, by multiplying it with a coefficient 
$R_{g}$ \cite{Kowalski:2006hc}, represented by
\bea
& & R_{g} = \frac{2^{2\lambda_{g} + 3}}{\sqrt{\pi}} \frac{\Gamma(\lambda_{g} + 5/2)}{\Gamma(\lambda_{g} + 4)},
\nonumber\\
& & \mbox{with}~\lambda_{g} = \frac{\partial \ln\!\Lb \xpom g(\xpom,\mu^{2}) \Rb}{\partial \ln\!\Lb 1/\xpom \Rb} .
\label{eq:skew}
\eea
Both corrections discussed in this section grow drastically in the large-$\xpom$ range outside the validity of theoretical models, 
where $\xpom > 0.01$. For this reason, the upper limit imposed on $\xpom$ in Sar{\em{t}}re is set to 0.01.

\section{The cross section scaling in production of exclusive diffractive vector mesons in $e+A$ scatterings at EIC kinematics}
\label{sec:Scaling}
In this section, we exhibit our results represented as pseudo-data from Sar{\em{t}}re-made vector meson production events. 
We show the data uncertainties obtained from a combined framework of using proposed detector resolutions and a smearing technique
\cite{Handbook}. We refer to~\ref{sec:AppII} for more details on the error analysis and producing pseudo-data.

\subsection{Coherent and incoherent diffractive cross section distributions}
\label{sec:White_paper}

Before discussing how the nuclear saturation effects modify the $A$ and $Q^{2}$ scaling properties of the exclusive vector meson 
production cross section \cite{Mantysaari:2017slo} in an EIC-like kinematical setup, it is relevant to show several  plots on 
$d\sigma/dt$ coherent and incoherent distributions for exclusive $J/\Psi$ and $\phi$ vector meson production in diffractive $e+Au$ 
scattering. This production is the experimentally cleanest tool in such processes, due to a small particle number in the final state,
by which one can systematically investigate the saturation physics. In this regard, Ref.~\cite{Krelina:2019gee} demonstrates a 
comprehensive analysis of the $A$, $Q^{2}$, energy, and momentum transfer dependencies of the exclusive vector meson 
coherent and incoherent cross sections.

\begin{figure}[h!]
\begin{center}
   \subfigure{\includegraphics[width=0.5\textwidth,height=0.430\textwidth]{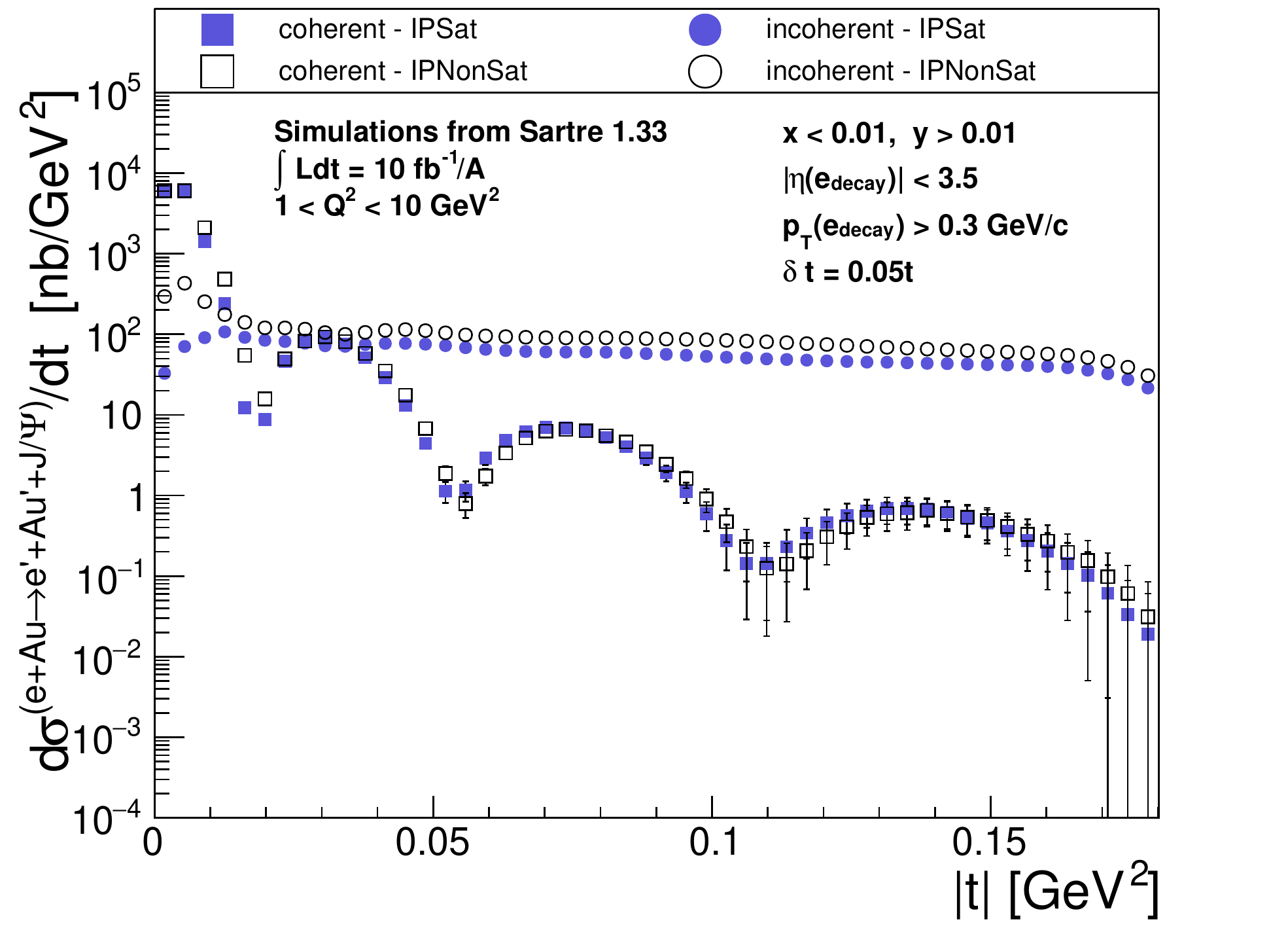}}
\hspace{0.0\textwidth}
   \subfigure{\includegraphics[width=0.5\textwidth,height=0.430\textwidth]{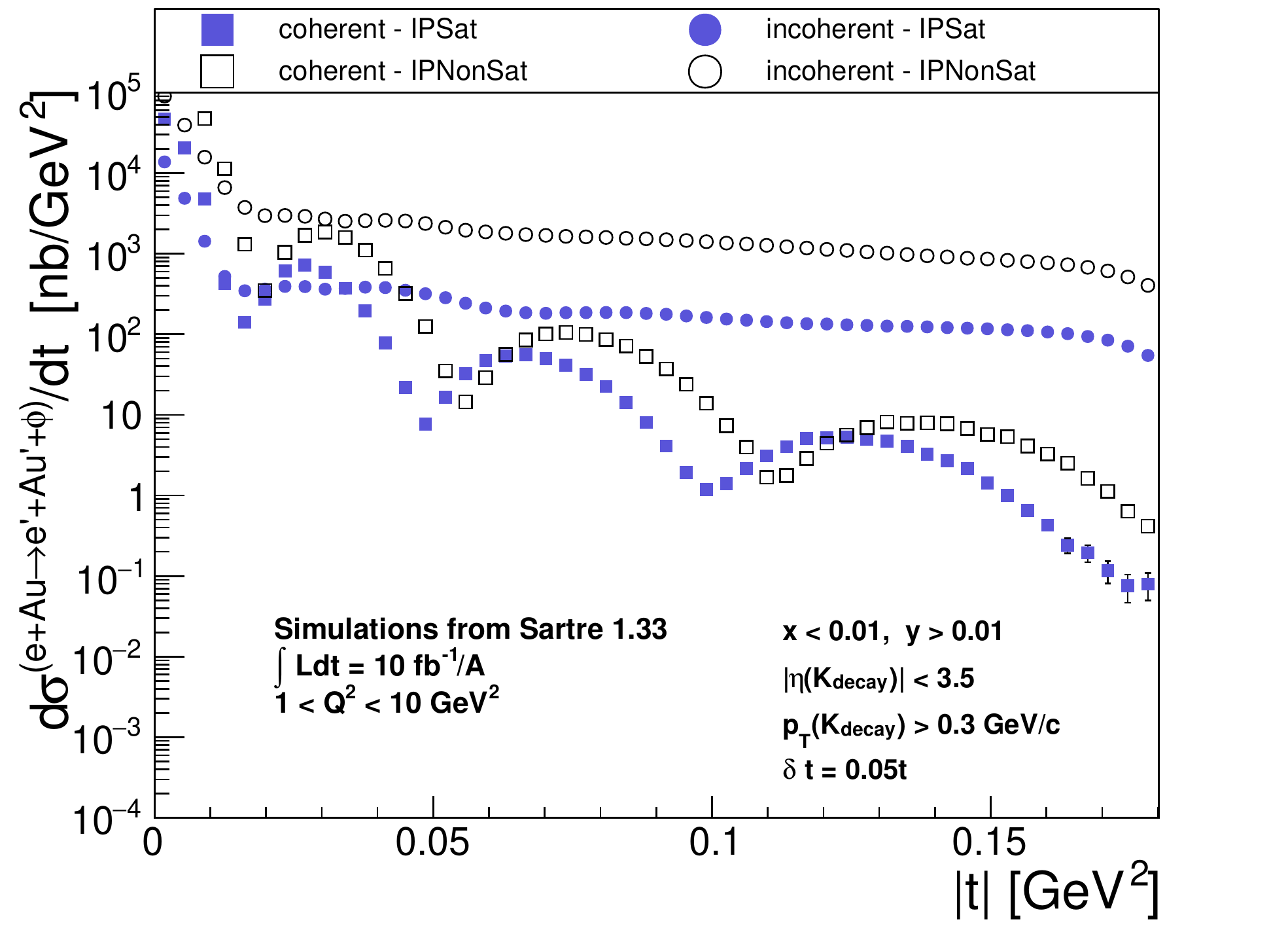}}
\end{center}
\vspace{-0.25cm}
\caption{(Color online) Coherent and incoherent cross sections, from exclusive $J/\Psi$ (top) and $\phi$ (bottom) production 
processes in diffractive $e+Au$ scattering, by having outputted the cross-section pseudo-data from Sar{\em{t}}re and passed it through 
detector resolutions described in the EIC Detector Handbook \cite{Handbook} with the smearing $\delta t = 0.05t$, within the validity range of the 
saturation (IPSat) and non-saturation (IPNonSat) models. The $\eta$ cut helps determine which direction a particle will move towards, 
and the $p_{T}$ cut helps find out if the particle will even reach the detectors.}
\label{fig:white_plot2}
\end{figure}

The coherent distribution depends on a target shape, by which one can study the nuclear spatial gluon distributions. The incoherent
distribution provides crucial information on geometric fluctuations of a target \cite{Mantysaari:2016ykx,Mantysaari:2016jaz,Mantysaari:2017dwh,Mantysaari:2020axf,Chang:2021jnu}. 
Experimentally, the clue to the spatial gluon distribution and fluctuations is thereby the measurements of the $d\sigma/dt$ distributions. 
Figure 54 of the EIC White Paper \cite{Accardi:2012qut} shows Sar{\em{t}}re-simulated $d\sigma/dt$ distributions 
for exclusive $J/\Psi$ and $\phi$ production, in both coherent and incoherent\footnote{It is shown in \cite{Lappi:2010dd} how 
the incoherent cross section is affected by saturation effects that is not negligible for $J/\Psi$ production.} events, in diffractive 
$e+Au$ scattering.
The simulations were restricted to $1\,{\rm GeV}^{2} < Q^2 < 10\,{\rm GeV}^{2}$ and $x < 0.01$. Besides, the produced events 
were passed through an experimental filter, and scaled for mimicking the integrated luminosity of $10\,{\rm fb}^{-1}/A$. Also, a simple 
5\% smearing in $t$ (meaning 5\% relative uncertainty in the event $t$ reconstruction) was performed on the simulated data. For the
comparison purpose, we replicate the two plots in that Figure~54 with the same smearing in $t$, which are shown in Fig.~\ref{fig:white_plot2}. 
The basic experimental cuts in these plots are listed in their legends, similar to those in Figure~54 of \cite{Accardi:2012qut} or in 
Figure~7.83 of \cite{AbdulKhalek:2021gbh}. We make the cut ${\rm p_{T}(e_{decay},K_{decay})} > 0.3\,{\rm GeV/c}$ to match the requirement 
of the EIC tracking detectors \cite{AbdulKhalek:2021gbh}.
\begin{figure}[h!]
\begin{center}
   \subfigure{\includegraphics[width=0.5\textwidth,height=0.430\textwidth]{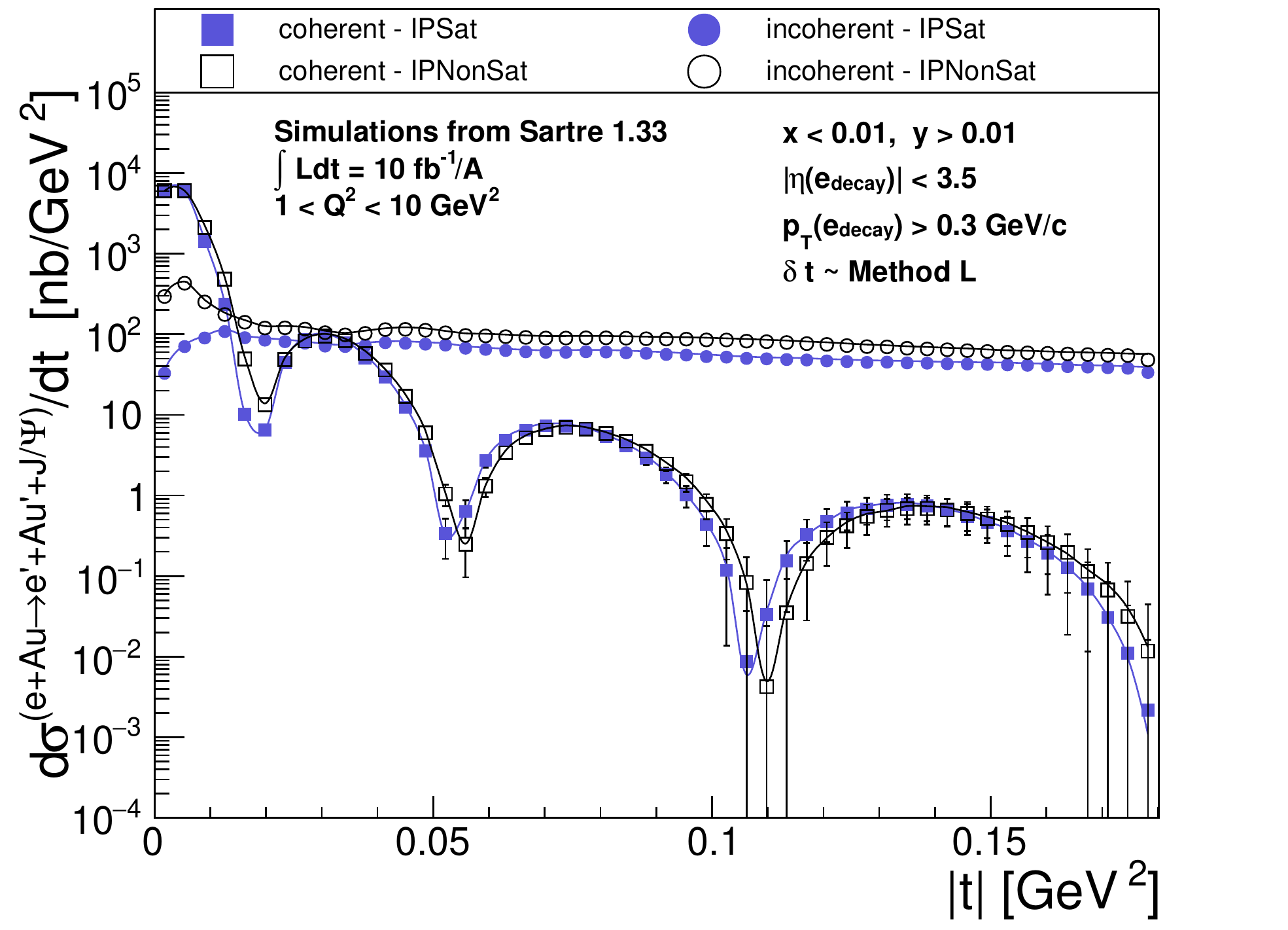}}
\hspace{0.0\textwidth}
   \subfigure{\includegraphics[width=0.5\textwidth,height=0.430\textwidth]{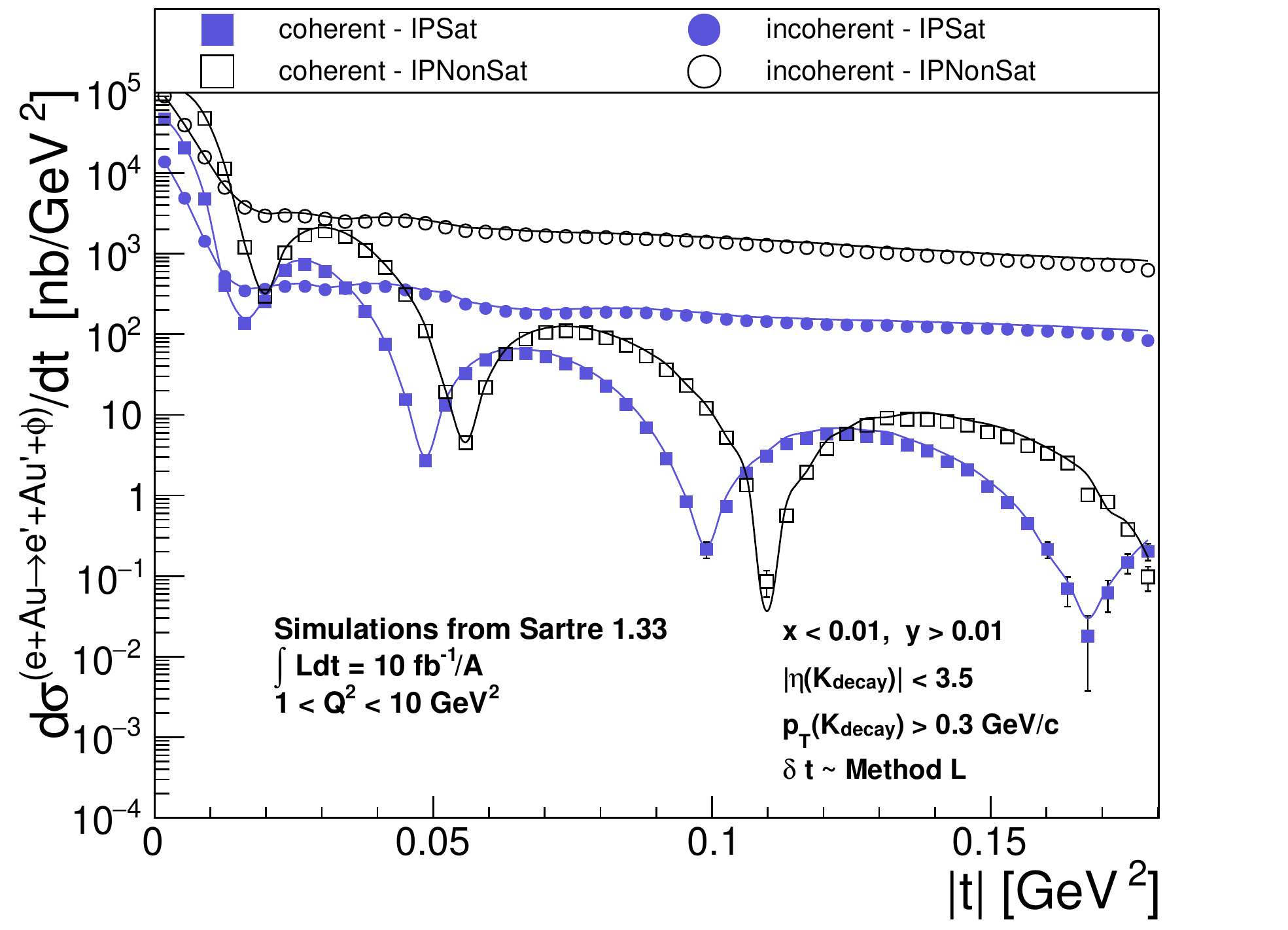}}
\end{center}
\vspace{-0.25cm}
\caption{(Color online) Similar cross-section distributions as in Fig.~\ref{fig:white_plot2} but obtained with the smearing 
$\delta t$ taken from the ``Method L" discussed in Sec.~8.4.6 of \cite{AbdulKhalek:2021gbh}. An addition here are the 
cross-section distributions represented as curves, made with the smearing and detector acceptance turned off (as they 
come out from Sar{\it t}re), and overlaid on top of the shown reconstructed $J/\psi$ and $\phi$ distributions (after the 
experimental effects have been accounted for).}
\label{fig:white_plot5}
\end{figure}

Generating plots with pseudo-data captures a realistic glance at an experimental measurement of the diffractive cross section. 
Our generated pseudo-data are obtained with the integrated luminosity of $10\,{\rm fb}^{-1}/A$. The sensitivity of such a 
measurement relies heavily on the resolution of $t$, which can be easily checked by gradually smearing diffractive peak for 
increasing relative $\delta t$ uncertainty. The pseudo-data are generated with the $e+Au$ collision energies\footnote{The ion 
beam energy here represents that of per nucleon.} of $10 \times 110\,{\rm GeV}$, taken from Table 10.3 of the Yellow Report 
\cite{AbdulKhalek:2021gbh}. In Fig.~\ref{fig:white_plot2}, the true event $t$ is smeared with a Gaussian whose width is some 
factor (written explicitly in the plot label) multiplied by $t$. In Fig.~\ref{fig:white_plot5}, the $t$-multiplying factor for obtaining 
the Gaussian width changes depending on the value of $t$ itself. Table~8.10 in \cite{AbdulKhalek:2021gbh}, produced for the 
$Q^{2}$ region of $1 < Q^{2} < 10\,{\rm GeV}^{2}$, gives six different ranges of $t$, placed sequentially from 
$0.00\,{\rm GeV}^{2}$ to $0.18\,{\rm GeV}^{2}$, in which one can find six values describing the effect of beam momentum 
spread and beam divergence on $t$-resolution, $\sigma_{t}/t$. The procedure whereby those values have been obtained is 
titled ``Method L"\footnote{The procedure is bottomed on an assumption leading to using a constraint when the invariant mass 
of the outgoing nucleus should be taken as $M_{A^{\prime}}^{2}$, to find the incoming electron's longitudinal momentum, 
instead of just assuming the nominal electron beam momentum (see Sec.~8.4.6 of \cite{AbdulKhalek:2021gbh}).}, which can 
also be used to improve the impact of momentum resolution effects on the $t$-resolution. Thus, for Fig.~\ref{fig:white_plot5}, 
depending on which range the true event $t$ falls within, the respective multiplicative factor is pulled from that Table~8.10 and 
used to calculate the width for smearing, based upon the ``Method L" that is currently the most promising procedure to improve 
the $t$-resolution as small $t$ values, such as $|t| = [0.000 - 0.001]\,{\rm GeV^{2}}$. The $t$-resolution numbers coming from 
the ``Method L" have been obtained for $J/\Psi$ production in \cite{AbdulKhalek:2021gbh}. Nevertheless, we assume the same 
numbers for $\phi$ production, and since this method relies on the reconstruction of a final-state decay, we plan to look into that 
thoroughly for the kaon decay mode in another paper.

Given that Fig.~\ref{fig:white_plot5} seems to be made with more accurate and relevant smearing, its top plot shows sufficient 
statistics in the first two minima of the coherent cross section of the produced $J/\Psi$ events for the simulation's integrated luminosity 
of $10\,{\rm fb}^{-1}/A$. Thereby, one can think of the first two minima as being experimentally measurable in terms of statistics. 
However, the largest problem here is related to the separation of coherent and incoherent processes. The coherent production 
dominates at low $t$, whereas the incoherent production starts to take over at $t \gtrsim 0.02\,{\rm GeV}^{2}$. The current 
experimental cuts are not yet sufficient to suppress the incoherent contribution to make the diffractive pattern of the coherent 
contribution measurable at the required level. Nonetheless, there is a prospect of improvement for such a separation (see the end 
of Sec.~8.4.6 in \cite{AbdulKhalek:2021gbh}). In addition to what is stated above, because of the $J/\Psi$ heavy mass (and 
therefore small size of its wavefunction) we see little difference between the IPSat and IPNonSat models for exclusive coherent 
and incoherent $J/\Psi$ electroproduction at moderate $Q^2$. The bottom plot of Fig.~\ref{fig:white_plot5} shows more of a 
favorable situation for $\phi$, with a clear separation between the events coming from the saturation and non-saturation scenarios 
for exclusive $\phi$ electroproduction. Thus, $J/\Psi$ is less sensitive to the non-linear effects than the much lighter 
$\phi$ meson, the contribution of which is enhanced in the saturation region relative to that of the heavier $J/\Psi$. The decay 
channel specifications for producing the $\phi$ and $J/\Psi$ mesons are shown at the beginning of Sec.~\ref{sec:Scaling1a}.

\subsection{$A$ and $Q^{2}$ scaling picture in the regime of high $Q^{2} > Q_{s,A}^{2}$}
\label{sec:Scaling1}
At EIC kinematics, the would-be measurable coherent $t$-spectrum, for example in Fig.~\ref{fig:white_plot2}, may be used to 
model-independently obtain the nuclear spatial gluon distribution, from exclusive $J/\Psi$ and $\phi$ production in $e+Au$ scattering, 
by measuring that distribution in impact parameter space, $F(b)$, through a two-dimensional Fourier transform of the square root 
of the coherent elastic cross section. Also, by measuring $F(b)$ for both $J/\Psi$ and $\phi$ mesons\footnote{The $\rho$ meson 
might be more sensitive to saturation effects at moderate $Q^2$ than $\phi$, nevertheless, there are large theoretical uncertainties 
in the knowledge of its wavefunction, making perturbative calculations less reliable at smaller $Q^2$ values. Perhaps it also
somewhat applies to $\phi$'s wavefunction but we prefer to work with $\phi$ and $J/\Psi$ following \cite{Accardi:2012qut} and \cite{Toll:2012mb}.}, 
we will be in a position to extract valuable information on how sensitive the corresponding measurement can be 
to saturation effects. Consequently, the studies of the exclusive vector meson production coherent cross section are quite important, 
so that in this section we will focus on its scaling properties, in terms of the nuclear mass number $A$ and virtuality $Q^{2}$, as 
has been done in Ref.~\cite{Mantysaari:2017slo}. However, now we will reproduce its main results in a realistic EIC setup by using 
Sar{\em t}re, in order to have better insight on whether the strong modification of the $A$ and $Q^{2}$ scaling of the coherent cross 
section, stemming from the gluon saturation in high-energy exclusive DIS off nuclei, can potentially be observed at EIC\footnote{The major
motivation, which the EIC and LHeC proposals are based upon, is to search for gluon many-body non-linear dynamics, and the gluon
saturation in particular.}. Such an observation shall be anchored upon measurements, which will allow to conspicuously identify 
different systematics in vector meson production, in the presence and absence of the gluon saturation in nuclei.

In the rest of Sec.~\ref{sec:Scaling} the figures are made for an integrated luminosity of $10\,{\rm fb}^{-1}/A$ with
beam energies $10 \times 110\,{\rm GeV}$ in diffractive $e + Au$ and $e+Ca$ scatterings, as well as for $100\,{\rm fb}^{-1}/A$ with
$10 \times 100\,{\rm GeV}$ in diffractive $e + p$. In this Sec.~\ref{sec:Scaling1}, we discuss exclusive vector $J/\Psi$ and $\phi$ 
meson production in the virtuality region $Q^{2} > 1\,{\rm GeV}^{2}$, which is the only relevant scale because the mass difference
between the two mesons is less appropriate in this case. The differential cross section tables included in Sar{\em t}re 1.33 reach 
a maximum $Q^{2}$ range of $20\,{\rm GeV^{2}}$ for $A > 1$ nuclear targets, with which we cannot address the whole extent 
of the $A$ and $Q^{2}$ scaling properties discussed in \cite{Mantysaari:2017slo}. Because of this reason we will use the expression 
``scaling onset", meaning that a lower $Q^{2}$ range is considered for showing the scaling trends of the cross sections and their ratios.

\subsubsection{The scaling onset at $|t| = 5\times 10^{-4}\,{\rm GeV^{2}}$}
\label{sec:Scaling1a}
For producing the scaling-related figures, \texttt{truth} information (see~\ref{sec:AppII}) from the event generator is used to separate out the
longitudinally and transversely polarized virtual photon events. Experimentally, separating out the individual polarization cross sections 
for these events involves a delicate technique known as Rosenbluth separation \cite{Defurne:2016eiy}. The ramifications of this technique 
are yet to be analyzed using the pseudo-data generated, thus making it a crucial focus in future work. In this 
Sec.~\ref{sec:Scaling}, we analyze the $\phi$ production cross section with the $e + A (p) \rightarrow \phi + A (p) \rightarrow K^{+}K^{-}$ 
decay channel, which is both statistically and experimentally practical. The $K^{+}K^{-}$ branching ratio for the $\phi$ meson, experimentally
measured to be $48.9\% \pm 0.5\%$ \cite{Zyla:2020zz}, is the largest of its decay modes. For the future EIC, the need for identification of 
light final-state mesons, such as kaons and pions, has been a vital focus to be carried out by exhaustive detector R\&D. For the heavier $J/\Psi$ 
production, the vast list of hadronic decays overwhelms its simple, yet less frequent leptonic decay channels ($e^{+}e^{-}$ and $\mu^{+}\mu^{-}$). 
For simplicity, we choose the $e^{+}e^{-}$ decay mode to the analysis, which has a branching ratio measured to be $5.94\% \pm 0.06\%$
\cite{Zyla:2020zzi}. Future studies may consider doubling the statistics by combining the contribution from the equally likely
$\mu^{+}\mu^{-}$ decay channel. 

Below are given the cross sections and number of events simulated for the three scattering types under consideration with the given branching
ratios:
\begin{itemize}
\item ~$e+Au$ $|$ $\phi \rightarrow K^{+}K^{-}$, ~total cross section = 80.20\,nb, ~$N_{\rm events} = 4.07 \times 10^{6}$;

\item ~$e+Ca$ $|$ $\phi \rightarrow K^{+}K^{-}$, ~total cross section = 6.21\,nb, ~$N_{\rm events} = 1.55 \times 10^{6}$;

\item ~$e+p$ $|$ $\phi \rightarrow K^{+}K^{-}$, ~total cross section = $6.8 \times 10^{-3}$\,nb, ~$N_{\rm events} = 6.80 \times 10^{5}$;

\item ~$e+Au$ $|$ $J/\Psi \rightarrow e^{+}e^{-}$, ~total cross section =  0.99\,nb, ~$N_{\rm events} = 5.04 \times 10^{4}$;

\item ~$e+Ca$ $|$ $J/\Psi \rightarrow e^{+}e^{-}$, ~total cross section =  0.053\,nb, ~$N_{\rm events} = 1.32 \times 10^{4}$;

\item ~$e+p$ $|$ $J/\Psi \rightarrow e^{+}e^{-}$, ~total cross section =  $37.2 \times 10^{-6}$\,nb, ~$N_{\rm events} = 3.72 \times 10^{3}$.
\end{itemize}

First, we consider the forward limit at $t \rightarrow 0$, by going back to and starting with Eq.~(\ref{eq:ttepamplitude}) but considering 
it for the case of $\gamma^{*}A\,\rightarrow\,V A$. However, in the limit of $Q^{2} \gg Q_{s,A}^{2}$, instead of the 
$\gamma^{\ast}$-$p$  amplitude (that can be used from the GBW model \cite{GolecBiernat:1998js}) one should use the 
$\gamma^{\ast}$-$A$ amplitude by making the substitution $Q_{s,p}^{2} \rightarrow Q_{s,A}^{2} \sim A^{1/3}Q_{s,p}^{2}$:
\beq
\frac{{\rm d}\sigma_{q\bar q}^{A}}{{\rm d}^{2} \bt} = 2\left[ 1 - \left( 1 - \rtt Q_{s,A}^{2} \right) \right] ,
\label{eq:scl_1}
\eeq
which is a simplified version of Eq.~(\ref{eq:analytical}). Consequently, the diffractive $\gamma^{*}A\,\rightarrow\,V A$ scattering 
amplitude will become
\bea
& & \mathcal{A}_{T,L}^{\gamma^{\ast}A\,\rightarrow\,V A} \sim 
\nonumber\\
& & ~~~~ \sim i \int{\rm d}^{2}{\bf r_{T}} {\rm d}^{2}\bt
\Lb \Psi_{\gamma^{\ast} \rightarrow q\bar{q}}\Psi_{q\bar{q} \rightarrow V}^{\ast} \Rb_{T,L} \rtt Q_{s,A}^{2} .
\label{eq:scl_2}
\eea
The ``Boosted Gaussian" parametrization is used for the vector meson wavefunction \cite{Kowalski:2006hc}.
At this point let us follow some of the argumentation of \cite{Mantysaari:2017slo} for addressing its principal scaling results.

\begin{figure}[h!]
\begin{center}
\includegraphics[width=0.475\textwidth,height=0.430\textwidth]{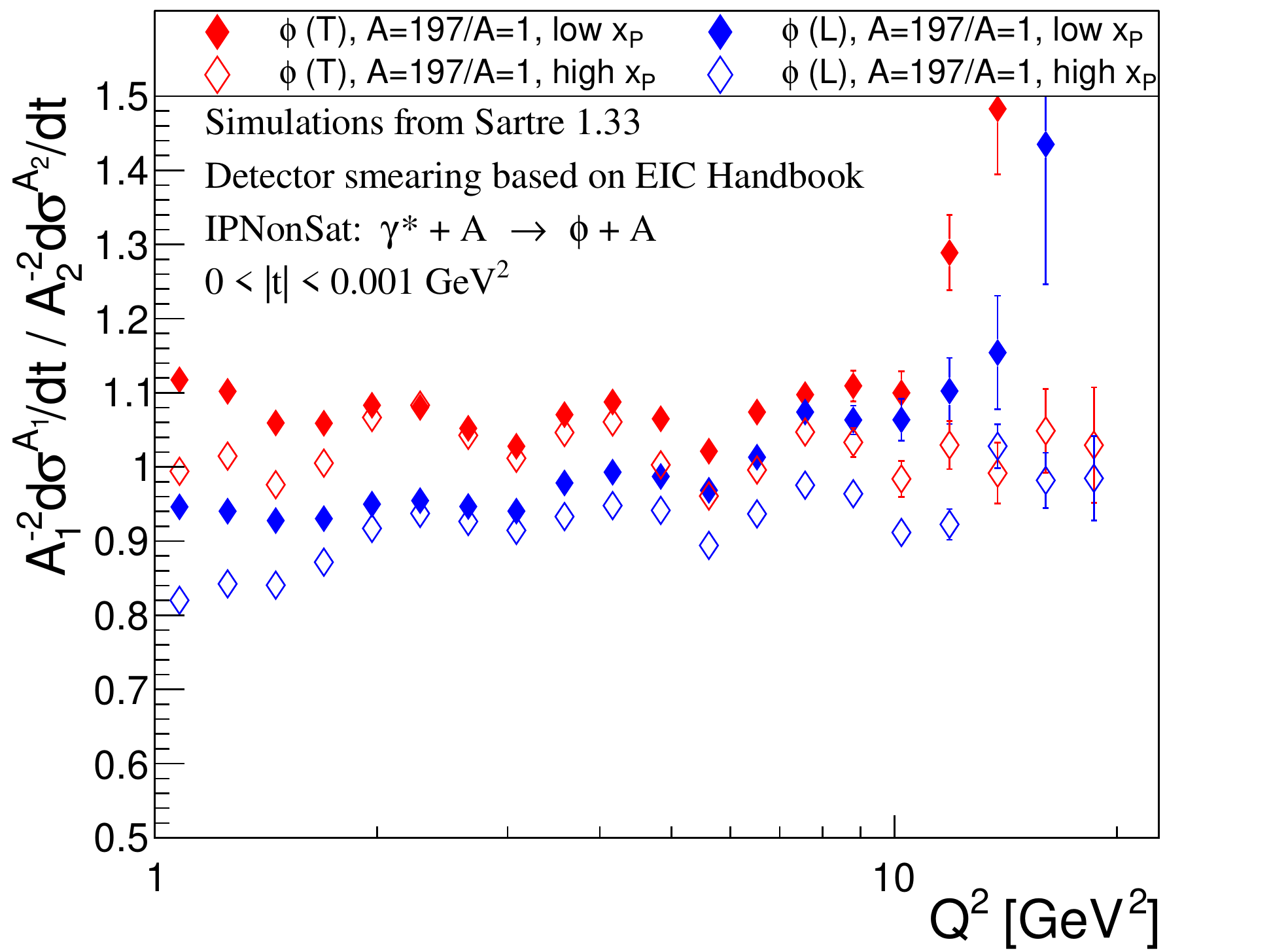} 
\end{center}
\vspace{0.0cm}
\caption{(Color online)   
The $A^{2}$-ratio at $|t| = 5\times 10^{-4}\,{\rm GeV^{2}}$ for the exclusive coherent $\phi$ electroproduction cross section 
in the IPNonSat model. $T$ and $L$ refer to transverse and longitudinal polarization, respectively. The vertical axis shows the 
$A^{-2}$ normalized ratio of the cross sections for the Gold over proton. The high $\xpom$ shows the range of pseudo-data obtained 
in $\xpom = [0.005 - 0.009]$, the low $\xpom$ shows the pseudo-data range of $\xpom = [0.001 - 0.005$].}
\label{fig:Scl_NonSat_fig}
\end{figure}

\begin{figure}[h!]
\begin{center}
   \subfigure{\includegraphics[width=0.475\textwidth,height=0.430\textwidth]{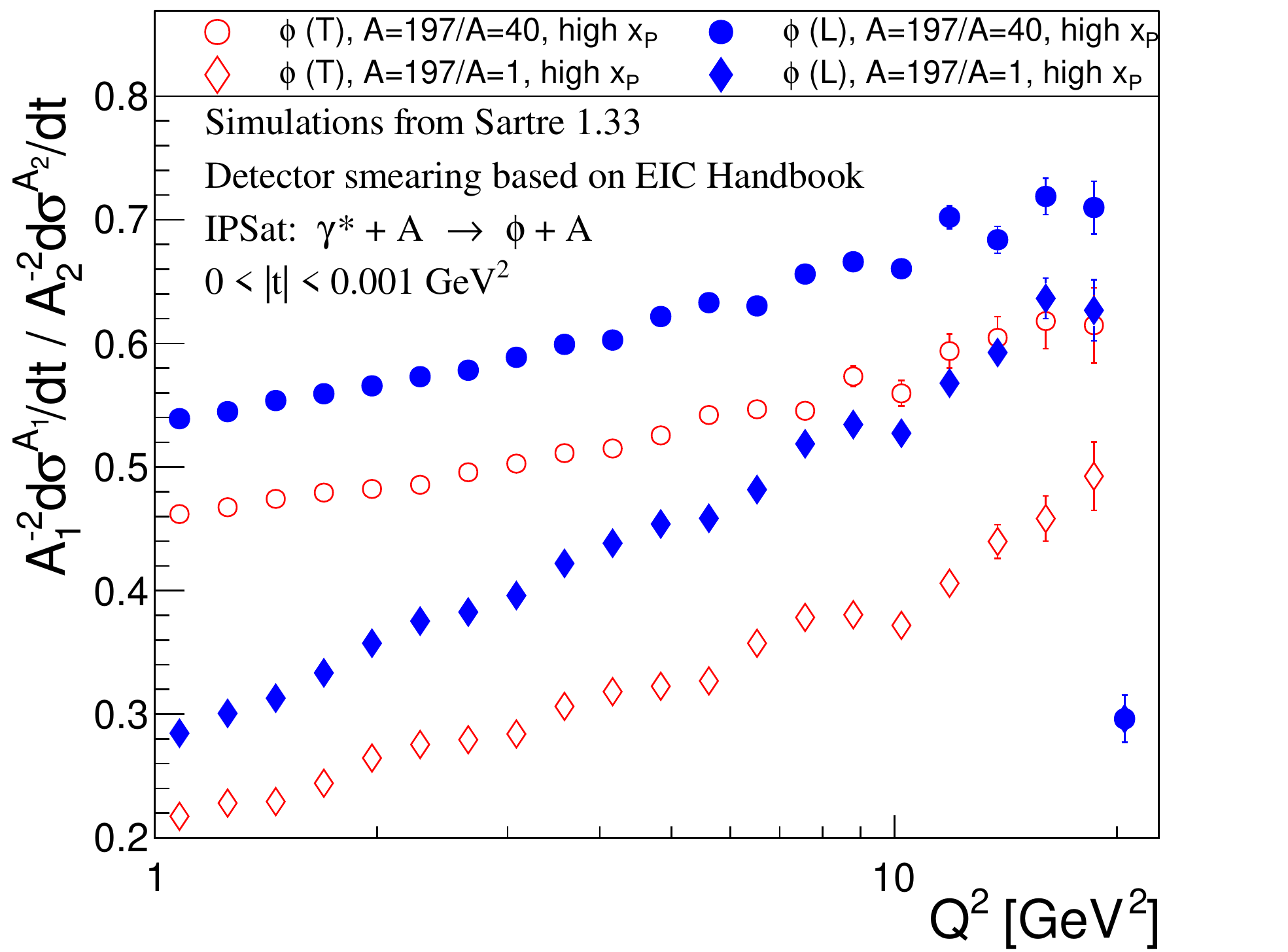}}
\hspace{0.0\textwidth}
   \subfigure{\includegraphics[width=0.475\textwidth,height=0.430\textwidth]{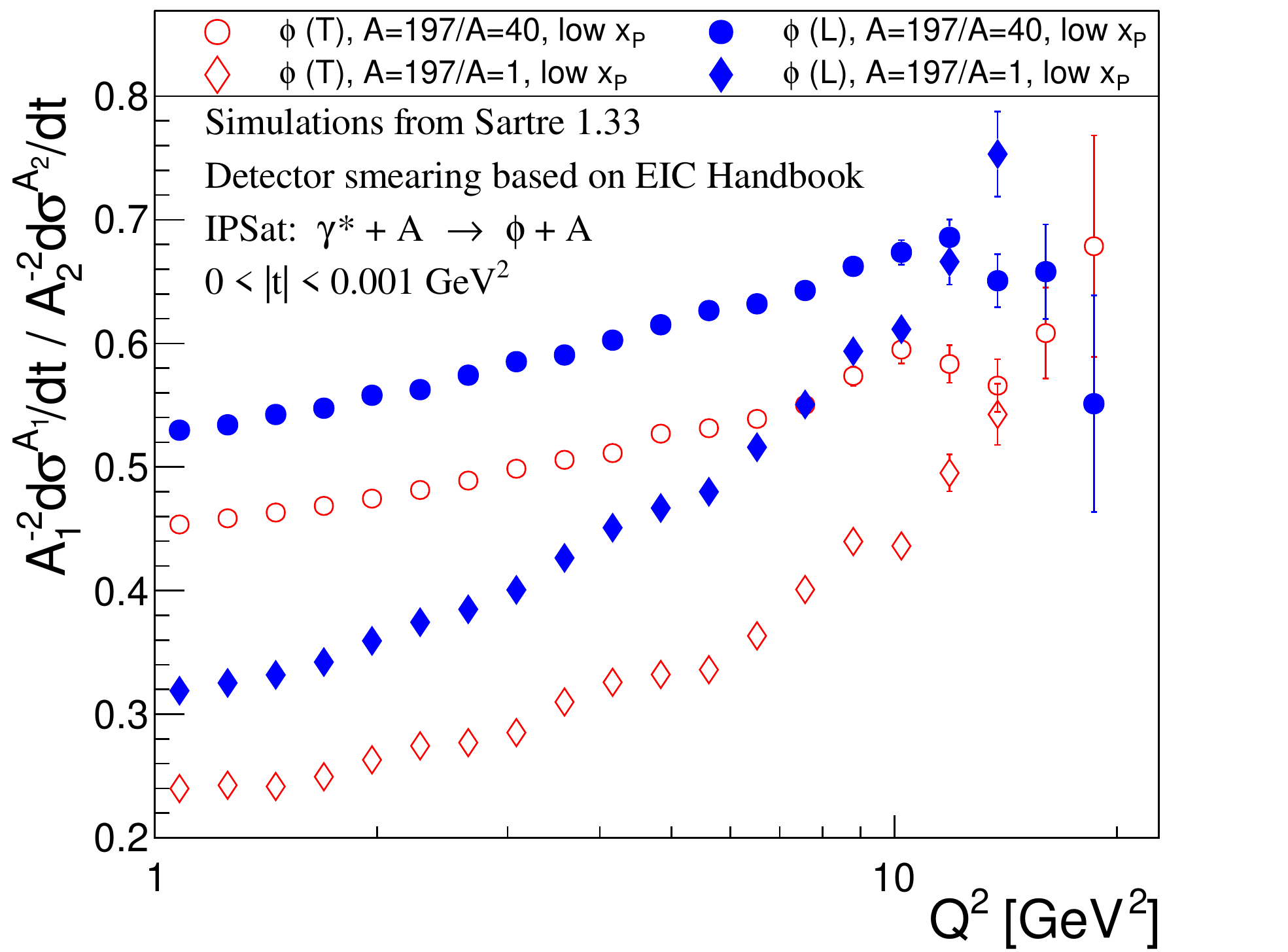}}
\end{center}
\vspace{-0.25cm}
\caption{(Color online)
The onset of the $A^{2}$ scaling at $|t| = 5\times 10^{-4}\,{\rm GeV^{2}}$ for the exclusive coherent $\phi$ electroproduction cross section 
in the IPSat model. The vertical axis shows the $A^{-2}$ normalized ratio of the cross sections for the Gold over Calcium and the 
Gold over proton. The nomenclature is the same as in Fig.~\ref{fig:Scl_NonSat_fig}. The top plot shows the pseudo-data at high $\xpom$,
the bottom plot shows the pseudo-data at low $\xpom$.}
\label{fig:Scl_Sat_fig1}
\end{figure}

\paragraph{$A^{2}$ scaling:} The $\bt$ integration of Eq.~(\ref{eq:scl_2}) gives the cross sectional area factor of $A^{2/3}$, 
resulting in
\bea
& & \mathcal{A}_{T,L}^{\gamma^{\ast}A\,\rightarrow\,V A} \sim 
\nonumber\\
& & ~~~~ \sim i A \int{\rm d}^{2}{\bf r_{T}}
\Lb \Psi_{\gamma^{\ast} \rightarrow q\bar{q}}\Psi_{q\bar{q} \rightarrow V}^{\ast} \Rb_{T,L} \rtt Q_{s,p}^{2} ,
\label{eq:scl_3}
\eea
which according to Eq.~(\ref{eq:gamma_A_cohdiff}) leads to the coherent diffractive cross section, such as
\bea
& & \Lb \frac{\dint\sigma^{\gamma^{*}A\,\rightarrow\,V A}_{T, L}}{\dint t}\bigg\vert_{t = 0} \Rb_{\rm coh.\,diff.} \equiv
\nonumber\\
& & ~~~~~~~~~~~~~~~~~~~~~~
\equiv \frac{\dint\sigma^{\gamma^{*}A\,\rightarrow\,V A}_{T, L} }{\dint t}\bigg\vert_{t = 0} \sim A^{2} .
\label{eq:scl_4}
\eea
This asymptotic $A^{2}$ is used in the normalization of the exclusive coherent $\phi$ and $J/\Psi$ cross section ratios for 
Gold over Calcium and Gold over proton at $|t| = 5\times 10^{-4}\,{\rm GeV^{2}}$ in
Figs.~\ref{fig:Scl_NonSat_fig}~and~\ref{fig:Scl_Sat_fig1}. 
As it is mentioned at the end of the introduction, in Sar{\em t}re simulations one cannot feasibly generate events at specific $t$ 
values, nevertheless, we are able to generate events in given $t$ bins: in this case, in the bin of $0<|t|<0.001\,{\rm GeV^{2}}$,
and by just referring to its central value instead of the bin. The same convention is adopted for the other $t$ bins discussed in 
the next Sec.~\ref{sec:Scaling1b}.

Fig.~\ref{fig:Scl_NonSat_fig} shows pseudo-data of some normalized ratios in the IPNonSat model, which fluctuate around unity. 
If there were existing Sartre look-up amplitude tables obtained from $e+Ca$ collisions for the IPNonSat model, then the ratio of 
$A=197/A=40$ (instead of Gold over proton) would be much closer to the perfect horizontal $A^{2}$ scaling that is expected to 
take place in the absence of gluon saturation effects.
Fig.~\ref{fig:Scl_Sat_fig1} shows pseudo-data of normalized ratios in the IPSat model, where one can see a substantial suppression 
due to gluon saturation at low-$Q^{2}$ region. It is obvious that the suppression is larger for transversely polarized photon case.
For simulations, the version 1.33 of Sar{\em{t}}re is used as in the previous figures. The cuts shown in 
Figs.~\ref{fig:white_plot2}~and~\ref{fig:white_plot5}, are used for making these two figures as well (and the subsequent figures).
Besides, the pseudo-data exhibited in Figs.~\ref{fig:Scl_NonSat_fig}~and~\ref{fig:Scl_Sat_fig1} are normalized by the scaling factor from 
Eq.~(\ref{eq:scl_4}). More details are presented in the corresponding captions. An analysis at a fixed $|t|$ value, although ideal, is impractical 
statistically for the actual experiment. The specific $|t| \sim 0$ range we selected provides us with ample experimental statistics to distinguish 
the growth of the cross section ratio to within statistical uncertainties. We argue that this range accurately reflects the theoretical behavior 
of the cross section at $|t| = 0$, upon direct comparison with the figures from \cite{Mantysaari:2017slo}.

\begin{figure}[h!]
\begin{center}
   \subfigure{\includegraphics[width=0.475\textwidth,height=0.430\textwidth]{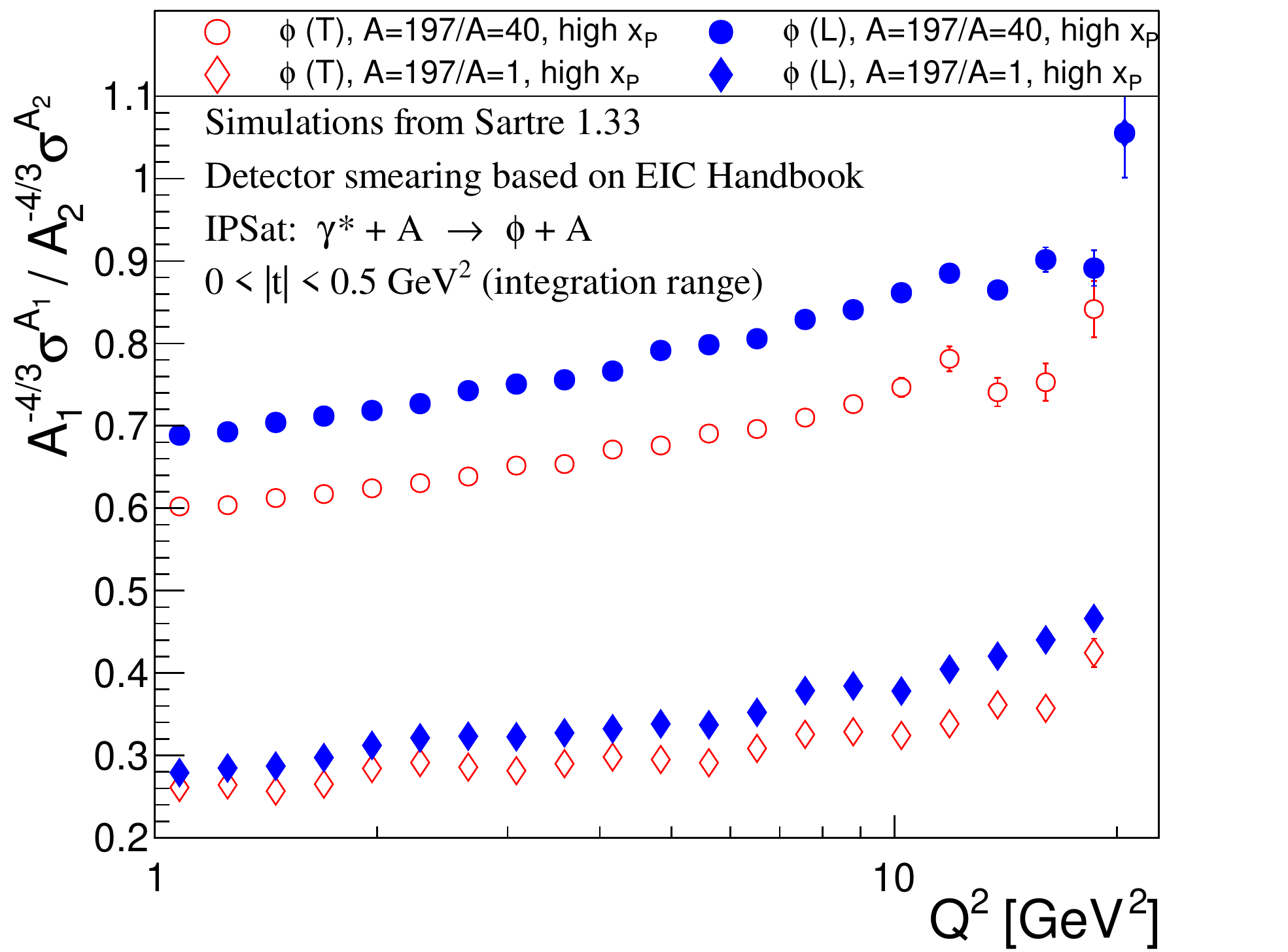}}
\hspace{0.0\textwidth}
   \subfigure{\includegraphics[width=0.475\textwidth,height=0.430\textwidth]{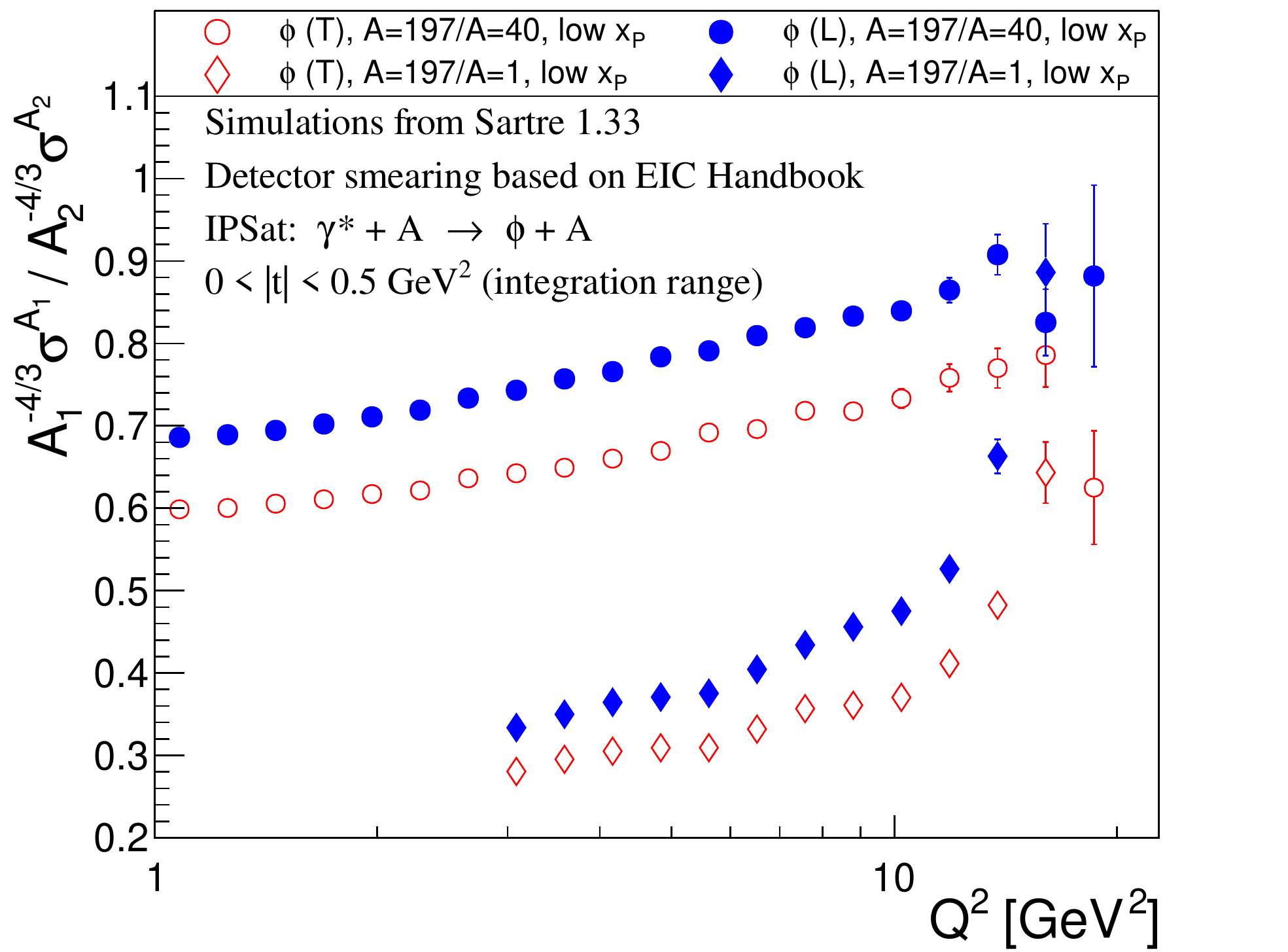}}
\end{center}
\vspace{-0.25cm}
\caption{The onset of the $A^{4/3}$ scaling for the total exclusive coherent $\phi$ electroproduction cross section in the IPSat model. 
The vertical axis shows the $A^{-4/3}$ normalized ratio of the cross sections for the Gold over Calcium and over proton. 
The nomenclature is the same as in Fig.~\ref{fig:Scl_NonSat_fig}. The top plot shows the pseudo-data at high $\xpom$,
the bottom plot shows the pseudo-data at low $\xpom$. Contrary to Fig.~\ref{fig:Scl_NonSat_fig} and Fig.~\ref{fig:Scl_Sat_fig1}, 
the range of $|t|$ here is the cross-section integration region. The integration used in Eq.~(\ref{eq:scl_5}) is valid in this range 
because we generate events only in $0 < |t| < 0.5\,{\rm GeV^{2}}$. }
\label{fig:Scl_Sat_fig2}
\end{figure}

\paragraph{$A^{4/3}$ scaling:} Fig.~\ref{fig:Scl_Sat_fig2} shows pseudo-data of the normalized ratios, for the total exclusive 
coherent cross section in the IPSat model, obtained after $t$-integration of Eq.~(\ref{eq:scl_4}):
\beq
\sigma^{\gamma^{*}A\,\rightarrow\,V A}_{T, L} \sim A^{2}A^{-2/3} \sim A^{4/3} ,
\label{eq:scl_5}
\eeq
where $A^{-2/3}$ comes from the width of the coherent peak. In Fig.~\ref{fig:Scl_Sat_fig1}, it is expected that the normalized 
cross section ratios go to unity at asymptotically large $Q^{2}$, as shown in \cite{Mantysaari:2017slo}\footnote{Currently, the 
look-up amplitude tables in Sar{\em t}re have maximum reach of $Q^{2}$ as 20\,{\rm GeV$^{2}$} for $e + Au$ and $e + Ca$, 
and as 200\,{\rm GeV$^{2}$} for $e + p$. We plan to update those tables to have a higher $Q^{2}$ reach, along
with making others for various $e + A$ beams that can be found in Table 10.3 of \cite{AbdulKhalek:2021gbh}.}. 
Meanwhile, the ratios in Fig.~\ref{fig:Scl_Sat_fig2} do not approach unity at large $Q^{2}$ because of oversimplification of the 
$t$-integration, based on the coherent peak assumption, which kind of changes the shape of the pseudo-data curves. But the ratio Gold 
over Calcium still shows suppression though not in a correct way\footnote{It should be emphasized that a better calculation gives 
an $A$-dependent parameter, standing in front of $A^{4/3}$ in the r.h.s of Eq.~(\ref{eq:scl_5}), which e.g., for the Gold nucleus 
is $C[197] \approx 1/2$ \cite{Mantysaari:2018nng}.}. On the other hand, the result in Eq.~(\ref{eq:scl_5}) is valid at asymptotically 
large $A$. Nonetheless, by looking also at the ratio Gold over Calcium and over proton in Fig.~\ref{fig:Scl_Sat_fig1}
and Fig.~\ref{fig:Scl_Sat_fig2}, one can conclude that generally, in normalized ratios of arbitrary $A_{1}/A_{2}$ and $A_{1}/A_{3}$ 
with three heavy nuclei $A_{1} > A_{2} > A_{3}$, the suppression would be stronger for $A_{1}/A_{3}$ rather than for
$A_{1}/A_{2}$. Note that a similar integrated cross section ratio is reported in \cite{AbdulKhalek:2021gbh,Lomnitz:2018juf} too.

In Figs.~\ref{fig:Scl_NonSat_fig},~\ref{fig:Scl_Sat_fig1}~and~\ref{fig:Scl_Sat_fig2}, for making pseudo-data we use the 
$Q^{2}$ interval $1 < Q^{2} < 20\,{\rm GeV}^{2}$ instead of $1 < Q^{2} < 10\,{\rm GeV}^{2}$. First, we have checked out 
all these figures with the exact same smearing ``Method L" (as in Fig.~\ref{fig:white_plot5}), the model numbers of which, 
made for $1 < Q^{2} < 10\,{\rm GeV}^{2}$ in totally six ranges of $t$, are shown in Table~8.10 of \cite{AbdulKhalek:2021gbh}. 
Then, our further cross-check has shown that the figures made in both $Q^{2}$ intervals are quite similar with each other 
quantitatively and qualitatively. Therefore, based upon this observation we assume that the current L model numbers are 
well applicable for making the pseudo-data in the range of $1 < Q^{2} < 20\,{\rm GeV}^{2}$. We also expand the last cell of 
Table~8.10, assuming all events with $|t| > 0.18\,{\rm GeV^{2}}$ have a $t$-resolution of $\sigma_{t}/t = 0.005$.

It is also relevant to emphasize that though the cross sections we are analyzing for all the figures in Sec.~\ref{sec:Scaling1} 
are beam energy-independent via the division of the virtual photon flux factor (see~\ref{sec:AppI} and~\ref{sec:AppII}), the 
kinematic phase space of the final-state particles are not. For the scattered electron, which is used to reconstruct the event 
kinematics such as $Q^2$ and $y$, it is favorable for its kinematic phase space to be as identical as possible when taking cross 
section ratios of $e+A$ and $e+p$ scatterings. For the beam energies of $e+Au$ and $e+Ca$ at $10 \times 110\,{\rm GeV}$, also $e+p$ 
at $10 \times 100\,{\rm GeV}$, the reconstruction quality of the event’s scattered electron is practically identical. In turn, we 
eliminate the need to consider different constraints (or difficulties) in event reconstruction after smearing at these beam 
energies. For Figs.~\ref{fig:white_plot2}~and~\ref{fig:white_plot5}, 
which are produced with the $e+Au$ beam at $10 \times 110\,{\rm GeV}$, the cross section is not beam-independent and should be 
expected to alter depending on the energies. For these plots of Sec.~\ref{sec:White_paper}, which do not intend to show 
comparisons between nuclear targets, we only place a cut to remove events with $y < 0.01$. This inelasticity cut, discussed 
frequently in \cite{AbdulKhalek:2021gbh}, alleviates the issue of large relative uncertainty in $y$ and $Q^{2}$ for high-energetic 
and low-angle scattered electrons. Without a beam energy difference, there is no need to increase the $y$ cut. But the sheer 
variability of the final-state electron's phase space can muddle our cross-section ratio figures, if we do not widen the $y$ cut. 
At small event $y$ below 0.05, the event reconstruction quality after smearing at low $Q^{2} \sim 1\,{\rm GeV}^{2}$, and even at 
moderate $Q^{2} \sim 20\,{\rm GeV}^{2}$, varies considerably between the $e+p$ and $e+A$ scatterings. To account for this, all 
the scaling figures (the ones shown later as well) contain an event cut, where any event with $y < 0.05$ is discarded. After this 
cut is placed, Figs.~\ref{fig:Scl_NonSat_fig},~\ref{fig:Scl_Sat_fig1}~and~\ref{fig:Scl_Sat_fig2} exhibit a clearer rise as a 
function of $Q^{2}$, qualitatively matching onto theoretical figures made in \cite{Mantysaari:2017slo}.

It is also necessary to mention that in the lower plot of Fig.~\ref{fig:Scl_Sat_fig2} made for the low-$\xpom$ range, the cross-section 
ratio of $\phi$(T) and $\phi$(L) vector mesons has a cut-off at $Q^{2} \sim 3$\,{\rm GeV$^{2}$}. The reason for doing so is that
there is an artificial bump seen below $Q^{2} < 3$\,{\rm GeV$^{2}$}. We have found that, due to the difference in final-state particle 
kinematics between $e+p$ at $10 \times 100\,{\rm GeV}$ and $e+Au$ at $10 \times 110\,{\rm GeV}$, the final-state lepton detection 
efficiencies (scattered electron and decay pair) for both $e+p$ and $e+Au$ differ noticeably enough within this specific phase space 
to create an imbalance in the cross-section ratio\footnote{This effect is absent in the lower plot of Fig.~\ref{fig:Scl_Sat_fig1} as 
the events are analyzed within a much narrower range of $|t|$.}. In particular, the placement of the $p_{T} > 0.3\,{\rm GeV}$ cut on all 
final-state leptons play a significant role in producing this imbalance. Regardless, our study shows that the scaling behavior at larger 
$Q^{2}$ can still be extracted without a careful consideration of the slight difference in the final-state kinematics between the EIC's 
$e+p$ and $e+Au$ energies. On the other hand, there is an overall caveat related to both top and bottom plots of Fig.~\ref{fig:Scl_Sat_fig2},
based on the coherent peak assumption in the $t$-integration of Eq.~(\ref{eq:scl_4}) leading to $A^{4/3}$ (as discussed below of 
Eq.~(\ref{eq:scl_5})).

\paragraph{$A^{2}Q^{-6}$ scaling:} The vector meson wavefunction has an overlap with the longitudinally polarized photon
wavefunction, given by the following functional form:
\beq
\Lb \Psi_{\gamma^{\ast} \rightarrow q\bar{q}}\Psi_{q\bar{q} \rightarrow V}^{\ast} \Rb_{L} \sim z(1 -z)\,Q\,K_{0}(\varepsilon\,r_{T})
\,\phi_{L}(r_{T}, z) ,
\label{eq:scl_6}
\eeq
where $\varepsilon = \sqrt{Q^{2}z(1 - z) + m_{q}^{2}} \approx Q$ at $Q^{2} \gg Q_{s}^{2}$, and 
$K_{0}(\varepsilon\,r_{T})$ is the modified Bessel function of the second kind of the 0th order. The scalar part of the vector meson 
wavefunction, $\phi_{L}(r_{T}, z) \sim z(1 - z) e^{-\rtt M_{V}^{2}}$, restricts contributions from dipoles with the sizes larger than 
$1/M_{V}$. Thereby, the diffractive longitudinal scattering amplitude reads as
\bea
& & \mathcal{A}_{L}^{\gamma^{\ast}A\,\rightarrow\,V A} \sim i \int \dint \rt\,\mathbf{r}_{\mathbf{T}}^{3}\,Q~K_{0}(Q\,r_{T}) 
\sim \frac{1}{Q^{3}} ,
\nonumber \\
& & {\mbox{leading~to}}~~~~
\frac{\dint\sigma^{\gamma^{*}A\,\rightarrow\,V A}_{L} }{\dint t}\bigg\vert_{t = 0} \sim \frac{1}{Q^{6}} .
\label{eq:scl_7}
\eea
\begin{figure}[hbt!]
\begin{center}
   \subfigure{\includegraphics[width=0.475\textwidth,height=0.430\textwidth]{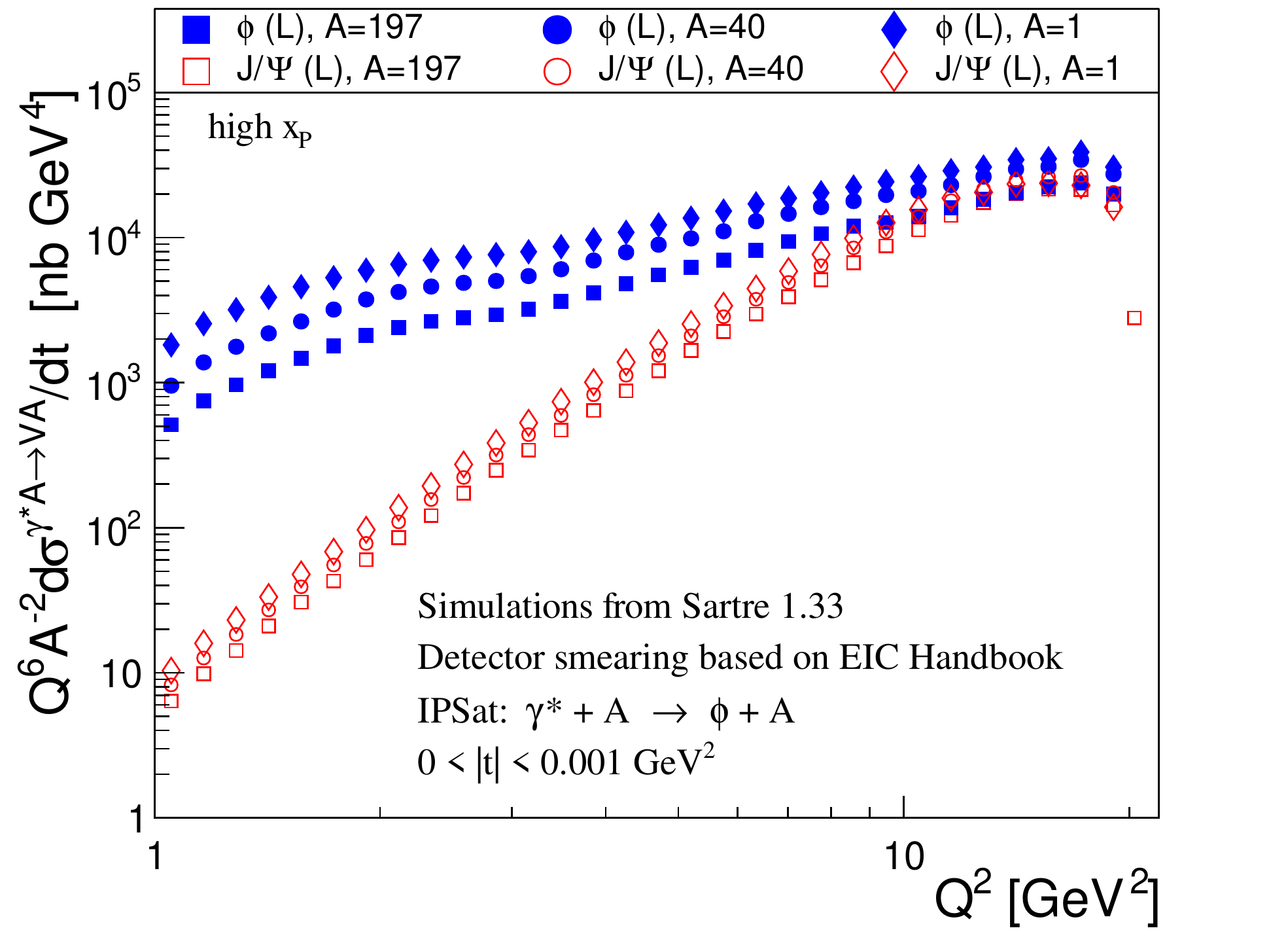}}
\hspace{0.0\textwidth}
   \subfigure{\includegraphics[width=0.475\textwidth,height=0.430\textwidth]{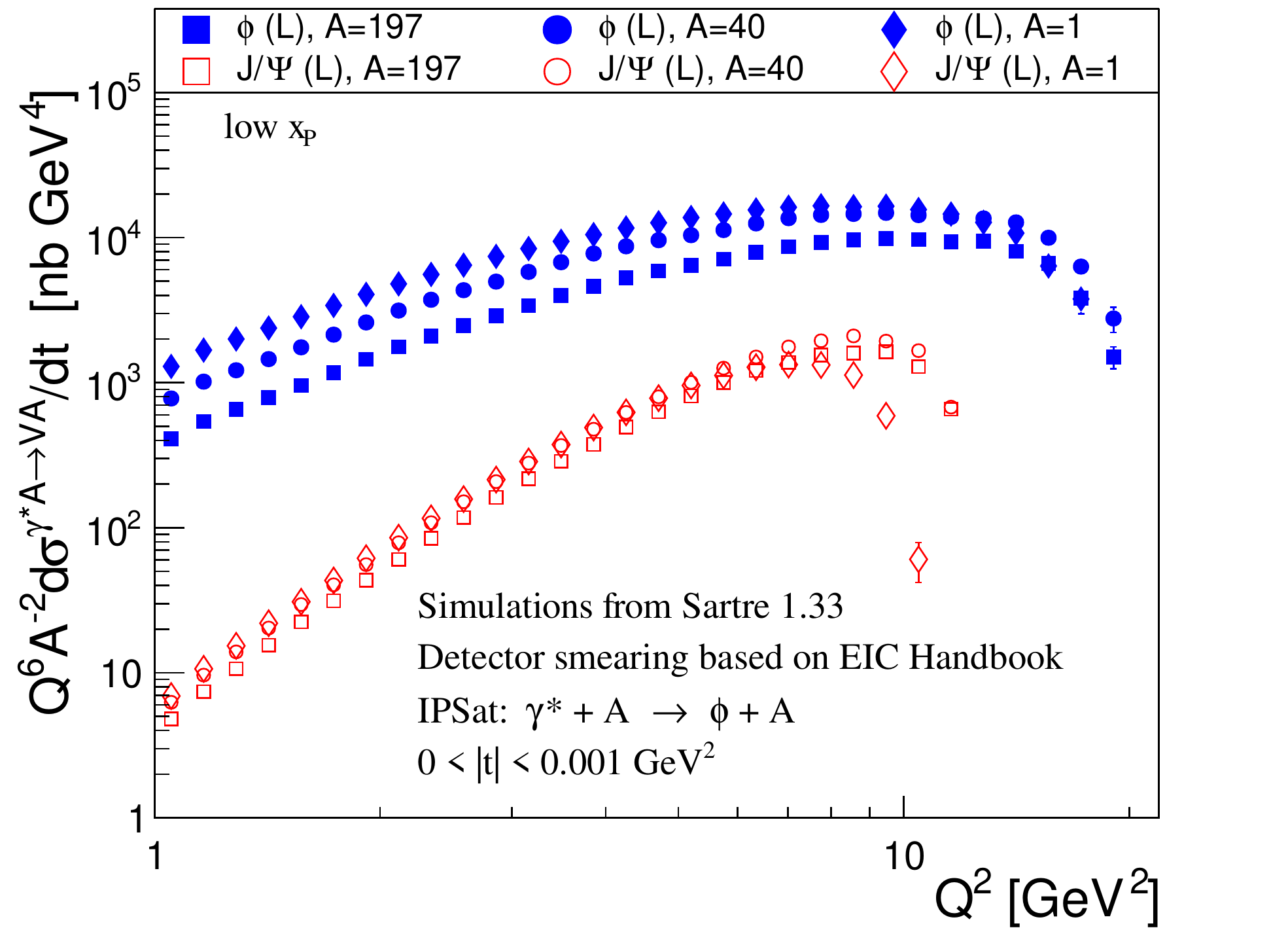}}
\end{center}
\vspace{-0.25cm}
\caption{The onset of the $A^{2}Q^{-6}$ scaling at $|t| = 5\times 10^{-4}\,{\rm GeV^{2}}$ for the exclusive coherent $\phi$ and $J/\Psi$  
longitudinal electroproduction cross section in the IPSat model for the Gold, Calcium and proton, multiplied by the scaling 
factor of $Q^{6}A^{-2}$. The nomenclature is the same as in Fig.~\ref{fig:Scl_NonSat_fig}. The top plot shows the pseudo-data 
at high $\xpom$, the bottom plot shows the pseudo-data at low $\xpom$.}
\label{fig:Scl_Sat_fig3}
\end{figure}

Fig.~\ref{fig:Scl_Sat_fig3} shows pseudo-data on exclusive coherent $\phi$ and $J/\Psi$ longitudinal electroproduction 
cross section at $|t| = 5\times 10^{-4}\,{\rm GeV^{2}}$ in the IPSat model.
For this specific case, the numerical calculations in \cite{Mantysaari:2017slo} show that the 
cross section becomes approximately flat in the region $Q^{2} \gtrsim 10^{2}$\,{\rm GeV$^{2}$} at $\xpom = 0.01$. In our case, 
the top plot of Fig.~\ref{fig:Scl_Sat_fig3} implies that the proton curves describing the vector meson production 
tend to have an approximately flat structure above $10^{2}$\,{\rm GeV$^{2}$}, for the events simulated in the range of high
$\xpom = [0.005 - 0.009]$. Besides, since the Gold and Calcium curves go along with the proton curves at low-$Q^{2}$, then one 
could expect that their $Q^{6}$ scaling trend would likewise continue above $\sim 10^{2}$\,{\rm GeV$^{2}$}.
The $A^{2}$ scaling is also visible in the simulated $Q^{2}$ ranges. This scaling is approximate for $\phi$ production 
given by the Gold and Calcium curves, however, it is better for $J/\Psi$ production.

As regards the bottom plot of Fig.~\ref{fig:Scl_Sat_fig3}, the curves decreasing at their largest $Q^{2}$ reach have one very 
plausible reason. That is the physical phase space of an event within low $\xpom = [0.001 - 0.005]$, which is also limited in 
Q$^{2}$. For example, if $Q^{2} = 2\,{\rm GeV^{2}}$ or so, then the event $\xpom$ is kinematically allowed to fall within the 
entire low-$\xpom$ range. However, as $Q^{2}$ increases, the kinematically allowed $\xpom$ range begins to shrink. This 
leads to a shrinking cross section at larger $Q^{2}$ since we are averaging over the entire low-$\xpom$ range. For the 
high-$\xpom$ range this effect is not noticeable, however, it occurs in a lower-$\xpom$ one.

\begin{figure}[h!]
\begin{center}
   \subfigure{\includegraphics[width=0.475\textwidth,height=0.430\textwidth]{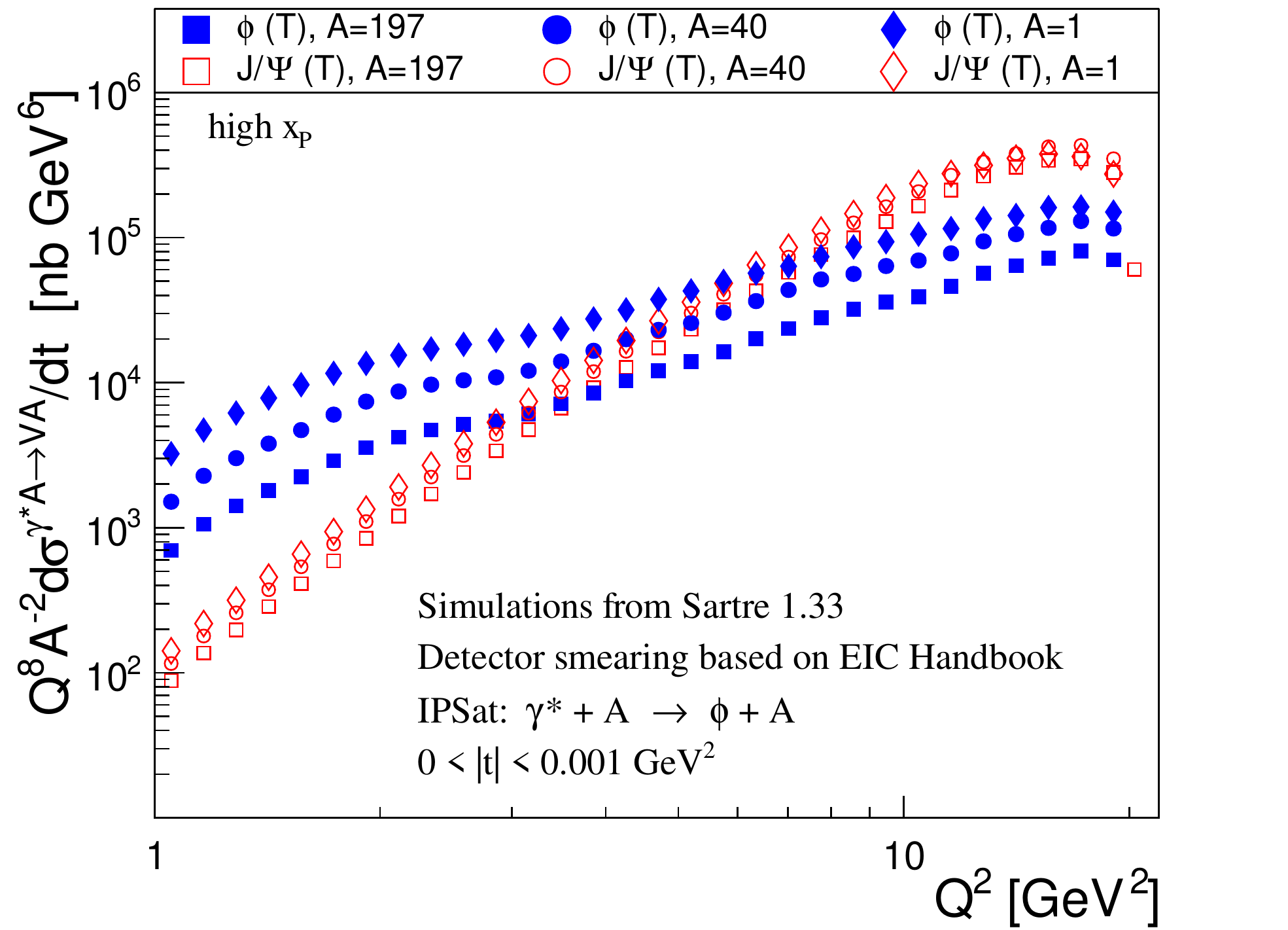}}
\hspace{0.0\textwidth}
   \subfigure{\includegraphics[width=0.475\textwidth,height=0.430\textwidth]{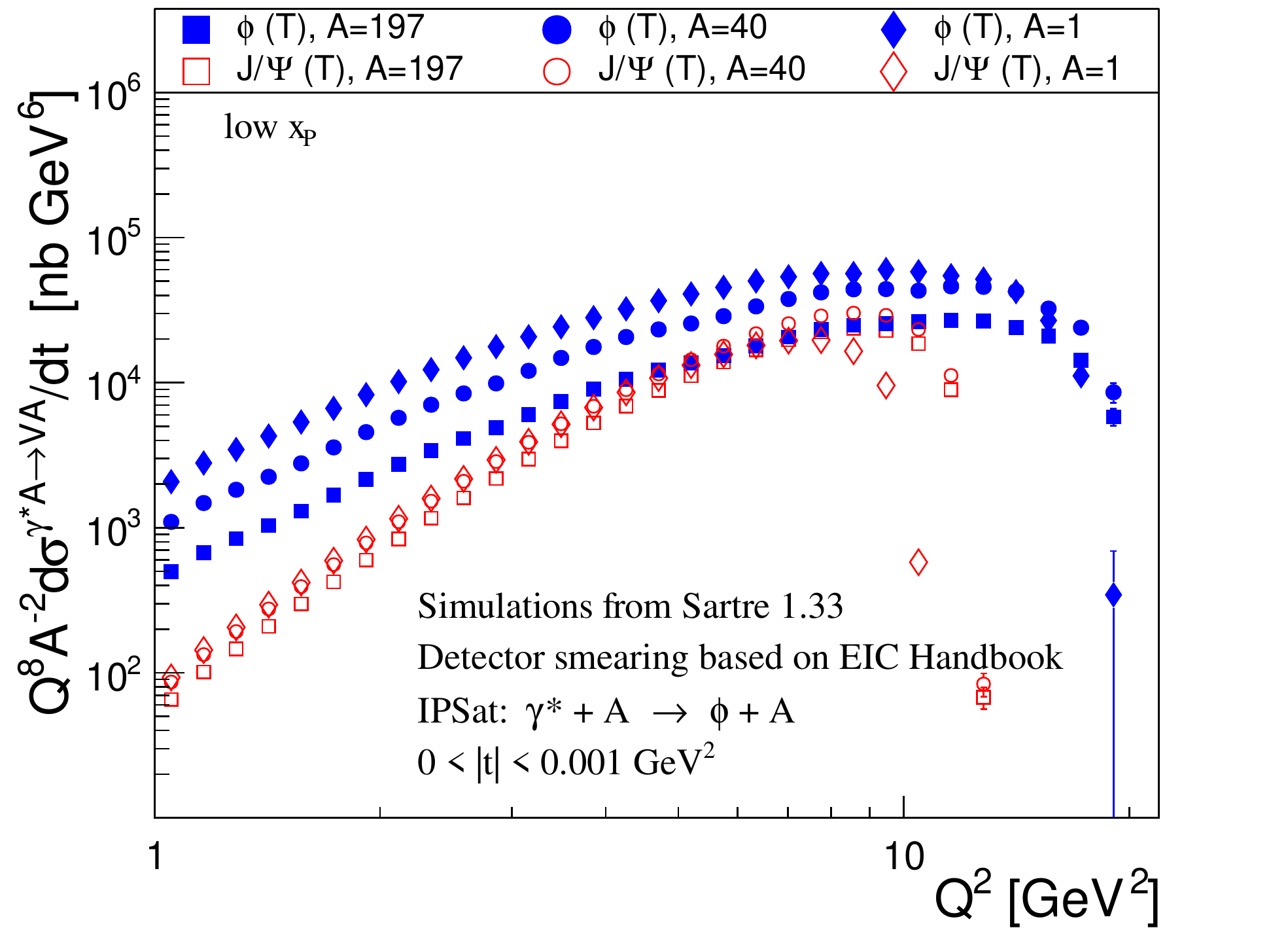}}
\end{center}
\vspace{-0.25cm}
\caption{The onset of the $A^{2}Q^{-8}$ scaling at $|t| = 5\times 10^{-4}\,{\rm GeV^{2}}$ for the exclusive coherent $\phi$ and $J/\Psi$ 
transverse electroproduction cross section in the IPSat model for the Gold, Calcium and proton, multiplied by the scaling factor 
of $Q^{8}A^{-2}$. The nomenclature is the same as in Fig.~\ref{fig:Scl_NonSat_fig}. The top plot shows the pseudo-data at high 
$\xpom$, the bottom plot shows the pseudo-data at low $\xpom$.}
\label{fig:Scl_Sat_fig4}
\end{figure}

One can see non-physical dips in the top plot of Fig.~\ref{fig:Scl_Sat_fig3} for the high-$\xpom$ $\phi$ cross 
sections, appearing at low $Q^{2} \sim 1\,{\rm GeV^{2}}$. These features of the $\phi$ cross section are also visible in
Figs.~\ref{fig:Scl_Sat_fig4},~\ref{fig:Scl_Sat_fig5}~and~\ref{fig:Scl_Sat_fig6}. The dips, which are absent when 
recreating the figures using unsmeared event generator data, arise in regions of particularly poor event reconstruction. When 
the beam electron scatters at very negative pseudorapidities $(\eta < -3.0)$, the equation for the event inelasticity, $y$, 
approaches the form of $y = 1 - E^{\prime}/E$. In this limit, the $y$ reconstruction is ultra-sensitive to events with 
$E^{\prime} \sim E$ (at $y \ll 1$), in comparison to events with $E^{\prime} \neq E$. The photon flux factors that construct our 
figures’ cross sections are functions of $y$ (see~\ref{sec:AppI}), so that when they are calculated with pseudo-data, unreliable 
results are produced. The future Electron Ion Collider’s hermetic detector system will precisely measure event kinematics in 
this unreliable regime with alternative techniques such as the Jacquet-Blondel method or Double Angle method \cite{Handbook}. 
An analysis of our vector meson production events with these techniques is an area of future work.

\paragraph{$A^{2}Q^{-8}$ scaling:} For transversely polarized photons, one can perform a similar estimation as shown for the 
longitudinal polarization case in Eq.~(\ref{eq:scl_6}), bringing up
\bea
& & \Lb \Psi_{\gamma^{\ast} \rightarrow q\bar{q}}\Psi_{q\bar{q} \rightarrow V}^{\ast} \Rb_{T} \sim 
\nonumber\\
& & ~~~~~~~~~~ \sim \frac{1}{z(1 -z)}\,
\varepsilon\,K_{1}(\varepsilon\,r_{T})\,\partial_{r}(\phi_{T}(r_{T}, z))\,,
\label{eq:scl_8}
\eea
where the scalar part of the vector meson wavefunction now is $\phi_{T}(r_{T}, z) \sim z^{2}(1 - z)^{2} e^{-\rtt M_{V}^{2}}$.
Then, the diffractive transverse scattering amplitude will be given by
\bea
& & \mathcal{A}_{T}^{\gamma^{\ast}A\,\rightarrow\,V A} \sim i \int \dint \rt\,\mathbf{r}_{\mathbf{T}}^{4}\,Q\,K_{1}(Q\,r_{T}) \sim \frac{1}{Q^{4}} ,
\nonumber\\
& & {\mbox{leading~to}}~~~~\frac{\dint\sigma^{\gamma^{*}A\,\rightarrow\,V A}_{T} }{\dint t}\bigg\vert_{t = 0} \sim \frac{1}{Q^{8}} .
\label{eq:scl_9}
\eea
Fig.~\ref{fig:Scl_Sat_fig4} shows pseudo-data on exclusive coherent $\phi$ and $J/\Psi$ transverse electroproduction cross section
at $|t| = 5\times 10^{-4}\,{\rm GeV^{2}}$ in the IPSat model. Some features observed in the two plots of this figure by explanation 
are similar to those seen in Fig.~\ref{fig:Scl_Sat_fig3}. Strong suppression is present in both figures, nevertheless, the $Q^{8}$ 
transverse scaling is less accurate than the $Q^{6}$ longitudinal scaling because the transversely polarized photon contribution 
from large dipoles is not suppressed by high $Q^2$, because of strong dependence of the wavefunction overlap on the $z \rightarrow 0,1$ 
limits.

\subsubsection{The scaling onset as a function of small $|t|$}
\label{sec:Scaling1b}

In Eq.~(\ref{eq:ttepamplitude}), for the factor $\exp\!{\Lb -i[\bt - (1 - z)\rt]\!\cdot\!\Deltat \Rb}$ we can also use 
an approximation based on applying $|\bt| \gg |\rt|$. However, one should note that in Sar{\em t}re this phase factor in the same formula
actually has the form of $J_{0}([1 - z]r_{T}\Delta)\,e^{-i\bt\cdot\Deltat}$ \cite{Toll:2013gda,Toll:2012mb}. Thereby, one may find out 
whether the scaling picture discussed in the previous section exists as a function of small values of $t$. This can be done if we derive
the $t$-dependent $A^{2}$ scaling. In this case the diffractive amplitude for $\gamma^{\ast}A\,\rightarrow\,V A$ in a simplified 
version will be given by
\begin{displaymath}
\mathcal{A}_{T,L}^{\gamma^{\ast}A\,\rightarrow\,V A} \sim i \int{\rm d}^{2}{\bf r_{T}} {\rm d}^{2}\bt
\Lb \Psi_{\gamma^{\ast} \rightarrow q\bar{q}}\Psi_{q\bar{q} \rightarrow V}^{\ast} \Rb_{T,L} \times
\end{displaymath}
\beq
~~~~~~~~~~~~~~~~~~~\times J_{0}([1 - z]r_{T}\Delta)\,e^{-i\bt \cdot \Deltat}\,\rtt Q_{s,A}^{2} ,
\label{eq:scl_10}
\eeq
which can also be written as
\bea
& & \mathcal{A}_{T,L}^{\gamma^{\ast}A\,\rightarrow\,V A} \approx  \mbox{const} \times i A^{1/3} \int \bt\,e^{-i\bt \cdot \Deltat} 
{\rm d} \bt \times 
\nonumber\\
& & \times \int {\rm d}^{2}{\bf r_{T}} \Lb \Psi_{\gamma^{\ast} \rightarrow q\bar{q}}\Psi_{q\bar{q} \rightarrow V}^{\ast} \Rb_{T,L}
J_{0}([1 - z]r_{T}\Delta)\,\rtt .
\label{eq:scl_11}
\eea
After the $\bt$ integration from 0 to $b_{0,j}A^{1/3}$ we will have
\bea
& & \mathcal{A}_{T,L}^{\gamma^{\ast}A\,\rightarrow\,V A} \approx \mbox{const} \times 
\nonumber\\
& & \times \,i A^{1/3}\,\frac{e^{b_{0,j}A^{1/3}(-i\Delta_{k})} \left[1 + b_{0,j}A^{1/3}(i\Delta_{k}) \right] - 1}{\Delta_{k}^{2}} \times
\nonumber \\
& & \times \int {\rm d}^{2}{\bf r_{T}} \Lb \Psi_{\gamma^{\ast} \rightarrow q\bar{q}}\Psi_{q\bar{q} \rightarrow V}^{\ast} \Rb_{T,L} 
J_{0}([1 - z]r_{T}\Delta)\,\rtt\,,
\label{eq:scl_11}
\eea
where
\begin{itemize}
\item[(i)] for $b_{0,j}$ we take \\
$b_{0,Au} = 1.2\,{\rm fm} \rightarrow 6.082\,{\rm GeV^{-1}}$, $b_{0,Ca} = 1.017\,{\rm fm} \rightarrow 5.155\,{\rm GeV^{-1}}$ \cite{AnMar:2013} 
and $b_{0,p} = 0.831\,{\rm fm} \rightarrow 4.212\,{\rm GeV^{-1}}$ \cite{Xiong:2019umf};

\item[(ii)] for $\Delta_{k}$ we take the values of $t_{k}$ at $|t_{0}| = 5\times 10^{-4}\,{\rm GeV^{2}}$, 
$|t_{1}| = 3.5\times 10^{-3}\,{\rm GeV^{2}}$, $|t_{2}| = 6.5\times 10^{-3}\,{\rm GeV^{2}}$, 
$|t_{3}| = 9.5\times 10^{-3}\,{\rm GeV^{2}}$ and $|t_{4}| = 12.5\times 10^{-3}\,{\rm GeV^{2}}$.
\end{itemize}
Thus, the final result for the amplitude is the following:
\bea
& & \mathcal{A}_{T,L}^{\gamma^{\ast}A\,\rightarrow\,V A} \sim
\nonumber\\
& & \sim i A^{1/3}\,\frac{e^{b_{0,j}A^{1/3}(-i\Delta_{k})} \left[ 1 + b_{0,j}A^{1/3}(i\Delta_{k}) \right] - 1}{\Delta_{k}^{2}} ,
\label{eq:scl_12}
\eea
and the coherent diffractive cross section is given by
\bea
& & \frac{\dint\sigma^{\gamma^{*}A\,\rightarrow\,V A}_{T, L} }{\dint t}\bigg\vert_{t = t_{k}} \sim 
\nonumber\\
& & \sim A^{2/3} \left| \frac{e^{b_{0,j}A^{1/3}(-i\Delta_{k})} \left[ 1 + b_{0,j}A^{1/3}(i\Delta_{k}) \right] - 1}{\Delta_{k}^{2}} \right|^{2} ,
\label{eq:scl_13}
\eea
which at $t \equiv t_{0} \rightarrow 0$ is proportional to $A^{2}b_{0,j}^{4}$.

\begin{figure}[h!]
\begin{center}
   \subfigure{\includegraphics[width=0.475\textwidth,height=0.430\textwidth]{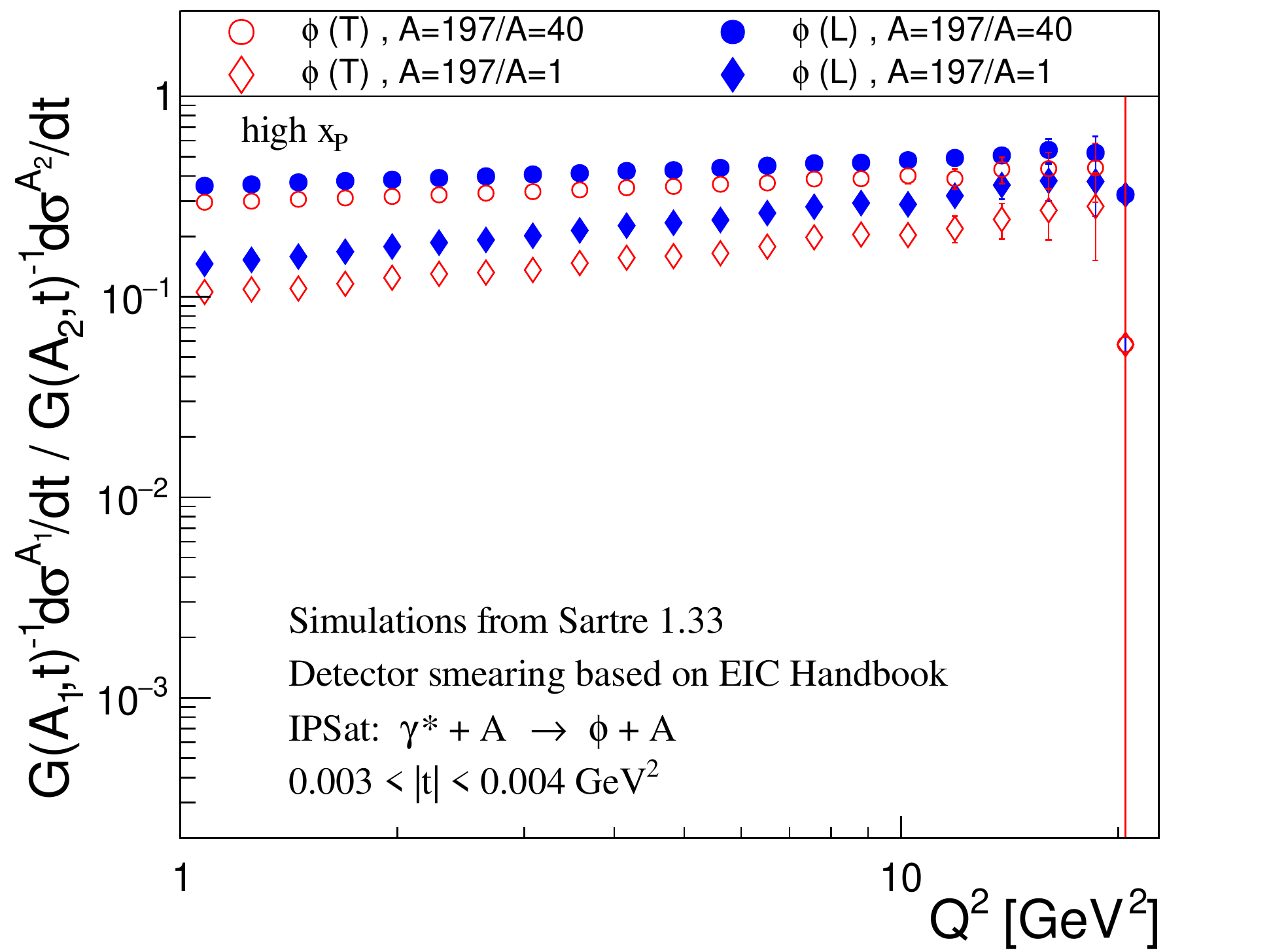}}
\hspace{0.0\textwidth}
   \subfigure{\includegraphics[width=0.475\textwidth,height=0.430\textwidth]{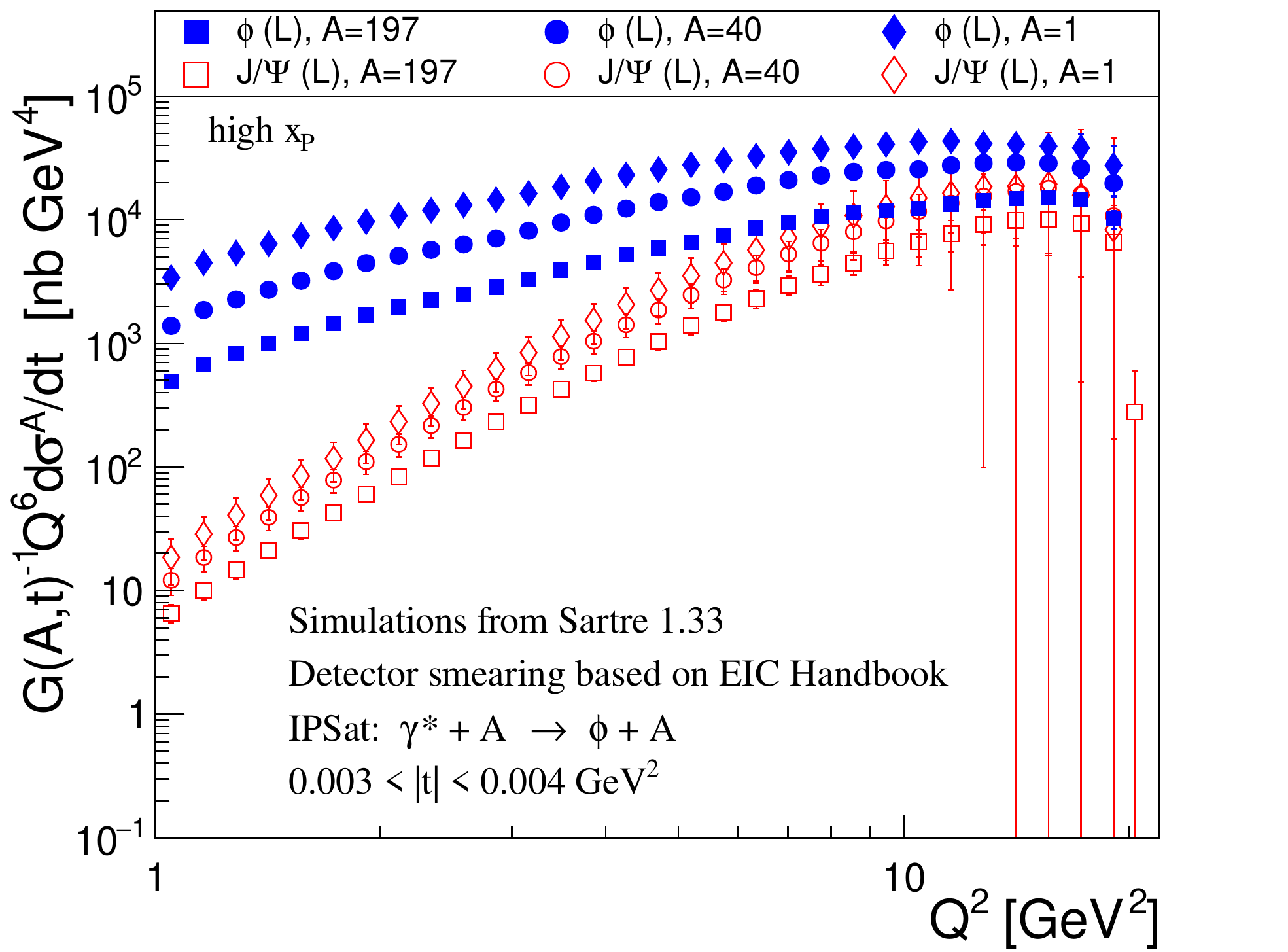}}
\end{center}
\vspace{-0.25cm}
\caption{(Color online)
The top plot shows the onset of the $G(A,t)$ scaling at $|t| = 3.5\times 10^{-3}\,{\rm GeV^{2}}$ for the exclusive coherent $\phi$
electroproduction cross section pseudo-data at high $\xpom$ in the IPSat model. The vertical axis shows the $G(A,t)^{-1}$ normalized 
ratio of the cross sections for the Gold over Calcium and the Gold over proton. The nomenclature is the same as in Fig.~\ref{fig:Scl_Sat_fig1}.
The bottom plot shows the onset of the $G(A,t)Q^{-6}$ scaling at $|t| = 3.5\times 10^{-3}\,{\rm GeV^{2}}$ for the exclusive coherent $\phi$ 
and $J/\Psi$ longitudinal electroproduction cross section pseudo-data at high $\xpom$ in the IPSat model for the Gold, Calcium and proton,
multiplied by the scaling factor of $Q^{6}G(A,t)^{-1}$. The nomenclature is the same as in Fig.~\ref{fig:Scl_Sat_fig3}.}
\label{fig:Scl_Sat_fig_t1}
\end{figure}
\begin{figure}[h!]
\begin{center}
   \subfigure{\includegraphics[width=0.475\textwidth,height=0.430\textwidth]{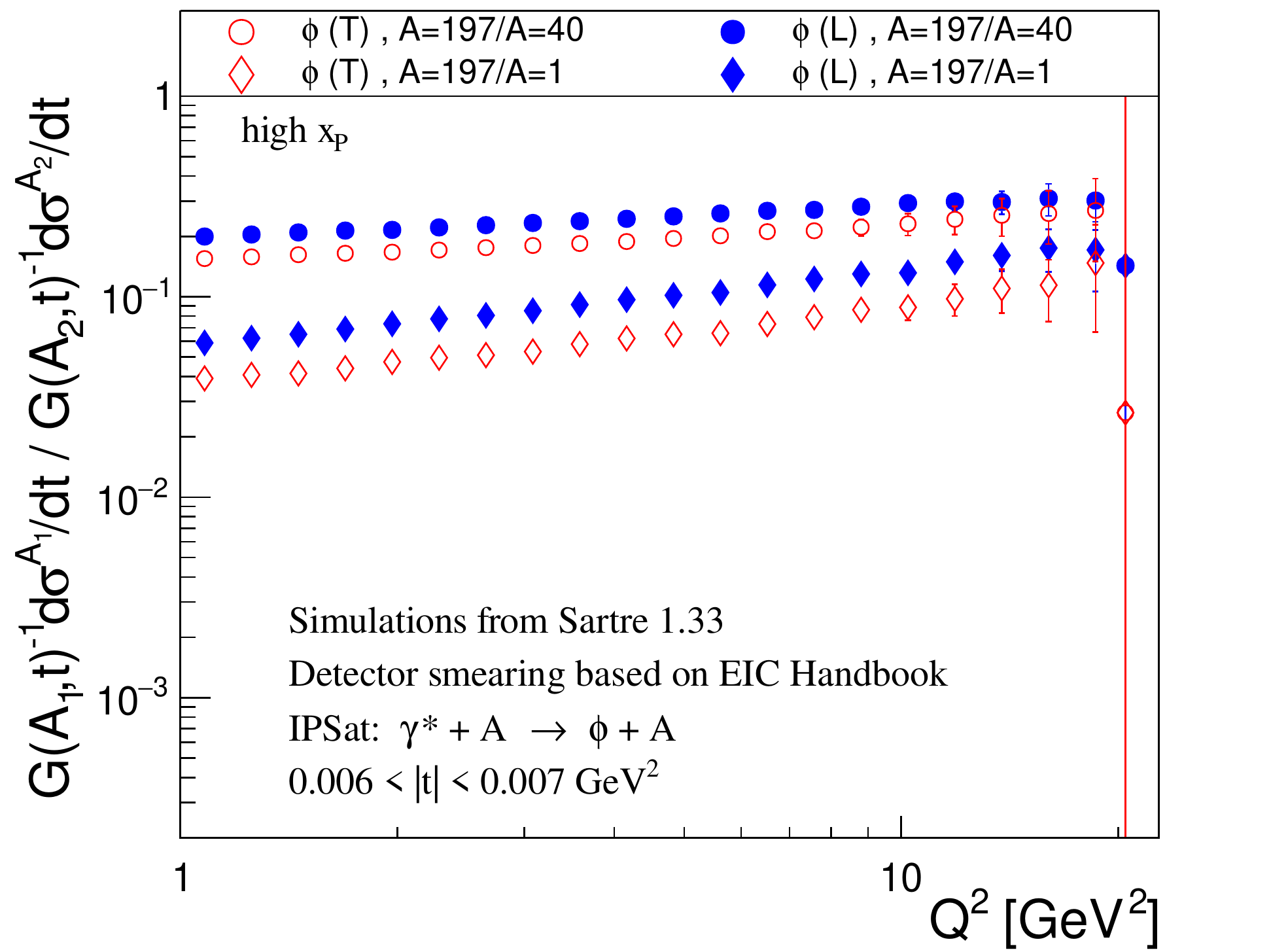}}
\hspace{0.0\textwidth}
   \subfigure{\includegraphics[width=0.475\textwidth,height=0.430\textwidth]{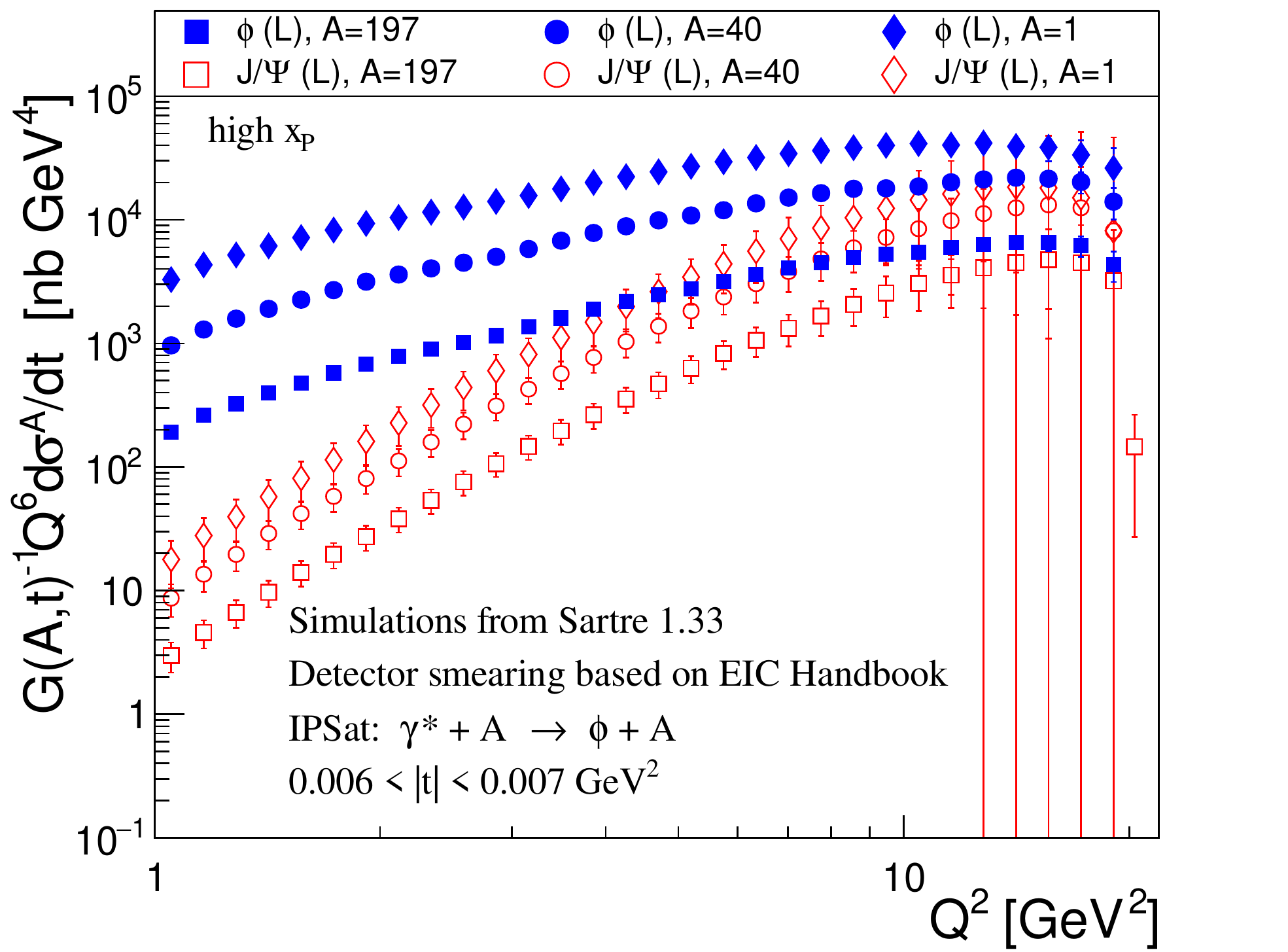}}
\end{center}
\vspace{-0.25cm}
\caption{(Color online)
Similar figure as in Fig.~\ref{fig:Scl_Sat_fig_t1} but at $|t| = 6.5\times 10^{-3}\,{\rm GeV^{2}}$.}
\label{fig:Scl_Sat_fig_t2}
\end{figure}

\begin{figure}[h!]
\begin{center}
   \subfigure{\includegraphics[width=0.475\textwidth,height=0.430\textwidth]{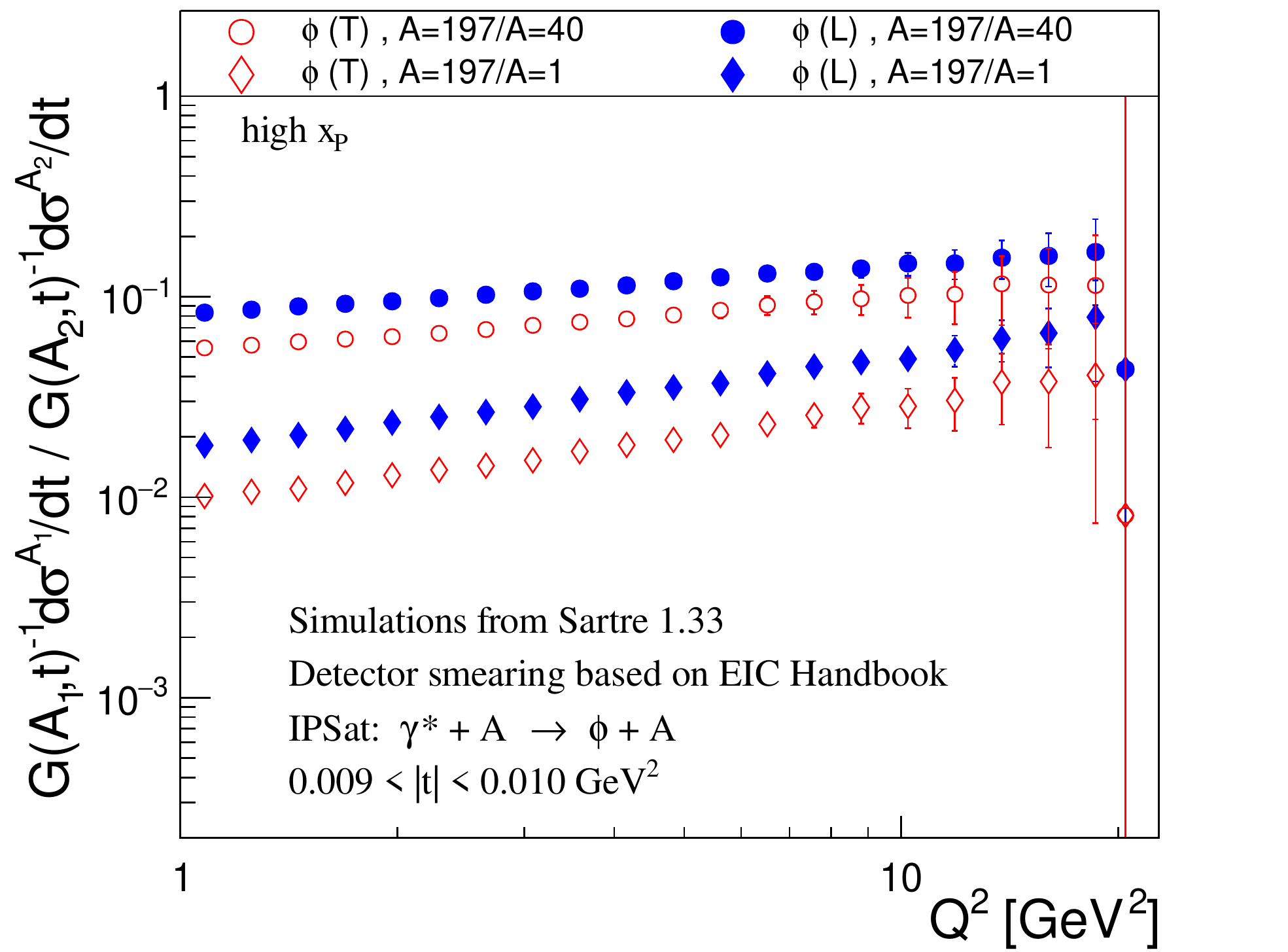}}
\hspace{0.0\textwidth}
   \subfigure{\includegraphics[width=0.475\textwidth,height=0.430\textwidth]{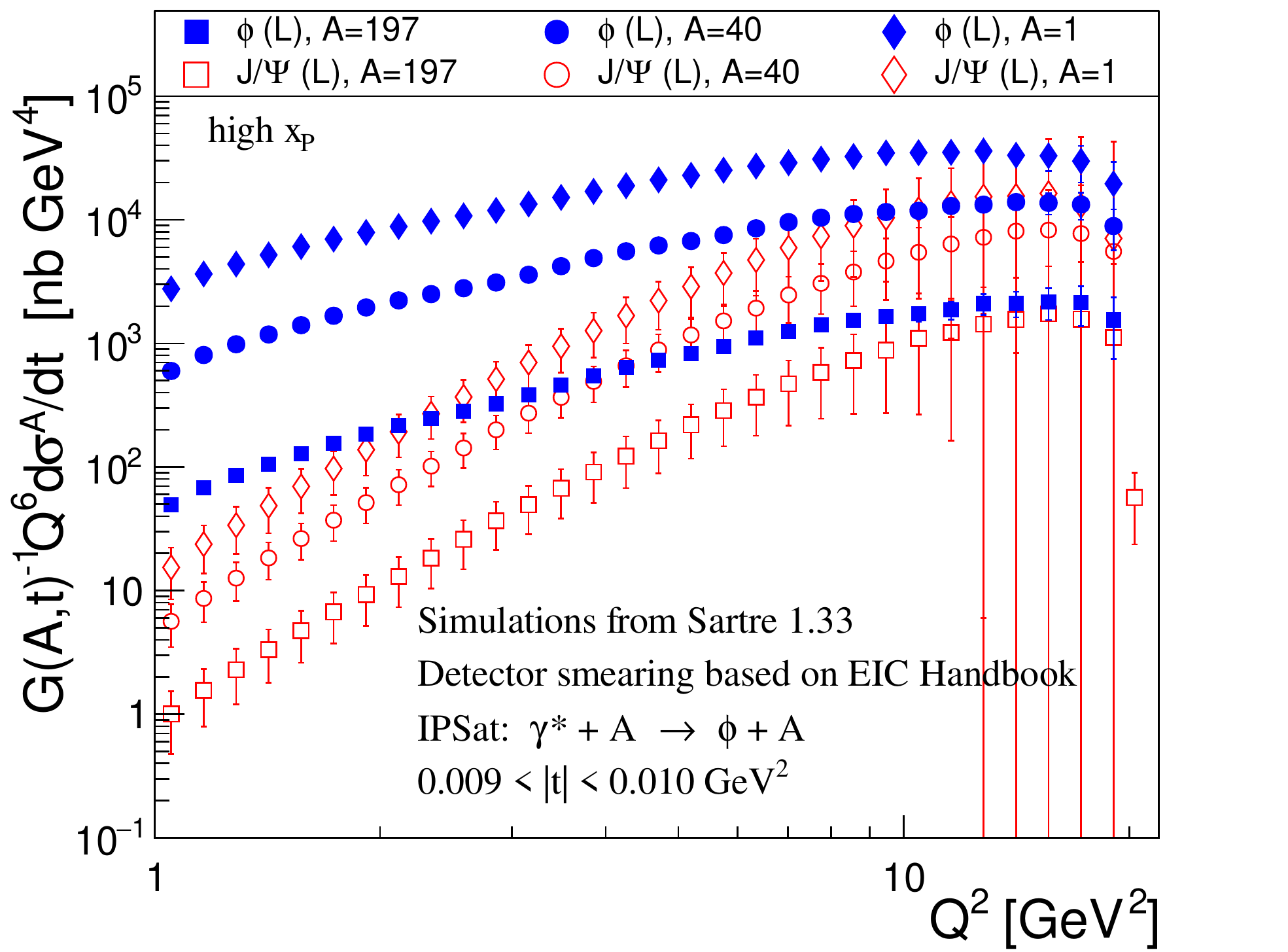}}
\end{center}
\vspace{-0.25cm}
\caption{(Color online)
Similar figure as in Fig.~\ref{fig:Scl_Sat_fig_t1} but at $|t| = 9.5\times 10^{-3}\,{\rm GeV^{2}}$.}
\label{fig:Scl_Sat_fig_t3}
\end{figure}

\begin{figure}[h!]
\begin{center}
   \subfigure{\includegraphics[width=0.475\textwidth,height=0.430\textwidth]{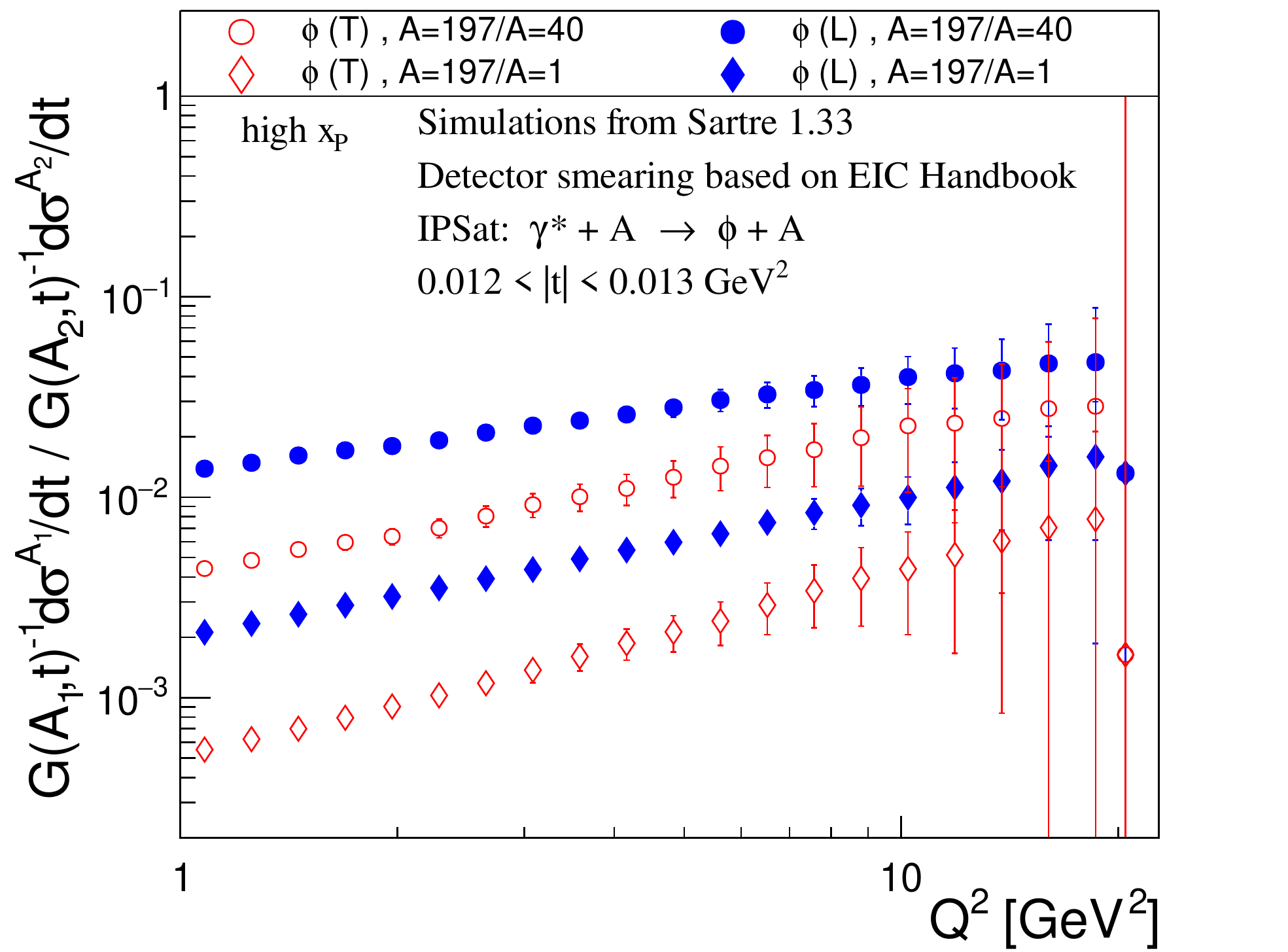}}
\hspace{0.0\textwidth}
   \subfigure{\includegraphics[width=0.475\textwidth,height=0.430\textwidth]{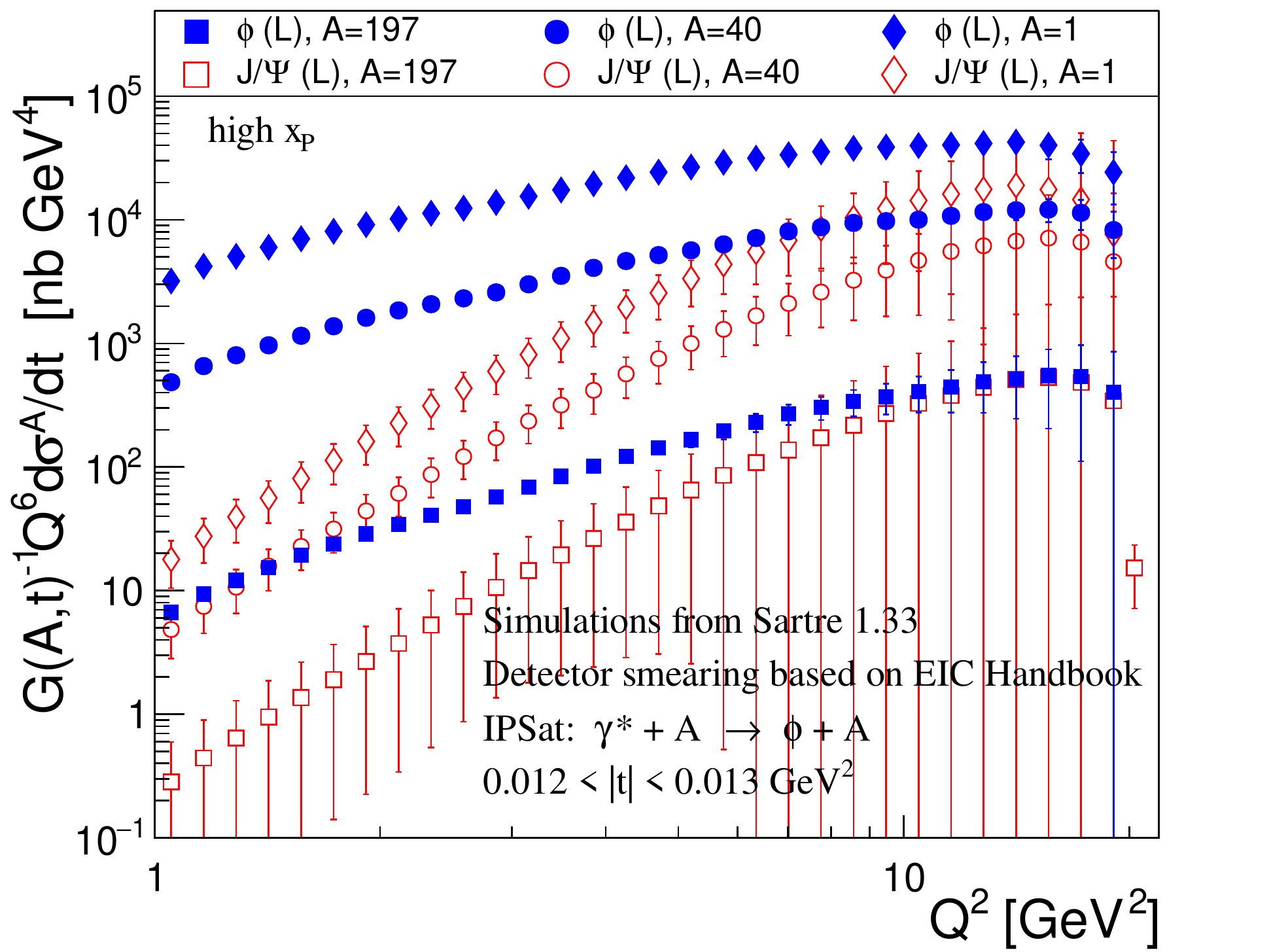}}
\end{center}
\vspace{-0.25cm}
\caption{(Color online)
Similar figure as in Fig.~\ref{fig:Scl_Sat_fig_t1} but at $|t| = 12.5\times 10^{-3}\,{\rm GeV^{2}}$.}
\label{fig:Scl_Sat_fig_t4}
\end{figure}

Let us designate the functional form in Eq.~(\ref{eq:scl_13}) divided by $b_{0,j}^{4}$ to be $G(A, t)$, which will be the normalization 
scaling factor for the cross section ratio or cross section at the above $t_{k}$ values. Therefore, the scaling factor in this formula 
is also a function of small $|t|$. In this case we can show, for example, the reproduction of the normalized cross section ratio of
Fig.~\ref{fig:Scl_Sat_fig1}, and the normalized cross section of Fig.~\ref{fig:Scl_Sat_fig3}, at the given four values of $t_{1}$, $t_{2}$,
$t_{3}$ and $t_{4}$. Thereby, we use the cross section asymptotic scaling factor defined as $G(A,t)$ that at $t_{0} \rightarrow 0$ gives 
$A^{2}$, which is the same as the factor in Eq.~(\ref{eq:scl_4}).
Thus, one can see how the behavior of the scaling patterns of the cross section ratio and the longitudinal cross section change as a 
function of small values of $|t|$ shown correspondingly in Figs.~\ref{fig:Scl_Sat_fig1}~and~\ref{fig:Scl_Sat_fig3} as well as in
Figs.~\ref{fig:Scl_Sat_fig_t1}-\ref{fig:Scl_Sat_fig_t4}. Contrary to the high-$\xpom$ and low-$\xpom$ ranges used in the previous section, 
we employ the entire range of $\xpom = [0.001 - 0.008]$ in producing these figures.
In particular, in the cross section ratio figures of the Gold over Calcium and the Gold over proton, we see the suppression in 
magnitude of the pseudo-data curves. Meanwhile, the shapes of the ratios, which have rising behaviors as a function of $Q^{2}$,
survive up to the case of $|t| = 12.5\times 10^{-3}\,{\rm GeV^{2}}$. Consequently, within the approximations used we may assume 
that the saturation pattern, which is observable from the pseudo-data in Fig.~\ref{fig:Scl_Sat_fig1} also occur as a function of 
small values of $|t|$, at least, in Figs.~\ref{fig:Scl_Sat_fig_t1}-\ref{fig:Scl_Sat_fig_t4}.

At the end of this section one should stress that in order to accurately reflect an experimental binning of $t$ in a given range, 
e.g., for the cross-section ratio in Fig.~\ref{fig:Scl_Sat_fig1}, we allow the event generator to produce events in a $t$ range larger 
than what is binned. For all the figures binned in $|t| = [0.000 - 0.001]\,{\rm GeV^{2}}$, we select our simulated $t$ range to 
extend to $|t| = [0.000 - 0.002]\,{\rm GeV^{2}}$. In doing so, we account for the realistic possibility that an event simulated outside 
the binning range will become smeared into the very same range. For instance, an event simulated with $|t| \sim 0.0011\,{\rm GeV^{2}}$ 
can be reconstructed as $|t| \sim 0.0009\,{\rm GeV^{2}}$, which places it into the $|t| = [0.000 - 0.001]\,{\rm GeV^{2}}$ binning range. 
Meanwhile, for the cross-section ratio in Fig.~\ref{fig:Scl_Sat_fig_t1}, we select the minimum and maximum $t$ to be separated by 
$\pm 0.0005\,{\rm GeV^{2}}$ on each side of that figure's $t$ binning range. Thereby, in this case we select our simulated $t$ range 
to be $|t| = [0.0025 - 0.0045]\,{\rm GeV^{2}}$. Then, an outside $t$ event, say, at $|t| \sim 0.0041\,{\rm GeV^{2}}$, can be reconstructed 
as $|t| \sim 0.0039\,{\rm GeV^{2}}$, placing it into the $|t| = [0.003 - 0.004]\,{\rm GeV^{2}}$ binning range. This procedure is also 
the case for the other figures (with other $t$ bins) displayed in this section.

\subsection{$A$ and $Q^{2}$ scaling picture in the regime of low $Q^{2} < Q_{s,A}^{2}$}
\label{sec:Scaling2}
Here we discuss the asymptotics in which one can approximate $\rtt Q_{s,A}^{2}$ in the limit of $Q^{2} \ll Q_{s,A}^{2}$.
In this limit the $\gamma^{\ast}$-$A$ cross section is represented by
\beq
\frac{{\rm d}\sigma_{q\bar q}^{A}}{{\rm d}^{2} \bt} = 2\left[ 1 - \exp{\!(- \rtt Q_{s,A}^{2})} \right] \rightarrow 2 ,
\label{eq:scl_1b}
\eeq
and at $t = 0$ the diffractive scattering amplitude from Eq.~(\ref{eq:ttepamplitude}) becomes
\bea
& & \mathcal{A}_{T,L}^{\gamma^{\ast}A\,\rightarrow\,V A} \sim 
\nonumber\\
& & ~~~~~\sim i \int{\rm d}^{2}{\bf r_{T}} {\rm d}^{2} 
\bt \Lb \Psi_{\gamma^{\ast} \rightarrow q\bar{q}}\Psi_{q\bar{q} \rightarrow V}^{\ast} \Rb_{T,L} \times 2 .
\label{eq:scl_2b}
\eea
In Eq.~(\ref{eq:scl_2b}) the $Q^{2}$ and $A$ dependence is determined by the dipole radius scale, and how it affects
the wavefunction overlap. Let us again follow Ref.~\cite{Mantysaari:2017slo} for looking into two scaling behaviors deep in the
saturation domain.

\begin{figure}[h!]
\begin{center}
   \subfigure{\includegraphics[width=0.465\textwidth,height=0.420\textwidth]{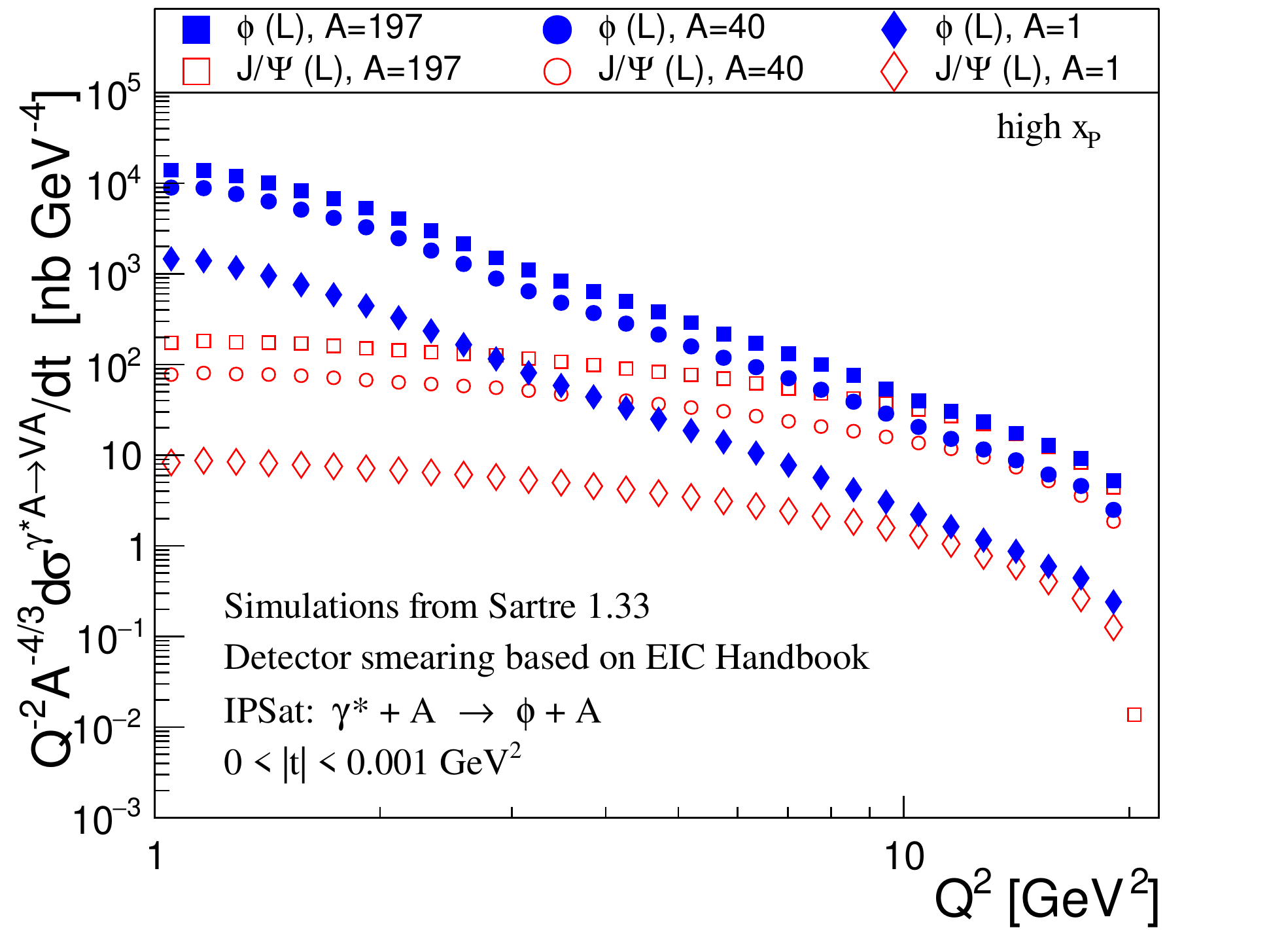}}
\hspace{0.0\textwidth}
   \subfigure{\includegraphics[width=0.465\textwidth,height=0.420\textwidth]{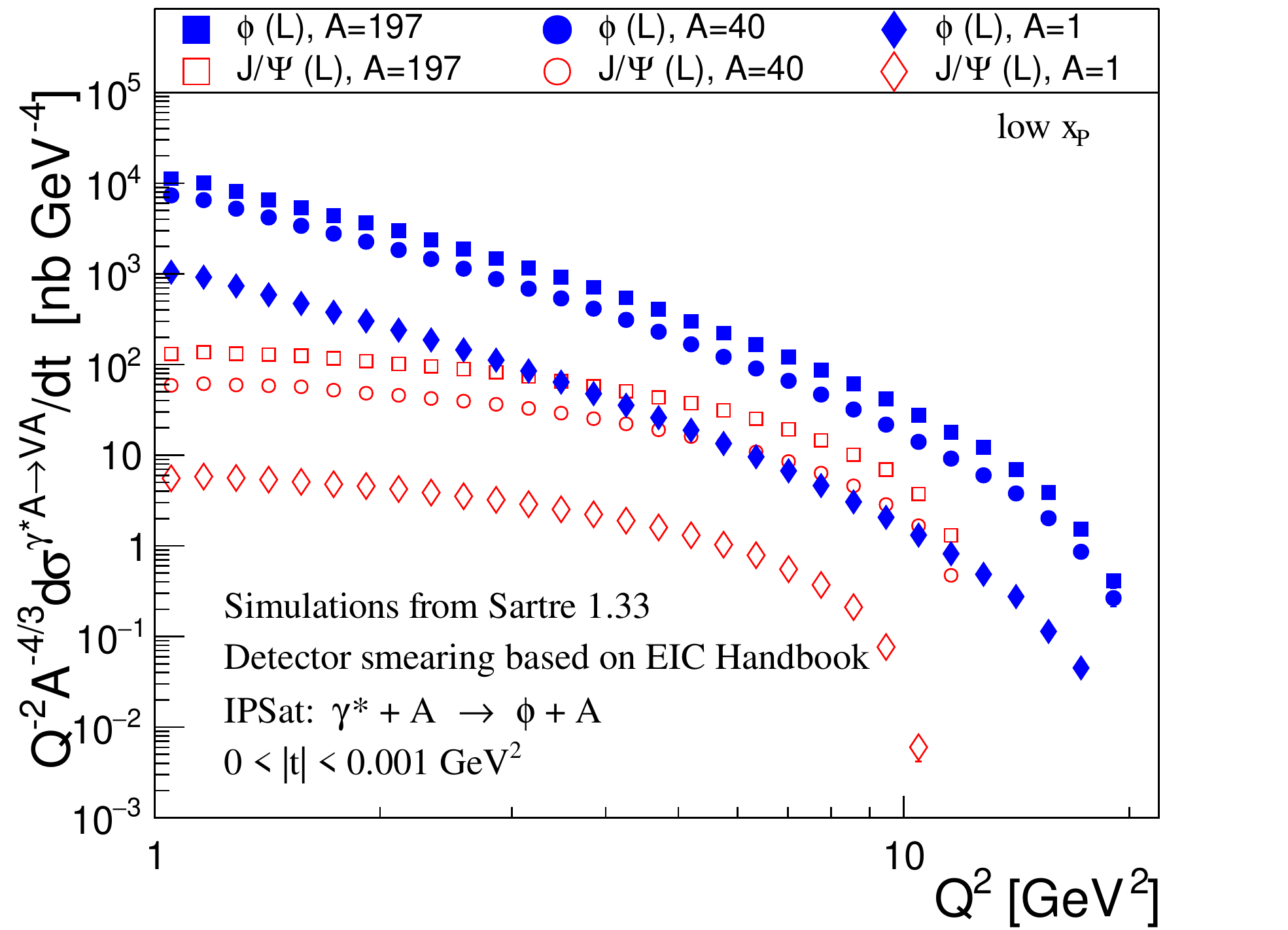}}
\end{center}
\vspace{-0.25cm}
\caption{The cross sections for the Gold, Calcium and proton at $|t| = 5\times 10^{-4}\,{\rm GeV^{2}}$ for the exclusive 
coherent $\phi$ and $J/\Psi$ longitudinal electroproduction in the IPSat model, multiplied by the factor $Q^{-2}A^{-4/3}$. 
The nomenclature is the same as in Fig.~\ref{fig:Scl_NonSat_fig}. The top plot shows the pseudo-data at high $\xpom$, the 
bottom plot shows the pseudo-data at low $\xpom$.}
\label{fig:Scl_Sat_fig5}
\end{figure}

\paragraph{$A^{4/3}Q^{2}$ scaling:}
Eq.~(\ref{eq:scl_6}) shows the overlap of the vector meson wavefunction with the longitudinally polarized photon wavefunction. 
One can use it in the saturation region but with $\varepsilon = \sqrt{Q^{2}z(1 - z) + m_{q}^{2}} \approx m_{q}$ for $Q \ll m_{q}$. 
Then, the diffractive longitudinal scattering amplitude reduces to
\beq
\mathcal{A}_{L}^{\gamma^{\ast}A\,\rightarrow\,V A} \sim i Q \int_{1/Q_{s,A}}^{1/m_{q}} \dint \rt\,\rt\,K_{0}(m_{q}\,r_{T}) .
\label{eq:scl_3b}
\eeq
The ultimate result in the limit of low-$Q^{2}$ is given by
\bea
& & \!\!\mathcal{A}_{L}^{\gamma^{\ast}A\,\rightarrow\,V A} \sim i Q\!\left( \mbox{const} + \mathcal{O}\left( \frac{m_{q}^{2}}{Q_{s,A}^{2}},
~\frac{M_{V}^{2}}{Q_{s,A}^{2}},~\frac{Q^{2}}{Q_{s,A}^{2}} \right) \right)
\nonumber\\
& & \Rightarrow ~\frac{\dint\sigma^{\gamma^{*}A\,\rightarrow\,V A}_{L} }{\dint t}\bigg\vert_{t = 0} \sim Q^{2}\,.
\label{eq:scl_4b}
\eea
Fig.~\ref{fig:Scl_Sat_fig5} shows pseudo-data of the exclusive coherent $\phi$ and $J/\Psi$ longitudinal electroproduction cross section
at $|t| = 5\times 10^{-4}\,{\rm GeV^{2}}$ in the IPSat model, where contrary to Fig.~\ref{fig:Scl_Sat_fig3} the cross section is 
now multiplied by the scaling factor $Q^{-2}$, and is also scaled in $A$ by the asymptotic analytical expectation $A^{4/3}$. For $J/\Psi$ 
production, the cross sections become flatter at low-$Q^{2}$ when scaled by $Q^{2}$. For $\phi$ production, this trend could be obtained 
at asymptotically small values of $Q^{2}$ (not shown here), which is beyond applicability of the approximations used.

\begin{figure}[h!]
\begin{center}
   \subfigure{\includegraphics[width=0.465\textwidth,height=0.420\textwidth]{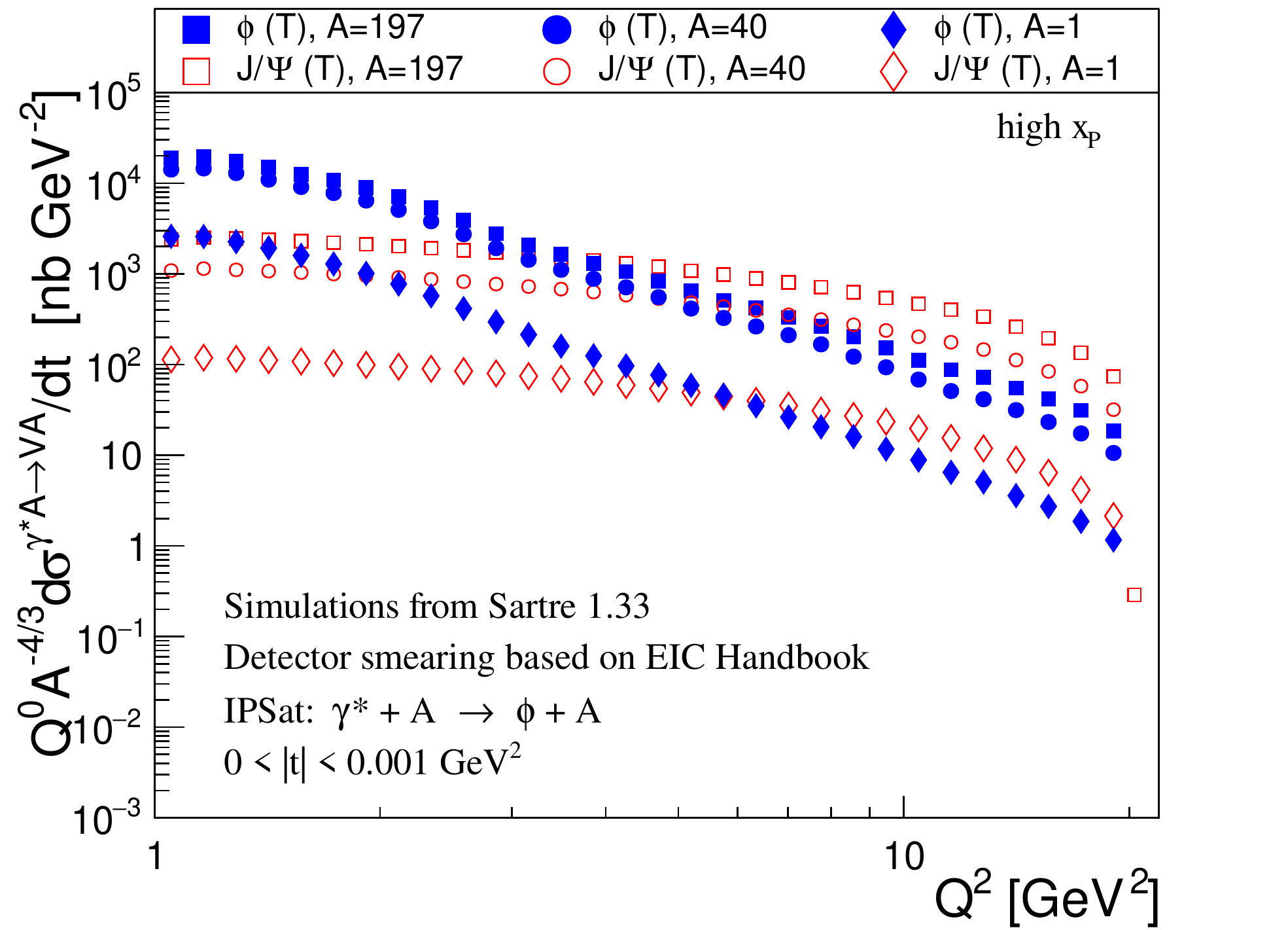}}
\hspace{0.0\textwidth}
   \subfigure{\includegraphics[width=0.465\textwidth,height=0.420\textwidth]{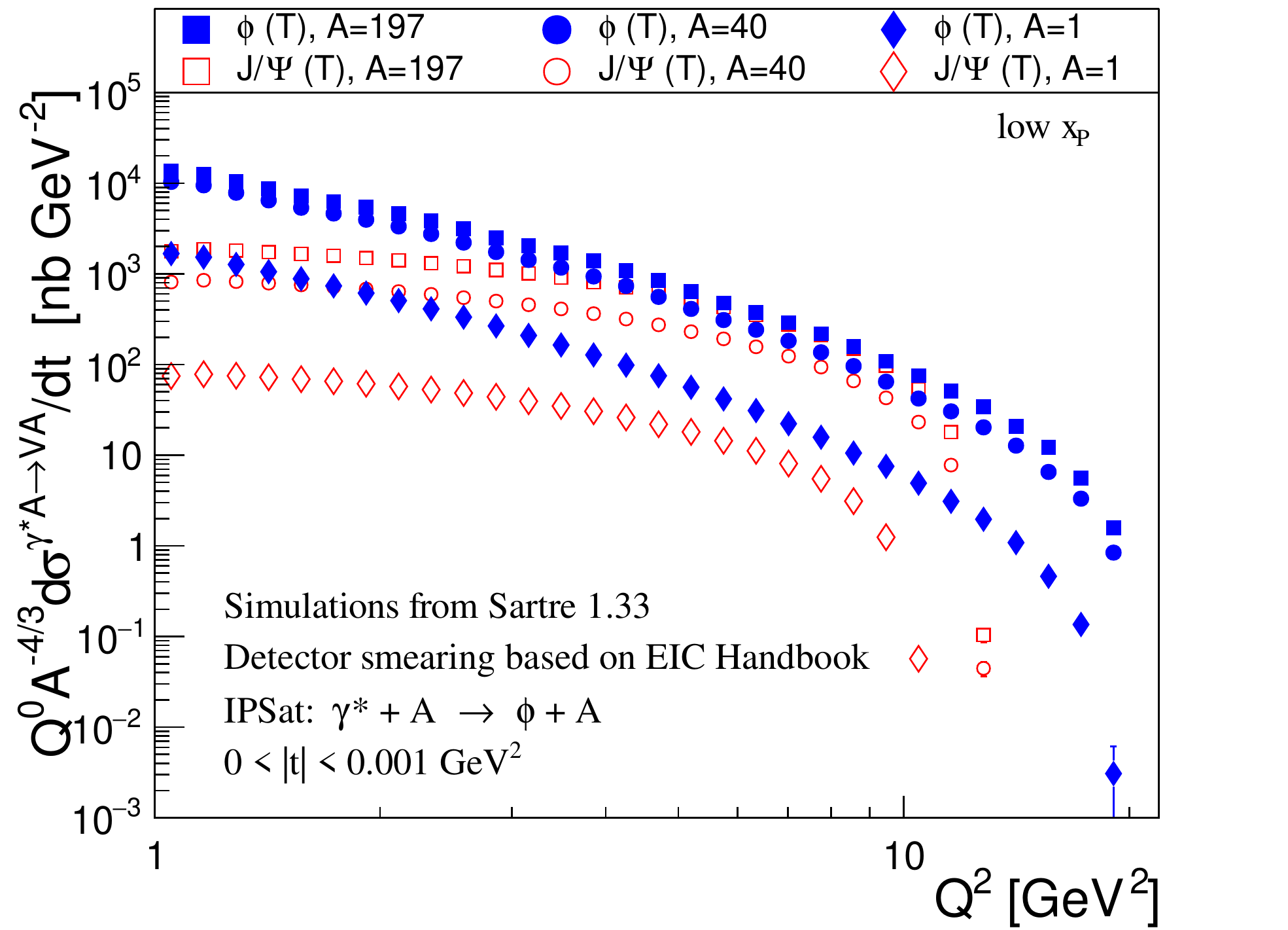}}
\end{center}
\vspace{-0.25cm}
\caption{The cross sections for the Gold, Calcium and proton at $|t| = 5\times 10^{-4}\,{\rm GeV^{2}}$ for the exclusive 
coherent $\phi$ and $J/\Psi$ transverse electroproduction in the IPSat model, multiplied by the factor $Q^{0}A^{-4/3}$. The 
explanation in Fig.~\ref{fig:Scl_Sat_fig5} is applied here as well, except for that the cross section is $Q$-independent. 
The nomenclature is the same as in Fig.~\ref{fig:Scl_NonSat_fig}. The top plot shows the pseudo-data at high $\xpom$, the 
bottom plot shows the pseudo-data at low $\xpom$.}
\label{fig:Scl_Sat_fig6}
\end{figure}

\paragraph{$A^{4/3}Q^{0}$ scaling:} Here is the case of transversely polarized photons, where one can make use of
Eq.~(\ref{eq:scl_8}) with the above approximation $\varepsilon \approx m_{q}$.
\beq
\mathcal{A}_{T}^{\gamma^{\ast}A\,\rightarrow\,V A} \sim i m_{q} \int_{1/Q_{s,A}}^{1/m_{q}} \dint \rt\,\rtt\,K_{1}(m_{q}\,r_{T})\,.
\label{eq:scl_5b}
\eeq
The ultimate result in the limit of low-$Q^{2}$ is given by
\begin{displaymath}
\mathcal{A}_{T}^{\gamma^{\ast}A\,\rightarrow\,V A} \sim i Q^{0}\!\left( \mbox{const} + \mathcal{O}\left( \frac{m_{q}^{2}}{Q_{s,A}^{2}},~
\frac{M_{V}^{2}}{Q_{s,A}^{2}},~\frac{Q^{2}}{Q_{s,A}^{2}} \right) \right)
\end{displaymath}
\beq
\Rightarrow ~\frac{\dint\sigma^{\gamma^{*}A\,\rightarrow\,V A}_{T} }{\dint t}\bigg\vert_{t = 0} \sim Q^{0} .
\label{eq:scl_6b}
\eeq
Fig.~\ref{fig:Scl_Sat_fig6} shows pseudo-data of the exclusive coherent $\phi$ and $J/\Psi$ transverse electroproduction cross 
section at $|t| = 5\times 10^{-4}\,{\rm GeV^{2}}$ in the IPSat model, where contrary to Fig.~\ref{fig:Scl_Sat_fig4} the 
cross section is now $Q$-independent, and is scaled in $A$ by the asymptotic analytical expectation $A^{4/3}$.

\section{Conclusions and outlook}
\label{sec:outlook}
The perturbative QCD cross section in exclusive vector meson production processes in high-energy DIS of electrons scattered off 
nuclei is proportional to the squared nuclear gluon distribution according to \cite{Ryskin:1992ui}, by which the measurements of 
exclusively produced vector mesons might be sensitive to gluon distributions at the regime of saturation (since these distributions
grow rapidly at small $x$). Consequently, the systematics that could have been determined from such measurements, based upon 
the presence of saturation, might be conspicuously different from the systematics of those measurements for which the saturation would 
be absent. With calculations performed to leading logarithmic accuracy, Ref.~\cite{Mantysaari:2017slo} has demonstrated that
the $A$ and $Q^{2}$ scaling properties of the cross section in exclusive vector meson production is substantially modified by 
saturation effects, taking place in the crossover region between the perturbative QCD and saturation regimes.

The results demonstrated in our paper seem to confirm this scaling picture, which one would expect from a realistic ``EIC setup" based upon 
the Monte Carlo event generator Sar{\em t}re. But before addressing the scaling problem, we first generated and analyzed pseudo-data 
to extract the coherent and incoherent cross section distributions of exclusive $J/\Psi$ and $\phi$ production in diffractive $e+Au$ 
scattering with an integrated luminosity of $10\,{\rm fb}^{-1}/A$ and the scattering energies $10 \times 110\,{\rm GeV}$, by gradually 
smearing the diffractive peak for increasing relative $t$ uncertainty. These results are given in 
Figs.~\ref{fig:white_plot2} and \ref{fig:white_plot5}. For obtaining all the pseudo-data along with the projected uncertainties, we have used
detector resolutions and smearing functions from the EIC Yellow Report \cite{AbdulKhalek:2021gbh} and the EIC Detector Handbook \cite{Handbook}.

Afterwards, based on the same framework, we qualitatively reproduced the cross section and cross-section 
ratio results of \cite{Mantysaari:2017slo}. Here we need to emphasize the word ``qualitatively'' because
\begin{itemize}
\item[$\bullet$] we used somewhat different kinematics than that used in \cite{Mantysaari:2017slo} to obtain pseudo-data on exclusive 
$J/\Psi$ and $\phi$ production for an integrated luminosity of $10\,{\rm fb}^{-1}/A$ with beam energies $10 \times 110\,{\rm GeV}$ in 
diffractive $e + Au$/$e+Ca$ scatterings, and for $100\,{\rm fb}^{-1}/A$ with $10 \times 100\,{\rm GeV}$ in diffractive $e + p$ scattering;
\item[$\bullet$] our considered $Q^{2}$ range is limited within $1\,{\rm GeV}^{2} < Q^2 < 20\,{\rm GeV}^{2}$ as compared to
much larger range used in \cite{Mantysaari:2017slo};
\item[$\bullet$] we studied $\phi$ production, instead of the $\rho$ production discussed in \cite{Mantysaari:2017slo};
\item[$\bullet$] in our Sar{\em{t}}re simulations we had to consider two regions of $\xpom$, instead of two fixed $\xpom$ values used in 
\cite{Mantysaari:2017slo}.
\end{itemize}
Our results, though obtained from a restricted kinematics in $Q^{2}$, in principle substantiate the main conclusions of \cite{Mantysaari:2017slo} 
within projected uncertainties. The scaling relations are visualized in Figs.~\ref{fig:Scl_NonSat_fig} - \ref{fig:Scl_Sat_fig4}, as well as in 
Figs.~\ref{fig:Scl_Sat_fig5} and \ref{fig:Scl_Sat_fig6}. In particular, the pseudo-data in Fig.~\ref{fig:Scl_Sat_fig1} shows a sign of gluon 
saturation, in terms of a low-$Q^{2}$ part of the asymptotic $A^{2}$ scaling at $|t_{0}| = 5\times 10^{-4}\,{\rm GeV^{2}}$ in the IPSat 
model. While if we integrate over $t$, we will have a low-$Q^{2}$ part of the $A^{4/3}$ scaling that is shown in Fig.~\ref{fig:Scl_Sat_fig2}. 
A larger $Q^{2}$ coverage for both scaling cases cannot be shown because of the lack of pseudo-data above $Q^{2} \sim 20\,{\rm GeV^{2}}$.
In the absence of the saturation effects (as in the IPNonSat model), the $A^{2}$ scaling should look like what is demonstrated in
Fig.~\ref{fig:Scl_NonSat_fig}, where the trend is expected to be similar at high $Q^{2}$ too. In the short \ref{sec:AppIII}, one can also see a 
comparison of some of our results (shown, e.g., in Figs.~\ref{fig:Scl_Sat_fig1}~and~\ref{fig:Scl_Sat_fig3}) with the IPSat model calculations 
from Sar{\it t}re without taking the experimental acceptance and resolution into account.

In addition, we discussed the scaling onset at values of $|t|$ equal to 
$|t_{1}| = 3.5\times 10^{-3}\,{\rm GeV^{2}}$, $|t_{2}| = 6.5\times 10^{-3}\,{\rm GeV^{2}}$, 
$|t_{3}| = 9.5\times 10^{-3}\,{\rm GeV^{2}}$, and $|t_{4}| = 12.5\times 10^{-3}\,{\rm GeV^{2}}$, where we see the extent to which 
the suppression pattern seen due to gluon saturation survives as a function of small values of $|t|$ within all the approximations used (see 
Figs.~\ref{fig:Scl_Sat_fig_t1}-\ref{fig:Scl_Sat_fig_t4}). Also, the $\phi$ and $J/\Psi$ production cross sections seem to have different $Q^{2}$ 
scaling behavior, which is to be expected as they probe different dipole sizes, which are directly dependent on $Q^{2}$. The $J/\Psi$ particle 
probes very small dipoles at all $Q^{2}$, such that the cross section is not very dependent on $Q^{2}$ as shown in 
Figs.~\ref{fig:Scl_Sat_fig5}~and~\ref{fig:Scl_Sat_fig6}. In these two figures the trend is that the pseudo-data curves become flatter at 
low-$Q^{2}$, which means that the expected scaling is observed. However, in this case, one should point out that due to small masses of 
the light vector mesons (like $\phi$), the results in the low-$Q^{2}$ region are at the edge of applicability of the employed weak-coupling 
framework \cite{Mantysaari:2017slo}.

In the next studies we may include more elaborate simulations/calculations of the scaling as a function of $|t|$ discussed in Sec.~\ref{sec:Scaling1b}, 
by including also pseudo-data on vector meson production for $e + d$, $e + {}^{4}He$, $e + C$, $e + Cu$ collision species, in addition to 
$e + Au$, $e + Ca$, and $e + p$, considered in this paper. This means that new look-up IPSat and IPNonSat tables (see~\ref{sec:AppI}) 
should be added to Sar{\em t}re, along with adding an IPNonSat look-up tables for $e + Ca$ that are currently absent in Sar{\em t}re\footnote{
It is anticipated that full-scale studies of gluon saturation at EIC will largely benefit from collisions, which include light-, medium- and large-sized nuclei.
Therefore, it would be highly desirable if the EIC community considers of having, e.g., $d$, $Ca$ and $Cu$ as parts of the baseline 
beam species, such as $p$ and $Au$.}. 
In new studies, it will also be important to incorporate the energy mode of $10 \times 41\,{\rm GeV}$, which is common for all those collision 
species according to Table~10.3 in \cite{AbdulKhalek:2021gbh}, thereby, the energy difference of the proton and nuclear beams will be zero. 
For the purpose of more precise simulations, one also needs to determine new $t$-resolution values from the ``Method L" 
discussed in Sec.~\ref{sec:White_paper}, at the $Q^{2}$ interval of $1 < Q^{2} < 20\,{\rm GeV}^{2}$ (or even larger), including the 
reconstruction of the final-state kaon decay too. This latter particular study should include extraction of those resolution values for both 
$J/\Psi$ and $\phi$ production. Besides, the $Q^{2}$ scaling itself is interesting and worth of additional detailed investigation, since it will
be a unique feature at EIC. Note that LHC and RHIC produce vector mesons in UPC at $Q^{2}$ = 0. At the end, let us mention that an 
interesting analysis could also be a generalization of our studies in the context of the geometric scaling phenomenon at small-$x$ region, 
where data on exclusive vector meson production and DVCS can be described with a universal scaling function \cite{Ben:2017xny}.

At the end, we wish to indicate ongoing issues stemming from a detector design we have used in this analysis, which makes us even more 
determined to keep our current project continuing with more developments and updates upcoming in the future. First of all, it is essential to 
compound the outputs of Sar{\it t}re simulations with the full detector geometry that will draw out unique challenges being absent in a 
simple smearing study used to make Fig.~\ref{fig:white_plot2}, and still being not present in the promising ``Method L"  that we used 
to make Fig.~\ref{fig:white_plot5} (and the scaling figures in this paper). These challenges include vetoing incoherent events in the
subsequent stage of the current analysis, working closely with the EIC community and paying attention on its current and future progress
related to this matter. We should also focus on having better particle identification by utilizing the tracking and electromagnetic calorimetry, 
beam angles, etc\footnote{Since our analysis was focused on and the scaling results for ``indirectly observing'' some hints of gluon saturation 
were produced in the momentum range of $1 < Q^{2} < 20\,{\rm GeV}^{2}$, then we would think of having a detector PID for such 
measurements in the future at EIC, at least, in that specified range of $Q^{2}$.}. 
Furthermore, it will be imperative to employ the experimental resolutions taken from a meticulously developed design of the 
ECCE (EIC Comprehensive Chromodynamics Experiment) Detector \cite{ECCE},  instead of using the handbook's somewhat outdated information 
taken from \cite{Handbook} (although the differences between the energy and/or tracking resolutions from these two documents seem to be minor).

It is also essential to emphasize that based on the work performed in the proposals of the ECCE Detector \cite{ECCE} and ATHENA Detector 
(A Totally Hermetic Electron Nucleus Apparatus) \cite{ATHENA:2022hxb}, it is a fact that the vetoing of incoherent events for both the ECCE 
and ATHENA designs is a major challenge (see also the Yellow Report \cite{AbdulKhalek:2021gbh}). This challenge is a driving force for guiding 
the detector development of the far-forward region since much of the exclusive physics depends on resolving the coherent peaks from an
exclusive production process. The assumption of a pure isolation of the coherent events is of course crude, especially in the region of interest 
(at moderate $|t|$) discussed in Sec.~\ref{sec:White_paper}, where the incoherent events for both $J/\Psi$ and $\phi$ production dominate 
their coherent counterparts. Nevertheless, since the main results of our paper are the scaling plots shown in 
Figs.~\ref{fig:Scl_NonSat_fig} - \ref{fig:Scl_Sat_fig4}, as well as in Figs.~\ref{fig:Scl_Sat_fig5} and \ref{fig:Scl_Sat_fig6}, all obtained near 
$|t|=0$, one can have real expectations for isolating here the coherent events from incoherent ones. This may very well be the case even for the 
nonzero fixed-$|t|$ scaling plots, given by  Figs.~\ref{fig:Scl_Sat_fig_t1} - \ref{fig:Scl_Sat_fig_t4}, because the considered $|t|$ regions there
are still quite small. If only the moderate to large $|t|$ region is unreliable due to detector limitations, then these low-$|t|$ results are promising 
candidates for a further analysis. On the other hand, the result in Fig.~\ref{fig:Scl_Sat_fig2} made in the integrated range of 
$0 < |t| < 0.5\,{\rm GeV^{2}}$ need some reconsideration since we extract physics from a region, where it is currently unclear 
how the dominating incoherent events may pollute the coherent scaling behavior.

\section*{Acknowledgments}
\label{sec:ack}
We are very grateful to Heikki M\"antysaari and Bj\"orn Schenke for the reading of the manuscript and giving us extremely 
useful comments, based on which the results of the paper have been improved. We are also thankful to Abhay Deshpande, 
Haiyan Gao, Barak Schmookler, Tobias Toll, Thomas Ullrich and Raju Venugopalan for fruitful and informative discussions on 
the subject matter. The work of Gregory Matousek is supported in part by Duke University. The work of Vladimir Khachatryan 
is supported in part by the U.S. Department of Energy, Office of Science, Offices of Nuclear Physics under contract DE-FG02-03ER41231. 
The work of Jinlong Zhang is supported by the Qilu Youth Scholar Funding of Shandong University.

\section*{Data Availability Statements}
\label{sec:data_avail}
The Monte-Carlo generated pseudo-data sets obtained and analyzed during the current study, including the analysis codes, root 
files, plot-making scripts, are available at the repository \cite{Matousek:2022}.

\appendix
\section{Calculating diffractive differential cross sections}
\label{sec:AppI}
An extremely complicated task is to calculate and generate total cross-sections, for which one has to evaluate complex four-dimensional 
integrals at each phase-space point. But Sar{\em{t}}re uses an approach based on computing the first and second moments of the 
scattering amplitudes separately, and then stores the results in three-dimensional look-up tables, in terms of $Q^{2}$, $W^{2}$ and 
$t$ independent variables. The ultimate outcome is a set of four look-up tables for each nuclear species, each final-state vector meson 
(and DVCS photon), each polarization, and each dipole model (either IPSat or IPNonSat).
\bea
~~~\left<\mathcal{A}_{T}(Q^2, W^2, t)\right>_{\Omega} ,&~~~&\left<\mathcal{A}_{L}(Q^2, W^2, t)\right>_{\Omega} ,
\nonumber\\
~~~\left<|\mathcal{A}_{T}(Q^2, W^2, t)|^2\right>_{\Omega} ,&~~~&\left<|\mathcal{A}_{L}(Q^2, W^2, t)|^2\right>_{\Omega} ,
\label{eq:lookup_tables}
\eea
These look-up tables contain all the physics information from both dipole models. The program also provides tables for calculating the
phenomenological corrections described in Sec.~\ref{sec:corrections}.

The master equation of Sar{\em{t}}re is the total diffractive differential cross section, which for electron-nucleus scattering has 
the following form:
\bea
& & \Lb \frac{{\rm d}^{3}\sigma^{\gamma^{*}A\,\rightarrow\,V A}}{{\rm d}Q^{2}\,{\rm d}W^{2}\,{\rm d}t} 
\Rb_{\rm tot.\,diff.} =
\nonumber\\
& & ~~~~~~~~~~~~= \sum_{T, L}\frac{R_{g}^{2}(1 + B^{2})}{16\pi}\frac{{\rm d}\Gamma_{T, L}}{{\rm d}Q^{2}\,{\rm d}W^{2}} \times
\nonumber\\
& & ~~~~~~~~~~~~\times \left< \left|\mathcal{A}_{T,L}^{\gamma^{*}A\,\rightarrow\,V A}(x_{I\!\!P},Q^{2},t,\Omega)\right|^{2} \right>_{\Omega} ,
\label{eq:master_totalCS}
\eea
where $B$ is given in Eq.~(\ref{eq:realpart_beta}), and $R_{g}$ in Eq.~(\ref{eq:skew}). The quantity $\Gamma_{T, L}$ is the flux of 
transversely $T$ and longitudinally $L$ polarized virtual photons \cite{Breitweg:1998nh,Adloff:1999kg,Smith:1992,Smith:1993,Hand:1963}, 
given by
\bea
& & \Gamma_{T} = \frac{\alpha}{2\pi} \left( \frac{1 + (1 - y)^{2}}{y\,Q^{2}} \right) ,
\nonumber\\
& & \Gamma_{L} = \frac{\alpha}{2\pi} \left( \frac{2(1 - y)}{y\,Q^{2}} \right) ,
\label{eq:flux}
\eea
where $y$ is the inelasticity defined as the fraction of the electron's energy lost in the nucleon rest frame. 
The averaging over configurations $\Omega$ is defined as
\bea
& & \left< \left|\mathcal{A}_{T,L}^{\gamma^{*}A\,\rightarrow\,V A}(x_{I\!\!P},Q^{2},t,\Omega)\right|^{2} \right>_\Omega =
\nonumber\\
& & ~~~~~= \frac{1}{N_{\rm max}}\sum_{j=1}^{N_{\rm max}} \left|\mathcal{A}_{T,L}^{\gamma^{*}A\,\rightarrow\,V A}
(x_{I\!\!P},Q^{2},t,\Omega_{j})\right|^{2} ,
\label{eq:average2}
\eea
where $N_{\rm max}$ is the number of configurations. If it is large enough, then the sum in Eq.~(\ref{eq:average2}) converges to a true
average. It is already shown in \cite{Toll:2012mb} that $2 \times 500$ configurations, 500 for $T$ polarized $\gamma^{\ast}$ and 500 for 
$L$ polarized $\gamma^{\ast}$, give a good convergence. Such that there are 1000 such integrals for each $(Q^{2}, W^{2}, t)$ 
phase-space point. 

The photon flux may emanate from electrons, as in the case of $e+p$ and $e+A$ scatterings, however, it may be also radiated from protons 
or nuclei \cite{Klein:1999gv}, as in the case of $A+A$ or $p+A$ UPC. A phase-space point together with a given beam energy fully 
determines the final state of a produced vector meson, except for its azimuthal angle, which is distributed uniformly.

The second moment of the amplitude in Eq.~(\ref{eq:average2}) for the nucleon configurations $\Omega_{j}$ can be calculated based 
upon \cite{Toll:2013gda,Toll:2012mb}:
\begin{displaymath}
\left< \left|\mathcal{A}_{T,L}^{\gamma^{*}A\,\rightarrow\,V A}(x_{I\!\!P},Q^{2},t,\Omega)
\right|^{2}\right>_{\Omega} =
\nonumber\\
\end{displaymath}
\bea
& & ~~~= \sum_{j=1}^{N_{\rm max}}\,\left| \int{\rm d}^{2}{\bf r_{T}} \int{\rm d}^{2}\bt
\int\frac{{\rm d} z}{4\pi} \,\times \right.
\nonumber\\
& & ~~~~~~~\times \left. \left[\Lb \Psi^{*}\Psi_{V} \Rb_{T,L}\!(Q^{2},\rt, z)\right] J_{0}([1 - z]r_{T}\Delta) \,\times \right.
\nonumber\\
& & ~~~~~~~\times \left. e^{-i\bt \cdot \Deltat}\,
\frac{{\rm d}\sigma_{q\bar q}^{A}}{{\rm d}^{2}\bt}(x_{I\!\!P}, \rt, \bt, \Omega_{j})\right|^{2} ,
\label{master_totalAmp}
\eea
where the last term ${\rm d}\sigma_{q\bar q}^{A}/{\rm d}^{2}\bt$ is defined in Eq.~(\ref{eq:bSateA}). 

Thus, Eqs.~(\ref{eq:master_totalCS}) and (\ref{master_totalAmp}) determine the total diffractive differential cross section. Its coherent 
part is given by
\bea
& & \Lb \frac{{\rm d}^{3}\sigma^{\gamma^{*}A\,\rightarrow\,V A}}{{\rm d}Q^{2}\,{\rm d}W^{2}\,{\rm d}t} \Rb_{\rm coh.\,diff.} = 
\nonumber\\
& & ~~~~~= \sum_{T, L} \frac{R_{g}^{2}(1 + B^{2})}{16\pi}\frac{{\rm d}\Gamma_{T, L}}{{\rm d}Q^{2}\,{\rm d}W^{2}} \times
\nonumber\\
& & ~~~~~~~~~\times \left| \left<\mathcal{A}_{T,L}^{\gamma^{*}A\,\rightarrow\,V A}(x_{I\!\!P},Q^{2},t,\Omega)\right>_{\Omega} \right|^{2} ,
\label{eq:coherentCS}
\eea
For the first moment of the amplitude, the integral to calculate will be
\bea
& & \!\!\left| \left<\mathcal{A}_{T,L}^{\gamma^{*}A\,\rightarrow\,V A}(x_{I\!\!P},Q^{2},t,\Omega)\right>_{\Omega} \right|^{2} =
\left| \int{\rm d}^{2}{\bf r_{T}} \int{\rm d}^{2}\bt \,\times \right.
\nonumber\\
& & ~~\left. \times \int\frac{{\rm d} z}{4\pi} \left[\Lb \Psi^{*}\Psi_{V} \Rb_{T,L}\!(Q^{2},\rt, z)\right] J_{0}([1 - z]r_{T}\Delta) \times \right.
\nonumber\\
& &  ~~\left. \times \,J_{0}(b_{T}\Delta) \left<\frac{{\rm d}\sigma_{q\bar q}^{A}}{{\rm d}^{2}\bt}(x_{I\!\!P}, \rt, \bt, \Omega)\right>_{\Omega} \right|^{2} ,
\label{eq:coherentAmp}
\eea
where the average $\left< {\rm d}\sigma_{q\bar q}^{A}/{\rm d}^{2}\bt \right>_{\Omega}$ in the last term is defined in 
Eq.~(\ref{eq:analytical}).

Thereby, as in Eq.~(\ref{eq:gamma_A_incohdiff}), the incoherent part of the total diffractive differential cross section is taken to be the 
difference between the total and coherent cross sections:
\bea
& & \!\!\!\Lb \frac{{\rm d}^{3}\sigma^{\gamma^{*}A\,\rightarrow\,V A}}{{\rm d}Q^{2}\,{\rm d}W^{2}\,{\rm d}t} \Rb_{\rm incoh.\,diff.} = 
\Lb \frac{{\rm d}^{3}\sigma^{\gamma^{*}A\,\rightarrow\,V A}}{{\rm d}Q^{2}\,{\rm d}W^{2}\,{\rm d}t} \Rb_{\rm tot.\,diff.} - 
\nonumber\\
& &~~~~~~~~~~~~~~~~~~~~~~~~~~~~~
- \Lb \frac{{\rm d}^{3}\sigma^{\gamma^{*}A\,\rightarrow\,V A}}{{\rm d}Q^{2}\,{\rm d}W^{2}\,{\rm d}t} \Rb_{\rm coh.\,diff.} .
\label{eq:incoherentCS}
\eea
The incoherent part directly gives the probability for the nuclear breakup.

\section{Error analysis of the pseudo-data using the EIC Detector Handbook}
\label{sec:AppII}
We use Sar{\em t}re 1.33, an exclusive event generator, to produce vector meson production data. A user-edited runcard is called at 
the initialization of a simulation to set beam energies, decay modes, and ranges on event kinematics such as $Q^{2}$ and $|t|$. The 
generator outputs the kinematics of final-state particles, such as their momentum and pseudorapidity. Additionally, important event 
information such as $Q^{2}$, $\xpom$, and $y$ are recorded. We refer to the event generator output as \texttt{truth} data. This data, 
while unobtainable in a physical experiment, allows us to perform perfect event identification and to create pseudo-data. Using detector 
resolutions outlined in the EIC Detector Handbook \cite{Handbook}, true particle kinematics are smeared to create pseudo-data. For instance, 
according to the handbook, the barrel ($|\eta|<1$) tracking resolution for electrons is $\sigma_p/p = 0.05\%p + 0.5\%$. By having such 
resolutions written out explicitly for the relevant final-state particles from vector meson production simulations allows us to calculate 
pseudo-data based on the handbook's projections. The smearing functions immediately produce smeared kinematics, such as momentum and energy. 
Event-by-event, we manually smear the final-state kinematics according to these functions. Subsequently, the pseudo-data reconstruction of 
the scattered electron is used to determine event kinematics such as $Q^{2}$. The event $|t|$ is the only quantity which we smear 
independently. One of the immediate effects of generating pseudo-data is the reduction of statistics. Final-state particles, which exit 
at pseudorapidities beyond the coverage of the detector system, render the exclusive event unrecoverable. Systematic errors introduced 
through particle/event mis-identification are not accounted for when analyzing the pseudo-data. Instead, we opt to use true, event 
generator information to identify the scattered electron and decay particles. Lastly, we use \texttt{truth} information to separate coherent vector 
meson production from the incoherent one, as well as events with transversely polarized virtual photons from longitudinally polarized 
virtual photons.

After the pseudo-data is generated, we are left with the reconstructed particle and event kinematics for a pile of production events. Then, 
the data is stored event-by-event in a \textit{ROOT TTree}, which is read and analyzed to produce plots. From the \textit{TTree}, the analyzed 
data is used to fill \textit{ROOT} histogram objects. These histogram objects, initially storing the number of events per bin, are scaled in 
a variety of ways. This includes scaling to the correct decay mode branching ratios, scaling to the desired luminosity, and scaling by bin 
size (to create differential cross sections). What follows is a detailed look at how the pseudo-data is analyzed, how this analysis is piped 
into histograms, how they are scaled, and how the error is propagated.


\paragraph{Pseudo-data and error propagation for Figs.~\ref{fig:white_plot2}~and~\ref{fig:white_plot5}.}
After the pseudo-data is generated, the following cuts are placed event-by-event:
\begin{itemize}
\item Reconstructed $\xpom < 0.01$.

\item Reconstructed $1.0 < Q^{2} < 10.0\,{\rm GeV}^{2}$.

\item True event coherent vector meson production.

\item Reconstructed final-state particle pseudorapidity between $-3.5 < \eta < 3.5$\footnote{In the original 
Figure~54 of the EIC White Paper \cite{Accardi:2012qut}, the final-state particle pseudorapidity is restricted 
between $-4.0 < \eta < 4.0$. In our case, the Handbook smearing functions from \cite{Handbook} are implemented 
to cover tracking up to $-3.5 < \eta < 3.5$.}.

\item Reconstructed final-state particle momentum greater than $1$\,GeV.
\end{itemize}
Then, using true event $|t|$ information outputted by Sar{\em t}re, we generate the reconstructed $|t|$ 
using the \textit{Gaus()} function from the \textit{ROOT TRandom} class. For each event, we select a 
random variable from a Gaussian centered at true $|t|$ with spread equal to true $|t|\times \mathit{tsmear}$. 
This random variable represents the reconstructed $|t|$ of the event. The quantity $\mathit{tsmear}$ 
is changed depending on our desired resolution in $|t|$. It ranges from $0$ all the way to $0.3\,{\rm GeV}^{2}$, 
or is parameterized by the ``Method L" in \cite{AbdulKhalek:2021gbh}. Once all of the events have been parsed through 
and either added or discarded, a \textit{TGraphErrors} object is created. This object will pick up information 
from the filled histogram and produce a plot.

Below are the individual modifications we make to the entries of the filled histogram, given in order. 
Beforehand, we define $\mathrm{N}_{\mathrm{entries}}$ to be the number of total entries in the histogram, 
and $w$ to be the amount of entries in an arbitrary bin.
\begin{enumerate}
\item First, we multiply each bin of the histogram by the quantity $\sigma \cdot L/A$, where $\sigma$ is 
the total event cross section (in nanobarns) outputted by Sar{\em t}re, and $L$ is the integrated luminosity. 
We also divide by the atomic number $A$ of a nucleus to isolate the scattering cross section of a single 
nucleon for a given nucleus beam.

\item Then, we multiply each bin by the branching ratio (BR) of the \texttt{truth} decay mode.

\item Next, we divide each bin by $\mathrm{N}_{\mathrm{entries}}$. And now for each bin, we are left with 
\textit{the number of events expected in an experiment with integrated luminosity $L$, within the 
reconstructed $|t|$ bin range}:
\begin{equation}
\mathrm{N}_{\mathrm{expected}} \equiv w' = \frac{\mathrm{BR} \times L\times \sigma \times w}{\mathrm{N}_{\mathrm{entries}} 
\times A}.
\label{eq:Nexp}
\end{equation}

\item We  further calculate the statistical uncertainty for each bin as the square root of its entries, $\sigma_{w'} = \sqrt{w'}$.

\item Afterwards, we divide each bin quantity $w$ by the branching ratio, the bin width $\Delta t$, and integrated 
luminosity. We repeat this for $\sigma_{w}$ as well.
\end{enumerate}
\bea
& & \frac{d\sigma}{dt}{\mbox{[pseudo-data]}} = \frac{A}{\mathrm{BR} \times \Delta t \times L}
\mathrm{N}_{\mathrm{expected}} , 
\nonumber \\
& & \frac{d\sigma}{dt}{\mbox{[pseudo-data~statistical~uncertainty]}} = 
\nonumber\\
& & ~~~~~~~~~~~~~~~~~~~~~~= \frac{A}{\mathrm{BR}\times \Delta t \times L}
\sqrt{\mathrm{N}_{\mathrm{expected}}} .
\label{eq:sigma_analysis}
\eea
This step completes the pseudo-data and error analysis for Figs.~\ref{fig:white_plot2}~and~\ref{fig:white_plot5}.

\begin{figure*}[h!]
\centering
\includegraphics[width=17.5cm]{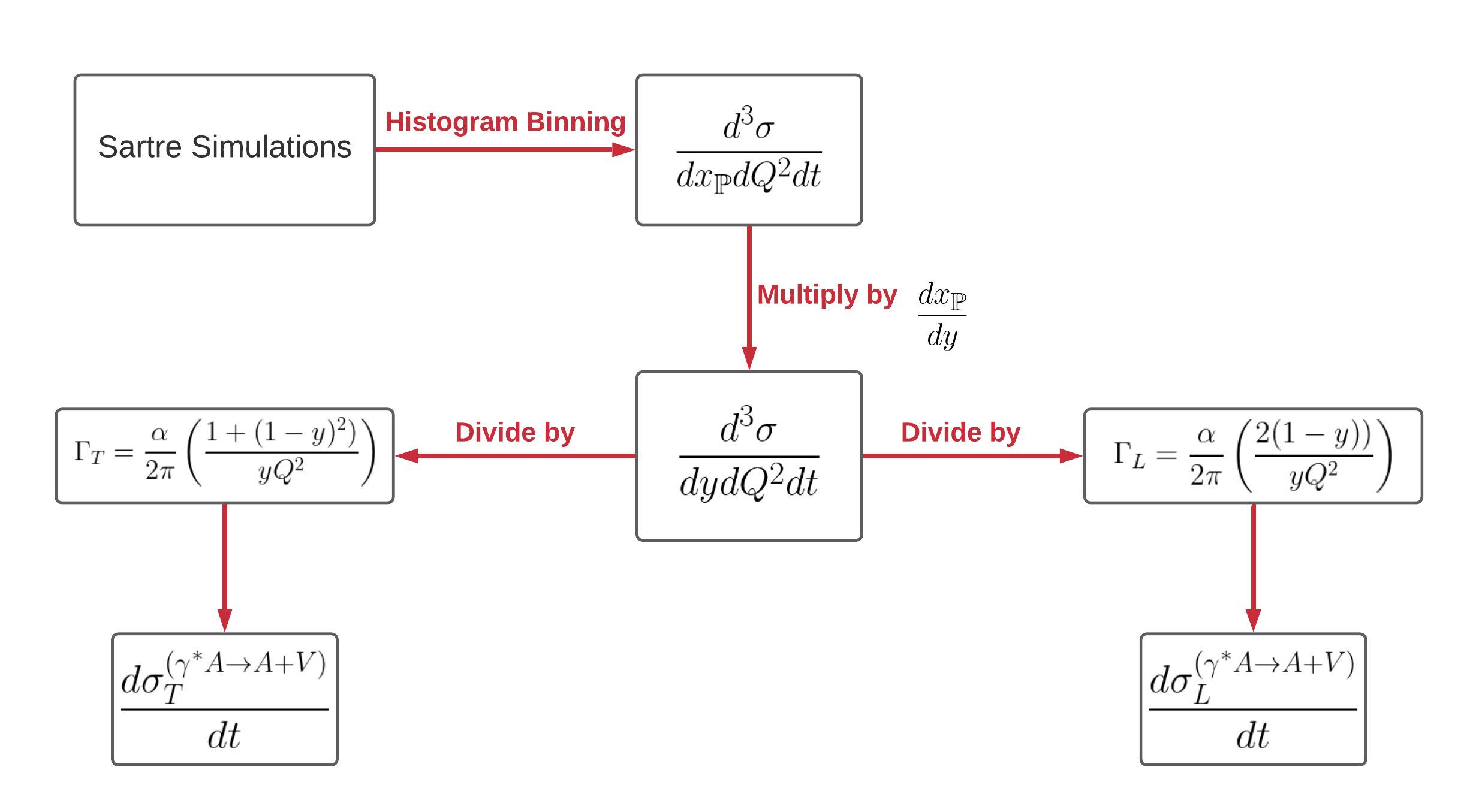}
\caption{Cross section conversion scheme for obtaining the scaling plots in Figs.~\ref{fig:Scl_NonSat_fig}-\ref{fig:Scl_Sat_fig6}
based upon Sar{\em t}re 1.33 simulations.}
\label{fig:cross-section}
\end{figure*}

\paragraph{Pseudo-data and error propagation for Figs.~\ref{fig:Scl_NonSat_fig}-\ref{fig:Scl_Sat_fig6}.}
The analysis of these figures differ from the above method in subtle ways\footnote{The bin size in these figures changes 
as a function of $Q^2$. When the pseudo-data are used to fill those asymmetrically-sized bins, the error bars in each bin 
are calculated accordingly.}. For each collision type ($e+p$ and/or $e+A$), 
vector meson produced ($J/\Psi$ or $\phi$), and dipole model used (IPSat or IPNonSat), we generate $N = 10^{8}$ exclusive 
events using Sar{\em t}re. At the beginning of the simulation, we save the total cross section of the simulation’s phase space. 
Using this cross section, the projected EIC luminosities ($100\,{\rm fb}^{-1}/A$ for $e+p$ at $10 \times 100\,{\rm GeV}$, 
$10\,{\rm fb}^{-1}/A$ for $e+A$ at $10 \times 110\,{\rm GeV}$), and the branching ratios of the studied vector mesons' decay 
modes, we calculate the number of expected events produced at the EIC. This number, we call $n$, can be divided by $N$ to 
tell us how much we should scale our simulation size to reflect the specific luminosity. We call this parameter $\mbox{Scale} = n/N$. 

Event by event, we use the true event generator output to determine if the event was coherent or incoherent. Skipping incoherent 
events, we then store the virtual photon polarization for an event, also given by Sar{\em t}re. We place a hard 
cut on the true event $y$ to minimize the impact of poor smearing in that kinematic regime. 
By using detector uncertainties projected in \cite{Handbook}, we smear the energy, momentum, and polar angle of the final-state electron 
and decay products. We skip events in which any one of these three-final state particles’ true pseudorapidity falls outside the 
reach of the detector system’s coverage. After the smearing is completed, we use the true beam energy of the incoming electron 
and nuclear/proton target to calculate event kinematics such as $Q^{2}$, $y$, and $\xpom$. We smear the event’s $t$ in a way described 
in Sec.~\ref{sec:Scaling}, skipping events with $y < 0.05$.

When filling our histograms, we weigh the fill by the factor
\begin{equation}
\mbox{Scale} \times \frac{d\xpom}{dy} \frac{Q^{m}}{\Gamma_{T, L}} ,
\label{eq:xpom_dy}
\end{equation}
where $m$ is determined by the specific $Q^{2}$ scaling being analyzed, the photon flux factor $\Gamma_{T, L}$ depends on the 
polarization of the event, and the derivative is given by
\begin{equation}
\frac{d\xpom}{dy} = \frac{s \times \xpom}{M_{V}^{2} + Q^{2} - t} .
\label{eq:xpom_dy2}
\end{equation}
where $s$ is the center-of-mass energy squared. The derivative $d\xpom/dy$ and $\Gamma_{T, L}$ factor effectively divide out the 
virtual photon flux element of the event cross section, leaving us with (see Fig.~\ref{fig:cross-section}) a beam-independent cross section 
$\sigma \Lb \gamma^{*} + p/A\,\rightarrow\,V + p^{\prime}/A^{\prime} \Rb$. The scale factor is included to reflect the number of 
events expected, given the experimental luminosity. If we had run Sar{\em t}re and generated exactly the number of events expected 
in a given luminosity, then the Scale would equal $1$. Each time a histogram’s bin is filled, ROOT updates the statistical error of 
the bin to be the summation in quadrature of all the bin’s individual weights. 

After all events have been simulated, we divide each bin of each histogram by the factor $\mathrm{BR} \times L/A$. We then divide 
out by the $Q^{2}$ bin width, $\xpom$ range, and proper nuclear size $A^{k}$ scaling, which leaves the bin content being equivalent 
to $A^{-k}\,Q^{m}\,\sigma^{T, L}$. To obtain the differential cross section, we finally divide by the $t$ binning, which is equal 
$0.001$. The proper nuclear size $A$ scaling and is applied to each histogram depending on $x$. For the non-zero $|t|$ scaling plots in 
Figs.~\ref{fig:Scl_Sat_fig_t1}-\ref{fig:Scl_Sat_fig_t4}, we additionally divide the vertical axis by $G(A,t)$ (see Eq.~(\ref{eq:scl_13})).

\section{Addendum to Sec.~\ref{sec:Scaling1a}}
\label{sec:AppIII}

\begin{figure}[h!]
\begin{center}
   \subfigure{\includegraphics[width=0.475\textwidth,height=0.430\textwidth]{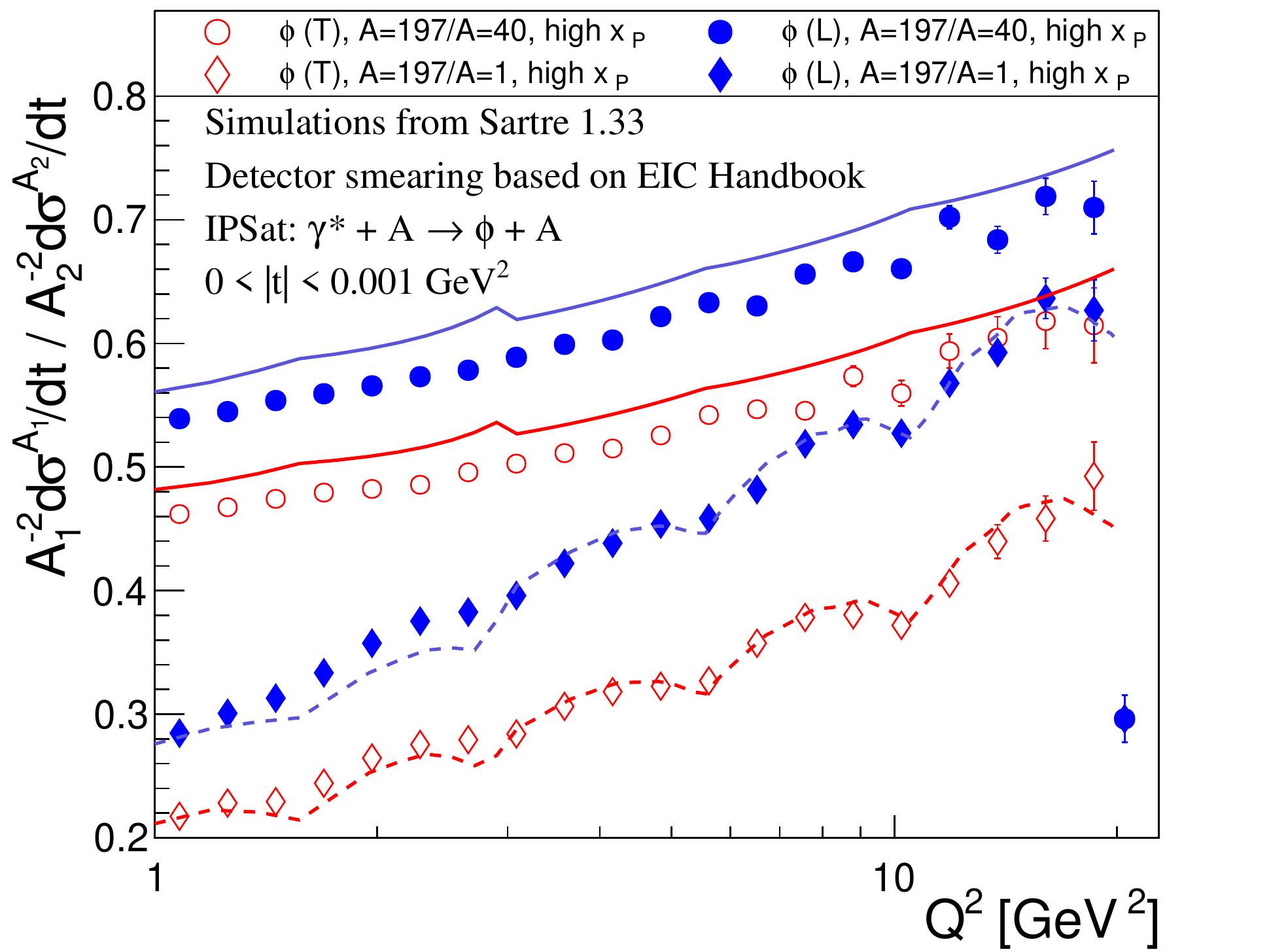}}
\hspace{0.0\textwidth}
   \subfigure{\includegraphics[width=0.475\textwidth,height=0.430\textwidth]{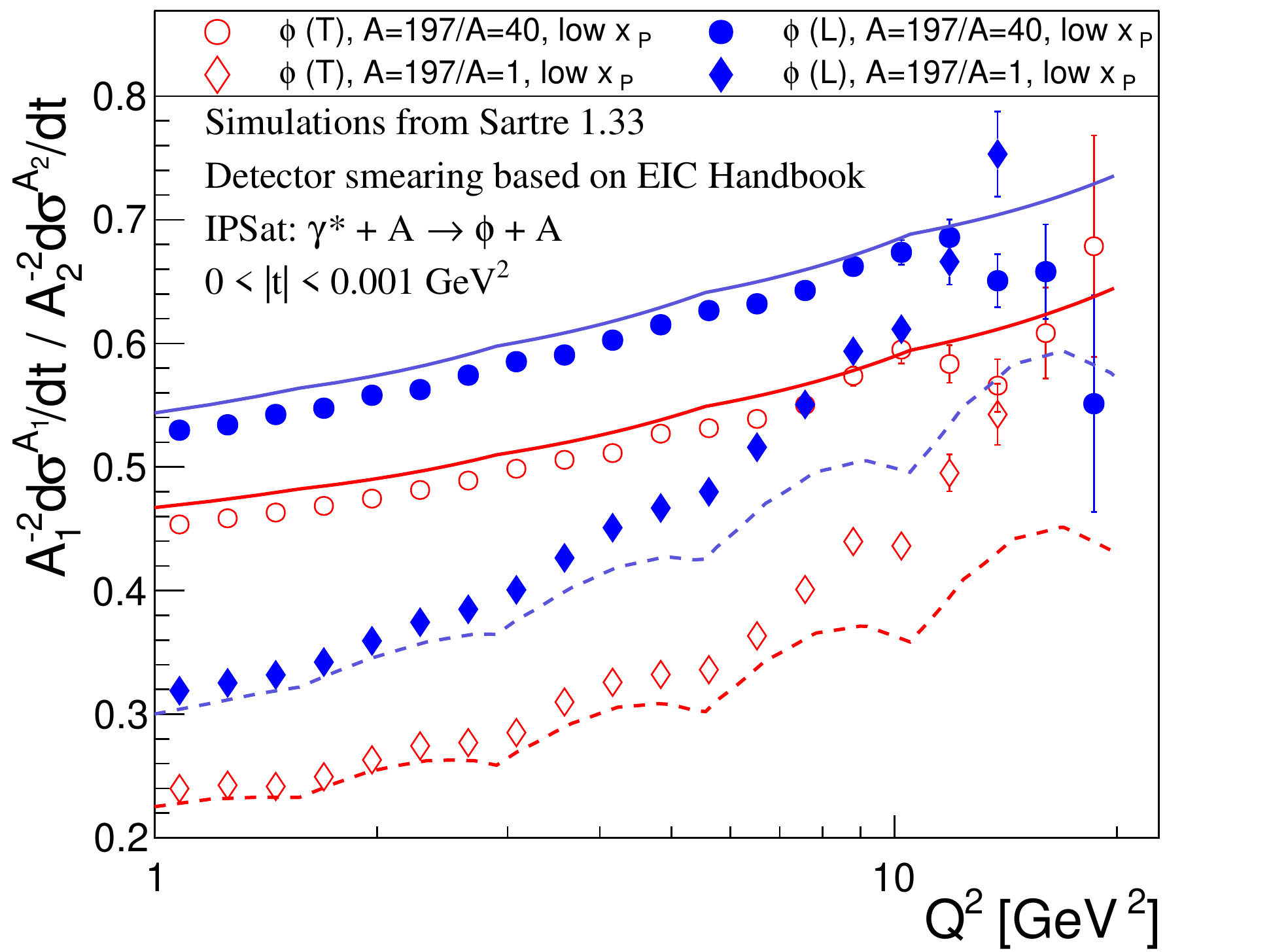}}
\end{center}
\vspace{-0.25cm}
\caption{(Color online)
In the top and bottom panels, the pseudo-data and labeling are exactly the same as in Fig.~\ref{fig:Scl_Sat_fig1}. In addition, there are 
also overlaid curves obtained directly from Sar{\it t}re's IPSat model, where any detector smearing/resolution and experimental acceptance 
considerations are excluded in the normalized cross-section ratio computations.}
\label{fig:Scl_Sat_fig1b}
\end{figure}

\begin{figure}[hbt!]
\begin{center}
   \subfigure{\includegraphics[width=0.475\textwidth,height=0.430\textwidth]{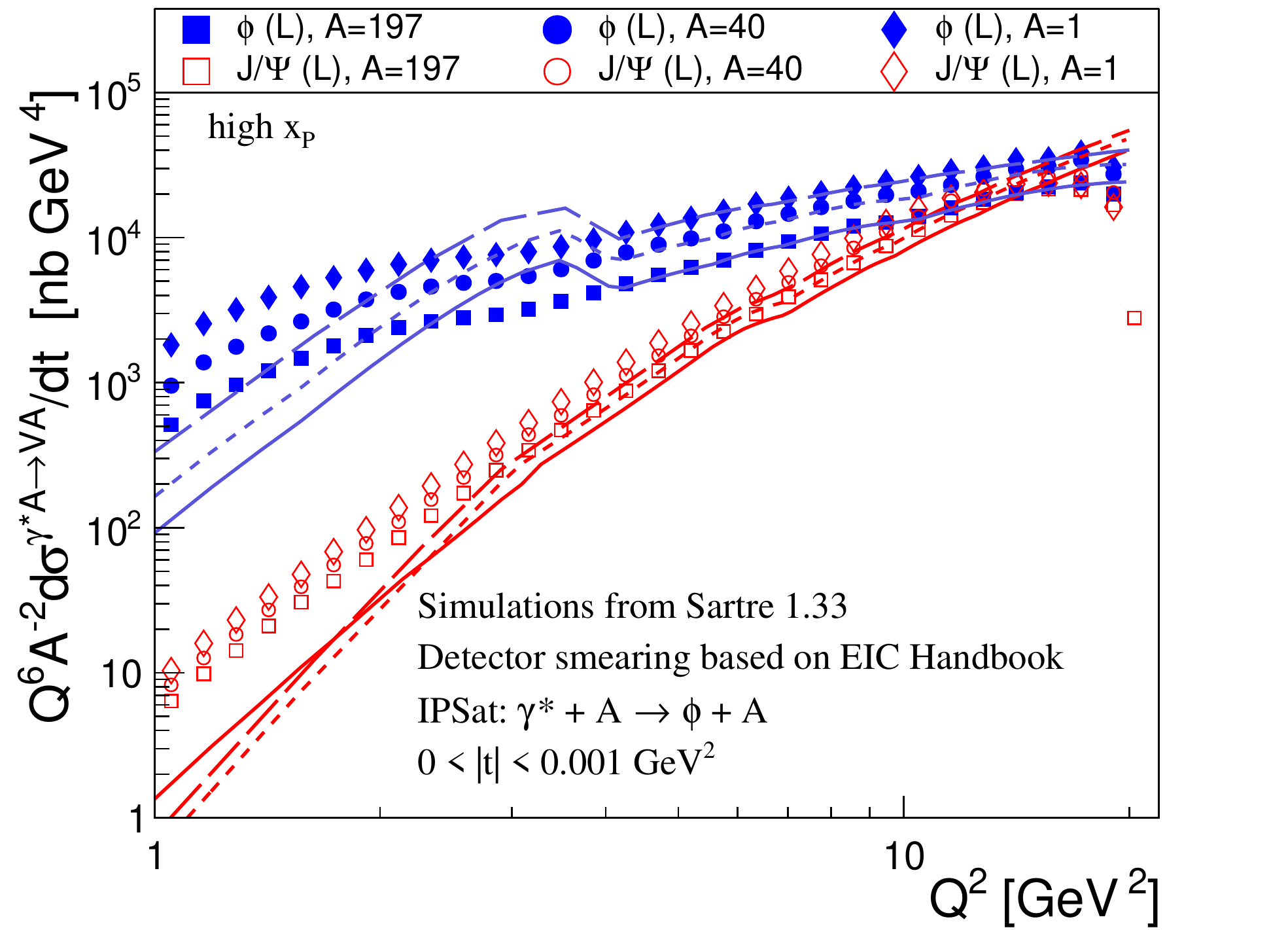}}
\hspace{0.0\textwidth}
   \subfigure{\includegraphics[width=0.475\textwidth,height=0.430\textwidth]{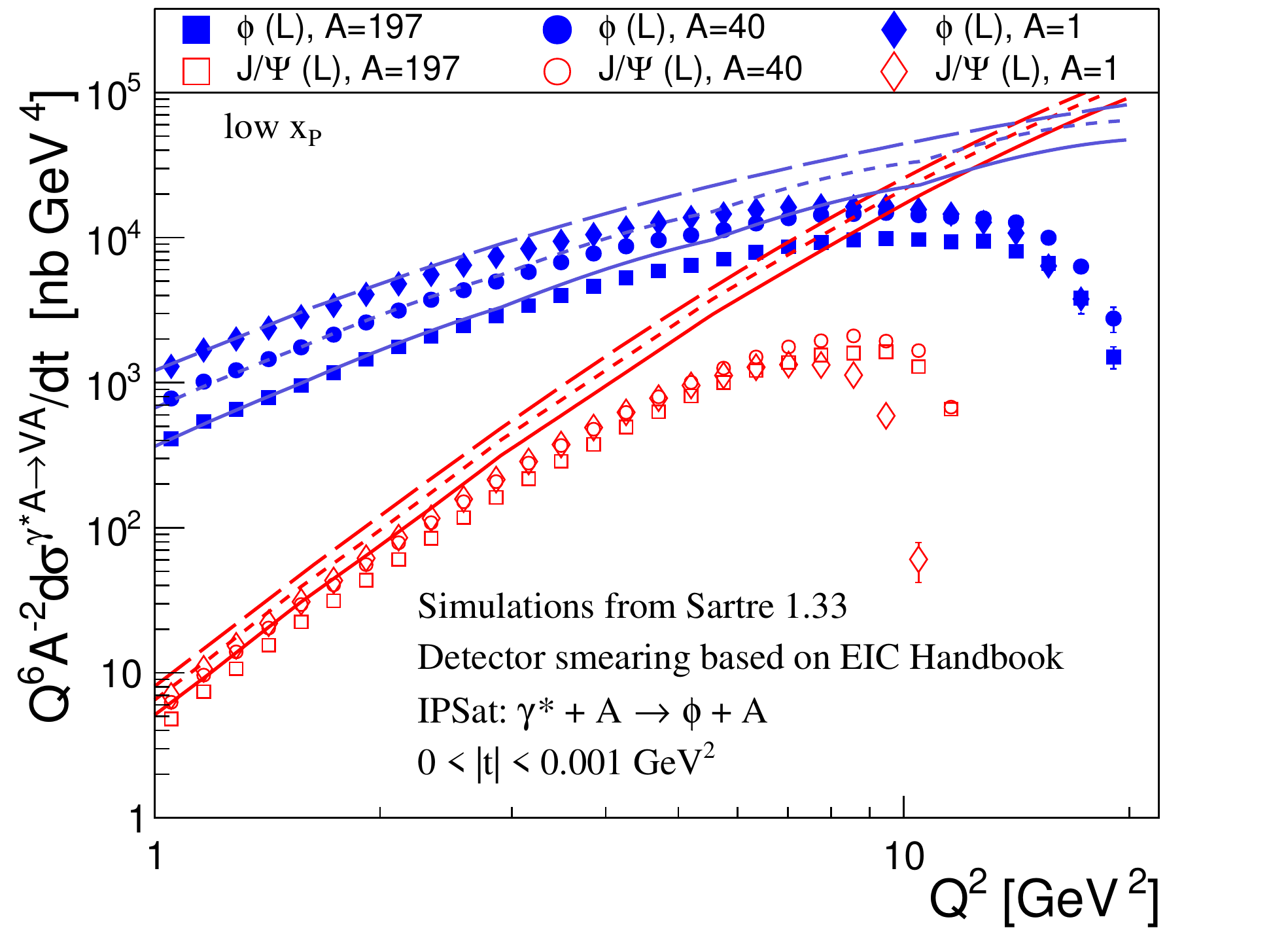}}
\end{center}
\vspace{-0.25cm}
\caption{(Color online)
In the top and bottom panels, the pseudo-data and labeling are exactly the same as in Fig.~\ref{fig:Scl_Sat_fig3}. The explanation for 
the overlaid curves is similar as in Fig.~\ref{fig:Scl_Sat_fig1b}.}
\label{fig:Scl_Sat_fig3b}
\end{figure}

In this appendix, we show qualitative comparisons between model calculations and simulated pseudo-data on the $A^{2}$ normalized cross-section 
ratio and normalized $A^{2}/Q^{6}$ cross section shown in Sec.~\ref{sec:Scaling1a}, in Fig.~\ref{fig:Scl_Sat_fig1} and Fig.~\ref{fig:Scl_Sat_fig3}, 
respectively. The pseudo-data is obtained by passing events generated by Sar{\it t}re through particle kinematics smearing functions with detector acceptance. 
In these plots, the high $\xpom$ denotes the range of pseudo-data made in $\xpom = [0.005 - 0.009]$, and the low $\xpom$ denotes the range of $\xpom = [0.001 - 0.005$]. 
So, Fig.~\ref{fig:Scl_Sat_fig1b} and Fig.~\ref{fig:Scl_Sat_fig3b} have their top and bottom panels being the same as in Fig.~\ref{fig:Scl_Sat_fig1} 
and Fig.~\ref{fig:Scl_Sat_fig3}, respectively. We also display \texttt{truth} curves corresponding to all sets of the shown pseudo-data that 
are calculated using the IPSat model in Sar{\it t}re. The ``kink'' structures seen in the curves around $\sim 3~{\rm GeV^{2}}$ in the top 
high-$\xpom$ plots, which describe the $\phi$ normalized cross-section ratio and cross section, partially come from the fact that these curves 
are actually calculated at fixed value of $\xpom = 0.008$. This region of $\xpom$ is close to 0.01, the point at which Sar{\it t}re's reliability starts to fail, especially 
for light vector meson production. Lastly, the low-$\xpom$ curves are calculated at 0.0012.

Overall, the need for additional reconstruction algorithms is evidenced by these figures. The comparison between the \texttt{truth} 
curves and pseudo-data should thus at this point stay qualitative. A future analysis will consider bin-by-bin corrections, such as detection efficiency and bin migration, 
to alleviate the discrepancies between the Monte Carlo and pseudo-data cross-section reconstructions. Although one may expect these 
systematic effects to ``divide themselves out" in the ratio plots, the differing center-of-mass energies planned for the $e+p$ and $e+A$ runs 
at the future EIC makes this assumption false. During this paper's composition, the EIC collaboration has begun narrowing down a single 
detector design for the first interaction region. Coined EPIC, the detector system and relevant software is currently in development, 
making our future analysis of detector effects for the continuation of the study in this paper extremely timely.


\end{document}